\documentclass[journal=jctcce, manuscript=article]{achemso}

\usepackage{chemformula} 
\usepackage[T1]{fontenc} 
 \usepackage[colorlinks=true,linkcolor=blue,urlcolor=blue,citecolor=blue]{hyperref}
\usepackage{array}
\usepackage{lipsum}
\usepackage{bm}
\usepackage{subfigure}
\usepackage{amssymb}
\usepackage{multirow}
\usepackage{tabularx}
\usepackage{amsmath}
\usepackage{braket}
\usepackage{caption}
\usepackage{float}
\usepackage{caption}
\usepackage{enumerate}
\usepackage{ulem}

\usepackage{caption}

\usepackage{MnSymbol}

\SectionsOn
\SectionNumbersOn

\author{Tobias Dornheim}
\email{t.dornheim@hzdr.de}
\affiliation{Center for Advanced Systems Understanding (CASUS), D-02826 G\"orlitz, Germany}
\alsoaffiliation{Helmholtz-Zentrum Dresden-Rossendorf (HZDR), D-01328 Dresden, Germany}
\author{Zhandos Moldabekov}
\affiliation{Center for Advanced Systems Understanding (CASUS), D-02826 G\"orlitz, Germany}
\alsoaffiliation{Helmholtz-Zentrum Dresden-Rossendorf (HZDR), D-01328 Dresden, Germany}
\author{Sebastian Schwalbe}
\affiliation{Center for Advanced Systems Understanding (CASUS), D-02826 G\"orlitz, Germany}
\alsoaffiliation{Helmholtz-Zentrum Dresden-Rossendorf (HZDR), D-01328 Dresden, Germany}
\author{Panagiotis Tolias}
\affiliation{Royal Institute of Technology (KTH) Stockholm, SE-100 44 Stockholm, Sweden}
\author{Jan Vorberger}
\affiliation{Helmholtz-Zentrum Dresden-Rossendorf (HZDR), D-01328 Dresden, Germany}

\title{Fermionic free energies from \textit{ab initio} path integral Monte Carlo simulations of fictitious identical particles}

\abbreviations{IR,NMR,UV}

\begin{document}

\abstract{
We combine the recent $\eta-$ensemble path integral Monte Carlo (PIMC) approach to the free energy [T.~Dornheim \textit{et al.}, \textit{Phys.~Rev.~B} \textbf{111}, L041114 (2025)] with a recent fictitious partition function technique based on inserting a continuous variable that interpolates between the bosonic and fermionic limits [Xiong and Xiong, \textit{J.~Chem.~Phys.}~\textbf{157}, 094112 (2022)] to deal with the fermion sign problem. As a practical example, we apply our set-up to the warm dense uniform electron gas over a broad range of densities and temperatures. We obtain accurate results for the exchange--correlation free energy down to half the Fermi temperature, and find excellent agreement with the state-of-the-art parametrization by Groth \textit{et al.}~[\textit{Phys.~Rev.~Lett.}~\textbf{119}, 135001 (2017)]. Our work opens up new avenues for the future study of a host of interacting Fermi-systems, including warm dense matter, ultracold atoms, and electrons in quantum dots.
}

\begin{figure}
\center
\includegraphics[width=7.5cm]{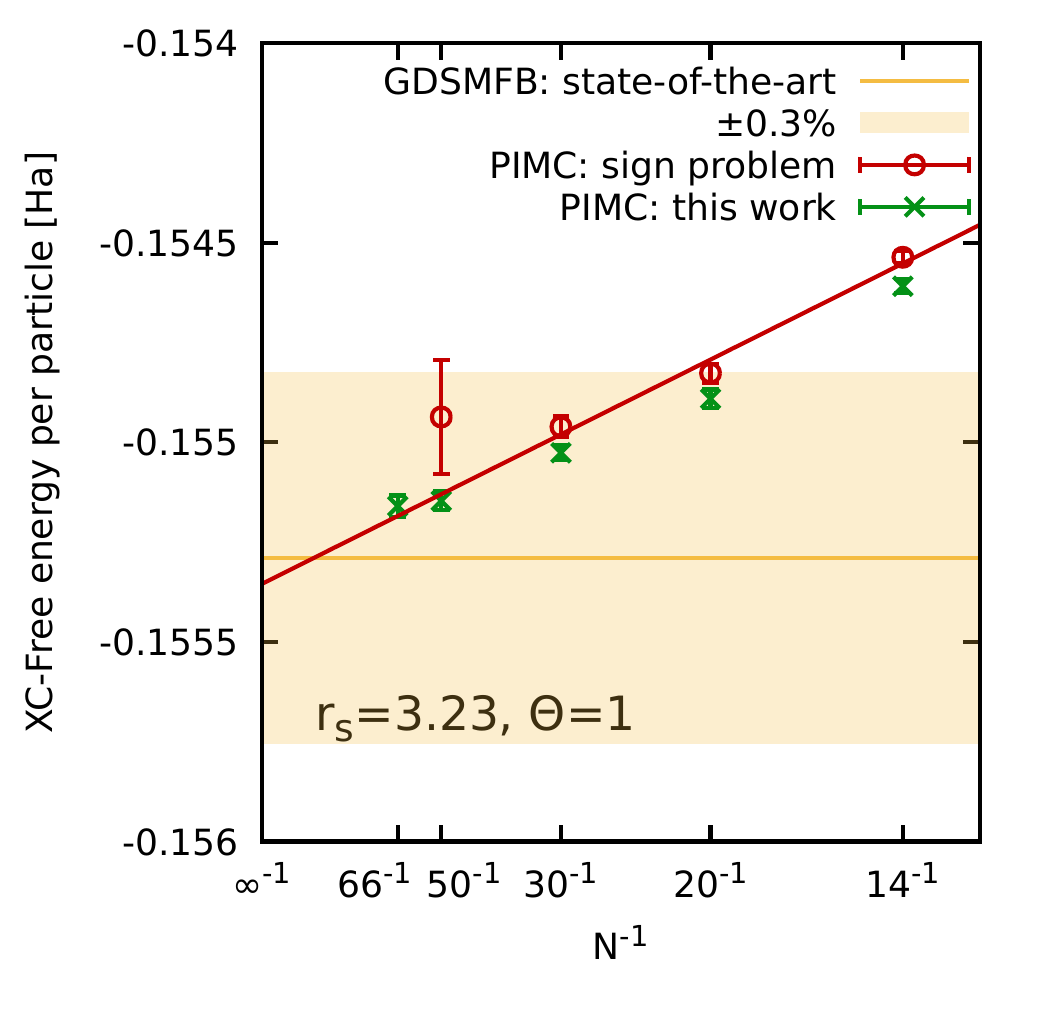}
\captionsetup{labelformat=empty}
\caption{ \label{fig:TOC} Table of Contents (TOC)/Abstract graphic. }
\end{figure} 
\addtocounter{figure}{-1}

\section{Introduction}

The \textit{ab initio} path integral Monte Carlo (PIMC) technique~\cite{cep} constitutes one of the most important methods for the simulation of interacting quantum many-body systems within physics, quantum chemistry and related disciplines. Based on Feynman's imaginary-time path-integral formalism~\cite{kleinert2009path}, modern sampling schemes facilitate quasi-exact (asymptotically exact within the given Monte Carlo error bounds) simulations of $N\sim10^4$ bosons~\cite{boninsegni1,boninsegni2}, which has substantially advanced our understanding of fundamental collective effects such as superfluidity and Bose-Einstein condensation~\cite{cep,Filinov_PRL_2010,morresi2025revisitingpropertiessuperfluidnormal}. In practice, PIMC simulations give one direct access to a host of physical observables, including pair correlation functions and static structure factors~\cite{cep,Brown_PRL_2013,dornheim_prl}, different energy contributions and pressures~\cite{Militzer2021EOS,Brown_PRL_2013,Bonitz_POP_2024}, as well as linear and non-linear static density response properties~\cite{dornheim_ML,Dornheim_JCP_ITCF_2021,Dornheim_PRL_2020,Dornheim_review}.
Moreover, the analytic continuation of various imaginary-time correlation functions allows, in principle, for the estimation of dynamic properties~\cite{Boninsegni_maximum_entropy,Filinov_PRA_2012,Ferre_PRB_2016,dornheim_dynamic,chuna2025dualformulationmaximumentropy,Vitali_PRB_2010}, although the required inversion, e.g.~of a two-sided Laplace transform, constitutes a notoriously difficult inverse problem~\cite{Jarrel_PhysReports_1996}. While the direct estimation of the free energy $F=-\beta^{-1}\textnormal{log}\left(Z\right)$, where $Z=Z(N,\Omega,\beta)$ and $\beta=1/k_\textnormal{B}T$ are the canonical partition function (assuming a cubical simulation cell of volume $\Omega=L^3$) and inverse temperature, is not possible, the introduction of generalized configuration spaces~\cite{Lyubartsev_JCP_1992,boninsegni1,Dornheim_PRBL_2025,dornheim2024etaensemblepathintegralmonte} allows for the straightforward estimation of free energy differences, e.g.~between the interacting system of interest and a non-interacting reference system for which the free energy is already known exactly~\cite{Zhou_2018}.

Unfortunately, PIMC simulations of quantum degenerate fermions are afflicted with the \textit{fermion sign problem} (FSP)~\cite{troyer,dornheim_sign_problem,Dornheim_JPA_2021}; it leads to an exponential increase in the required compute time e.g.~with increasing $N$ or decreasing $T$. Thus, the FSP often prevents the direct utilization of PIMC to study the wealth of interesting physical phenomena in interacting many-fermion systems, such as the formation of Wigner molecules~\cite{Egger_PRL_1999,Reimann_PRB_2000} and Wigner crystals at low densities~\cite{Clark_PRL_2009,Azadi_PRB_2022} and the exchange--correlation (XC) induced incipient roton feature in strongly coupled Fermi liquids at intermediate wavenumbers~\cite{Takada_PRB_2016,koskelo2023shortrange,dornheim_dynamic,Dornheim_Nature_2022,Godfrin2012}. Due to the pressing need to accurately describe quantum degenerate fermions, a number of methodological advances~\cite{Brown_PRL_2013,Militzer_PRL_2015,Schoof_PRL_2015,Dornheim_NJP_2015,Malone_PRL_2016,Joonho_JCP_2021,Hirshberg_JCP_2020,Dornheim_Bogoliubov_2020,Yilmaz_JCP_2020,Filinov_PRE_2023} have been presented over the last decade that exhibit different strengths and weaknesses; see, e.g., Ref.~\cite{Bonitz_POP_2024} for a topical overview.

Here we focus on and combine two interesting recent ideas. The first idea by Xiong and Xiong concerns the circumvention of the FSP in path integral molecular dynamics simulations of fermions based on the controlled extrapolation over a continuous partition function variable $\xi$~\cite{Xiong_JCP_2022,Xiong_PRE_2023,Xiong_PRE_2024}. This idea was subsequently adapted to fermionic PIMC simulations~\cite{Dornheim_JCP_xi_2023} and successfully applied to various systems such as the uniform electron gas (UEG)~\cite{Dornheim_JCP_xi_2023,Dornheim_JPCL_2024}, warm dense hydrogen and beryllium~\cite{Dornheim_JCP_2024,Dornheim_Science_2024,dornheim2024modelfreerayleighweightxray}, as well as ultracold $^3$He~\cite{morresi2025normalliquid3hestudied}. On the one hand, this $\xi$-extrapolation method allows for large fermionic PIMC simulations of weakly to moderately degenerate fermions without the exponential scaling with respect to the system size, culminating in simulations of up to $N=1000$ electrons~\cite{Dornheim_JPCL_2024}. On the other hand, the application of the $\xi$-extrapolation method to strongly degenerate phase diagram regions has proven to be less straightforward~\cite{Dornheim_JCP_xi_2023,Xiong_PRE_2023}. The second idea by the present authors concerns the utilization of the extended $\eta$-ensemble scheme that allows for the estimation of the free energy of quantum many-fermion systems by combining a number of bosonic, sign-problem free PIMC simulations with the correct fermionic expectation value of $F$ via the \textit{average sign} $S$~\cite{Dornheim_PRBL_2025,dornheim2024etaensemblepathintegralmonte}. Here, the main limitation is the sufficiently accurate resolution of the average sign $S$ that is known to decrease, in leading order, exponentially with decreasing temperature or increasing particle number~\cite{dornheim_sign_problem}.

In the present work, we show that it is possible to accurately estimate the average sign $S$, and, consequently, the corresponding difference between the Fermi-Dirac and Bose-Einstein systems of interest, in the fermionic limit of $\xi=-1$ based on PIMC simulations for $|\xi|<1$ for which the sign problem is substantially less severe. As a result, we are able to obtain accurate results for the XC-free energy of the warm dense uniform electron gas (UEG)~\cite{review,groth_prl,ksdt,status} down to half the Fermi temperature, which was not possible before with coordinate space implementations without using approximate nodal restrictions~\cite{Brown_PRL_2013}. We find excellent agreement with the parametrization by Groth \textit{et al.}~\cite{groth_prl}, which further substantiates the high quality of current UEG representations. Our work opens up new avenues for the future investigation of a gamut of interacting Fermi-Dirac systems, including warm dense matter~\cite{wdm_book}, ultracold atoms~\cite{morresi2025normalliquid3hestudied,Dornheim_SciRep_2022,Ceperley_PRL_1992}, and electrons in quantum dots~\cite{Egger_PRL_1999,Dornheim_NJP_2015,Xiong_JCP_2022}.

The paper is organized as follows: In Sec.~\ref{sec:theory}, we introduce the relevant theoretical background, including the basic idea behind the PIMC method~\cite{cep}, the $\xi$-extrapolation technique~\cite{Xiong_JCP_2022,Dornheim_JCP_xi_2023}, and the $\eta-$ensemble PIMC approach to the free energy~\cite{Dornheim_PRBL_2025,dornheim2024etaensemblepathintegralmonte}. Sec.~\ref{sec:results} contains our new simulation results, starting with an investigation of the ideal Fermi gas (\ref{sec:ideal}), proceeding with the UEG (\ref{sec:UEG}) and culminating in new estimates for the XC-free energy $F_\textnormal{xc}$ (\ref{sec:Fxc}). The paper is concluded by a summary and outlook in Sec.~\ref{sec:outlook}.

\section{Theory}\label{sec:theory}

The \textit{ab initio} PIMC method is based on the celebrated quantum-to-classical isomorphism~\cite{Chandler_JCP_1981} that allows one to map the non-ideal quantum many-body system of interest onto an effectively classical system of interacting ring polymers. Accessible introductions to PIMC have been presented elsewhere~\cite{cep,boninsegni1,Dornheim_permutation_cycles}; here we restrict ourselves to a discussion of the most important relations.
As a starting point, we consider the canonical partition function
\begin{eqnarray}\label{eq:Z}
    Z(N,\Omega,\beta)=\sum_{\sigma}\int \textnormal{d}\mathbf{X}\ \xi^{p_l(\sigma)} W(\mathbf{X})\quad ,
\end{eqnarray}
where $\mathbf{X}$ is a so-called path configuration that contains the coordinates of all $N$ particles on $P$ discrete imaginary-time slices and $W(\mathbf{X})$ is the corresponding configuration weight, an analytically known function that is easy to evaluate in practice.
In addition, we have to evaluate the sum over all permutations of particle coordinates $\sigma$ with $\xi=1$ and $\xi=-1$ corresponding to Bose-Einstein or Fermi-Dirac statistics~\cite{Dornheim_permutation_cycles}. Evidently, when $\xi=-1$, contributions to Eq.~(\ref{eq:Z}) can be negative for an odd number of pair exchanges $p_l(\sigma)$, precluding the interpretation of $P(\mathbf{X})= \xi^{p_l(\sigma)} W(\mathbf{X})/Z(N,\Omega,\beta)$ as a proper probability distribution for fermions. 

As a practical workaround, it is common practice to perform a bosonic PIMC simulation enabled by setting $\xi=1$ in Eq.~(\ref{eq:Z}) and subsequently to extract fermionic expectation values of any observable $\hat{A}$ as~\cite{dornheim_sign_problem}
\begin{eqnarray}\label{eq:ratio}
    \braket{\hat{A}}_{\xi} = \frac{\braket{\hat{A}\hat{S}}_{|\xi|}}{\braket{\hat{S}}_{|\xi|}}\quad .
\end{eqnarray}
The denominator in Eq.~(\ref{eq:ratio}) is known as the \textit{average sign} $S$ and constitutes a straightforward measure for the amount of cancellation between negative and positive contributions to Eq.~(\ref{eq:Z}).
In fact, it directly corresponds to the ratio of fermionic and bosonic partition functions,
\begin{eqnarray}\label{eq:sign}
    S(\xi) = \frac{Z_{\xi}(N,\Omega,\beta)}{Z_{|\xi|}(N,\Omega,\beta)}\ .
\end{eqnarray}
In practice, a vanishing sign indicates a vanishing signal-to-noise ratio of Eq.~(\ref{eq:ratio}), which is known as the \textit{fermion sign problem} in the literature~\cite{dornheim_sign_problem,troyer}. It constitutes an exponential computational bottleneck with respect to important system parameters such as the number of particles $N$ or the inverse temperature $\beta$, restricting the application of fermionic PIMC simulations to relatively moderate temperatures~\cite{Bonitz_POP_2024}.

Very recently, Xiong and Xiong~\cite{Xiong_JCP_2022} have proposed to circumvent the sign problem by performing path integral molecular dynamics simulations of fictitious identical particles with continuous values of the $\xi$-variable $-1\leq\xi\leq1$. In particular, the simulations are FSP-free for $\xi\geq0$, and the expectation value of any observable should be a smooth function of $\xi$. The basic idea of the $\xi$-extrapolation method is then to perform a polynomial fit of $\braket{\hat{A}}_\xi$ to extrapolate to the true fermionic limit of $\xi=-1$. Subsequently, this idea has been adapted to fermionic PIMC simulations~\cite{Dornheim_JCP_xi_2023}, where it has been demonstrated to be capable of giving highly accurate and essentially unbiased results for large systems at weak to moderate degrees of quantum degeneracy~\cite{Dornheim_JPCL_2024}. In particular, it removes the exponential scaling with respect to the number of simulated fermions $N$, which is very important in its own right. Consequently, the $\xi$-extrapolation method has been used to compute a gamut of observables including energies, structural properties and even imaginary-time correlation functions, with its applications ranging from electrons in quantum dots to warm dense quantum plasmas and ultracold atoms~\cite{morresi2025normalliquid3hestudied,Dornheim_Science_2024,Dornheim_JCP_2024,dornheim2024modelfreerayleighweightxray,Xiong_PRE_2023,Xiong_PRE_2024}.

In the present work, we focus on the free energy
\begin{eqnarray}
    F(N,\Omega,\beta) = -\frac{1}{\beta}\textnormal{log}\left(Z(N,\Omega,\beta)\right)\quad ,
\end{eqnarray}
which, being equivalent to the partition function itself, is not a straightforward observable for PIMC calculations. Instead, PIMC is generally limited to the estimation of free energy differences between two Hamiltonians $\hat{H}_1$ and $\hat{H}_2$~\cite{Lyubartsev_JCP_1992}. A common approach is to relate the interacting system of interest with a non-interacting reference system at the same conditions for which the free energy can be computed semi-analytically in absolute terms~\cite{Zhou_2018}. This, however, is tremendously impractical in the case of fermions as PIMC simulations of ideal Fermi systems are afflicted with a stifling sign problem~\cite{dornheim_sign_problem}. As a practical workaround, Dornheim \textit{et al.}~\cite{Dornheim_PRBL_2025,dornheim2024etaensemblepathintegralmonte} have recently suggested a different route where the free energy of the interacting Bose system is first evaluated with respect to the ideal Bose gas and then the free energy of the interacting Fermi system of interest is recovered from a fermionic PIMC simulation, where the FSP is substantially less severe. In this approach, the free energy of the interacting Fermi system is given by~\cite{Dornheim_PRBL_2025}
\begin{eqnarray}\label{eq:F_total}
    F_\textnormal{F} = F_{\textnormal{B},0} - \frac{1}{\beta}\left\{
\sum_{i=1}^{N_\eta} \textnormal{log}\left(r(\eta_{i-1},\eta_i)\right)
+ \textnormal{log}\left(S\right)
    \right\}\quad ,
\end{eqnarray}
where $F_{\textnormal{B},0}$ denotes the free energy of a noninteracting Bose system at the same conditions, the first sum in the curly brackets connects the noninteracting (i.e., $\eta=0$) with the interacting (i.e., $\eta=1$) limits in the bosonic sector with $r(\eta_{i-1},\eta_i)$ being the ratio of partition functions for adjacent $\eta$-values (with $\dots<\eta_{i-1}<\eta_i<\eta_{i+1}<\dots$) and the second sum in the curly brackets relates the interacting free energies of fermions and bosons~\cite{Dornheim_PRBL_2025,dornheim2024etaensemblepathintegralmonte}. 
From the perspective of the fermion sign problem, the only problematic contribution in Eq.~(\ref{eq:F_total}) is given by the sign $S=S(-1)$ [cf.~Eq.~(\ref{eq:sign})], which vanishes within the given Monte Carlo error bars for too large systems or too low temperatures.

The aim of the present work is to generalize the spirit behind the $\xi$-extrapolation method to the computation of fermionic free energies via Eq.~(\ref{eq:F_total}). To this end, and without loss of generality, we make a simple exponential ansatz for $S(\xi)$,
\begin{eqnarray}\label{eq:sign_vs_xi}
    S(\xi\leq0) = e^{a(\xi\leq0)\xi}\ \Rightarrow\ a(\xi\leq0) = \frac{\textnormal{log}\left[S(\xi\leq0)\right]}{\xi}\ .
\end{eqnarray}
This is motivated by the recent observation that the sign exponentially decays with $|\xi|$ at some parameters~\cite{Dornheim_JCP_xi_2023}, in which case the function $a(\xi\leq0)$ will be a simple constant. Crucially, any known functional dependence of $a$ on $\xi$ would allow us to extrapolate to the true fermionic limit of the sign $S(-1)$ from PIMC results for $S(\xi<0)$ at smaller $|\xi|$, for which the sign can be resolved with feasible computational effort.

\begin{figure*}[t!]
\center
\includegraphics[width=0.499\textwidth]{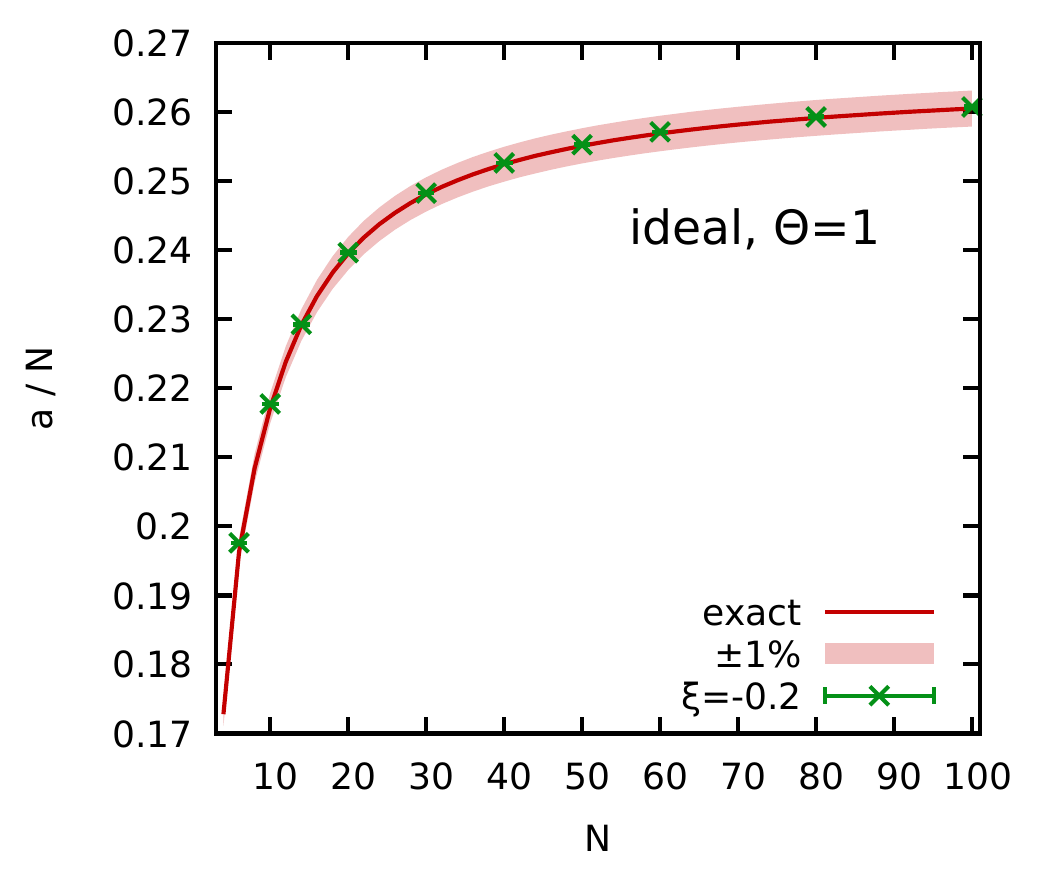}\includegraphics[width=0.499\textwidth]{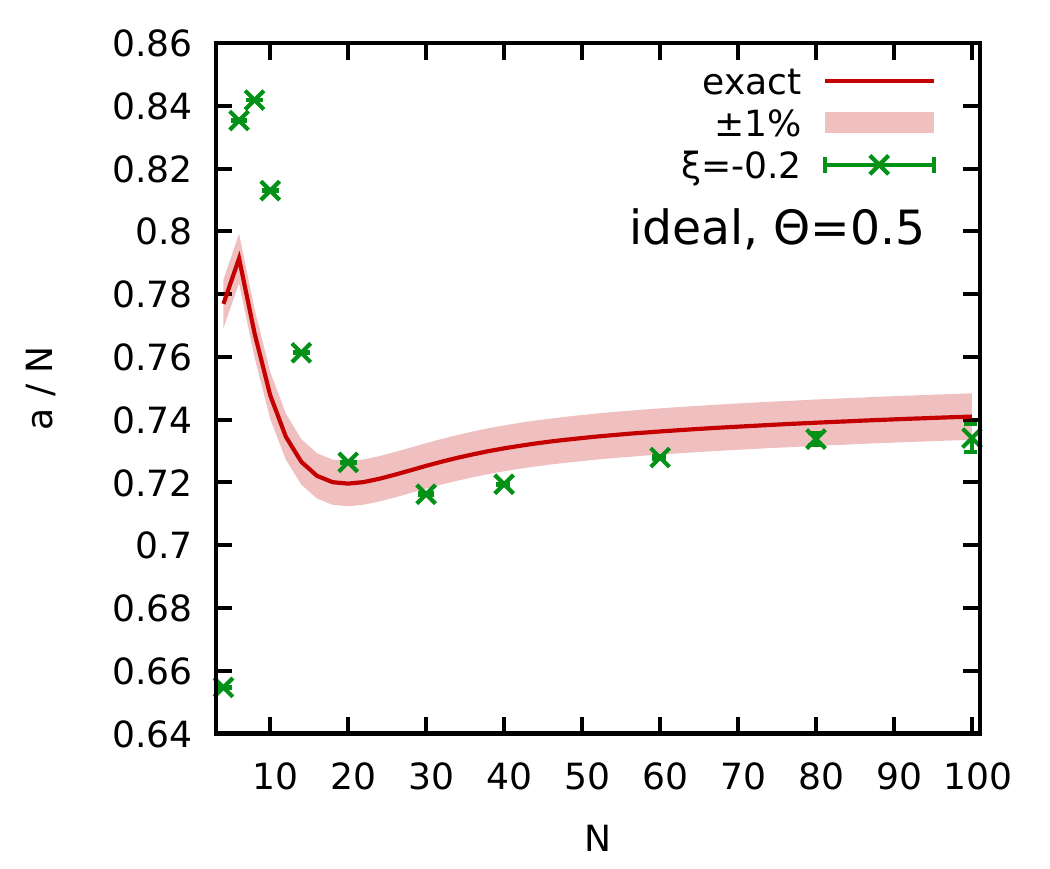}
\caption{ \label{fig:ideal_a_theta0p5} 
Dependence of the sign exponential scaling factor per particle $a/N$ of the unpolarized ideal Fermi gas on the number of particles $N$ at $\Theta=1$ (left) and $\Theta=0.5$ (right). The red lines depict the exact semi-analytical results for the true fermionic limit $a(\xi=-1)$ computed from the noninteracting fermionic average sign $S_0$~\cite{dornheim2024etaensemblepathintegralmonte}, while the green crosses correspond to PIMC results for $a(\xi=-0.2)$ at the same conditions. In both panels, the shaded red area indicates an interval of $\pm1\%$ around the exact fermionic limit that has been included as a guide-to-the-eye.
}
\end{figure*}

\section{Results}\label{sec:results}

All PIMC results that are presented in this work have been obtained using the open-source \texttt{ISHTAR} code~\cite{ISHTAR}, and are freely available online~\cite{repo}.
We use $P=200$ primitive imaginary-time propagators throughout, and the convergence with $P$ has been carefully checked.
Note that we assume Hartree atomic units.

\subsection{Ideal Fermi gas}\label{sec:ideal}

We start our investigation with the ideal Fermi gas, which, as we shall see below, exhibits the same qualitative trend as the interacting UEG. In particular, in Fig.~\ref{fig:ideal_a_theta0p5}, we show $a(\xi)/N$ as a function of the system size $N$ for $\xi=-1$ and $\xi=-0.2$ at $\Theta=1$ (left) and $\Theta=0.5$ (right). We note that, for the ideal system at constant $\Theta$, neither the average sign $S(\xi<0)$ nor the sign exponential scaling factor $a(\xi<0)$ depend on the density. The red solid curves correspond to the exact fermionic limit of $a(-1)/N$, which have been obtained by combining the non-interacting limit of Eq.~(\ref{eq:sign}) with semi-analytical results for the partition function of the ideal Bose gas and the ideal Fermi gas~\cite{dornheim2024etaensemblepathintegralmonte,Zhou_2018}. We observe a monotonic and non-monotonic dependence on $N$ for $\Theta=1$ and $\Theta=0.5$, respectively, which is an immediate consequence of the corresponding finite-size effects in the noninteracting free energy. In fact, it is easy to see that the value $a(-1)$ directly relates to the free energy difference between the bosonic and fermionic limits,
\begin{eqnarray}\label{eq:F_from_a}
    F_\textnormal{F}-F_\textnormal{B} = \frac{a(-1)}{\beta}\quad .
\end{eqnarray}
The green crosses show PIMC results for $a(-0.2)/N$. For the moderately quantum degenerate case of $\Theta=1$, we find excellent agreement between $a(-0.2)$ and $a(-1)$, which substantiates the previous observation that $a(\xi<0)=a=const$ at these conditions. In the spirit of Eq.~(\ref{eq:F_from_a}), the combination of Eqs.~(\ref{eq:ratio},\ref{eq:F_total}) with Eq.~(\ref{eq:sign_vs_xi}) implies that, for any state point, the free energy difference between its fermionic sector $\xi<0$ image and its bosonic section $-\xi>0$ image is a linear function of the $\xi$ variable.
\begin{eqnarray}\label{eq:F_from_a_new}
    F(\xi<0)-F(-\xi>0)=-\frac{a(\xi<0)}{\beta}\xi\simeq-\frac{a}{\beta}\xi\quad .
\end{eqnarray}
In practice, this gives us an exponential acceleration of our PIMC simulations as we only have to resolve $S(\xi)$ for small absolute values of $\xi$ in order to estimate its true fermionic limit of $S(-1)$ with very high precision. For $\Theta=0.5$, on the other hand, we observe significant deviations between $a(-0.2)$ and $a(-1)$, which are particularly pronounced for small $N$. A possible explanation for this trend is based on the related length scales: the physical behaviour of the system can be expected to intrinsically change with the variable $\xi$ when the box length $L$ is comparable to the thermal wavelength $\lambda_\beta=\sqrt{2\pi\beta}$, which would explain the seeming convergence of $a(-0.2)$ towards $a(-1)$ at large $N$.

\begin{figure*}[t!]
\center
\includegraphics[width=0.499\textwidth]{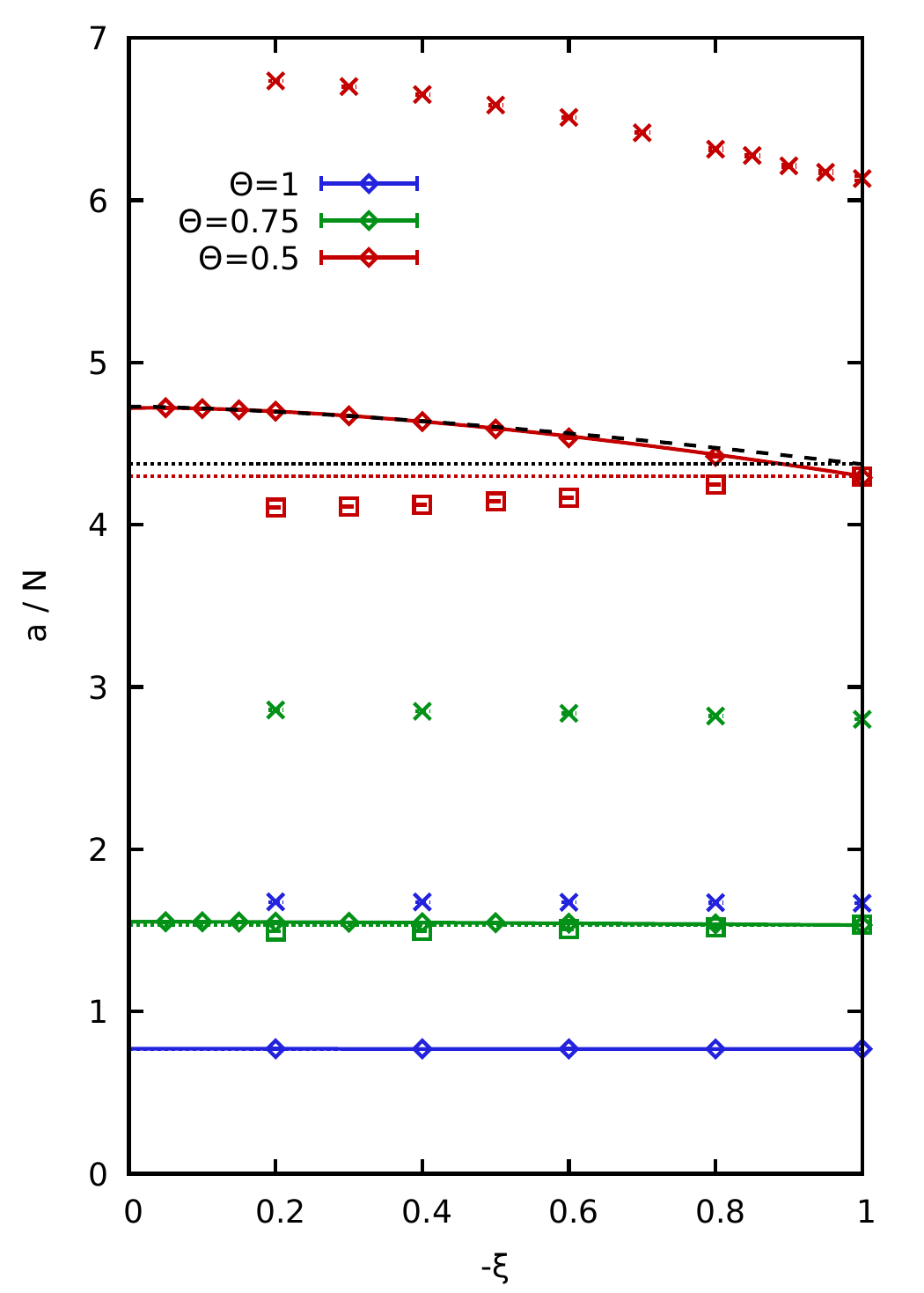}\includegraphics[width=0.499\textwidth]{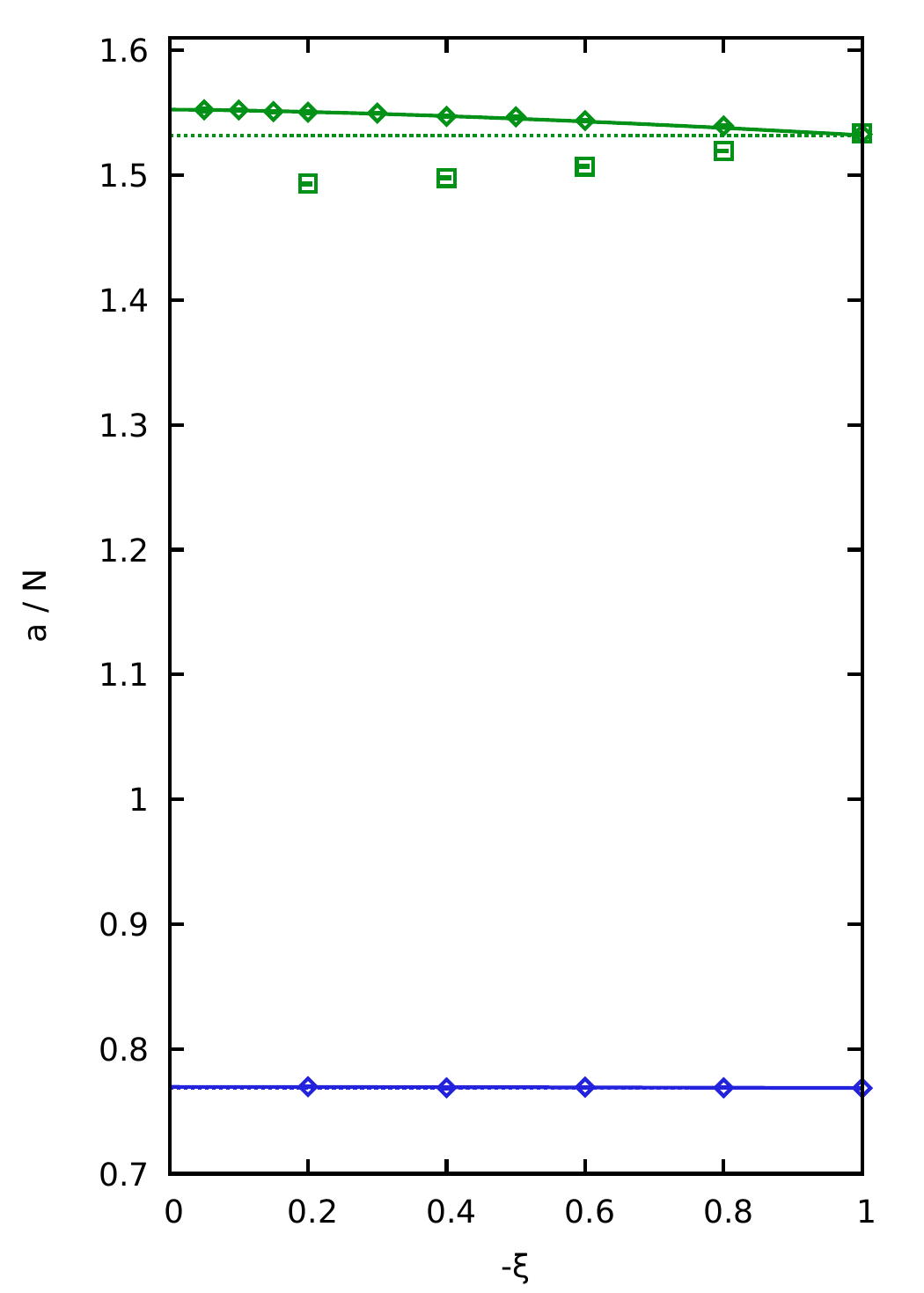}
\caption{ \label{fig:I_want_ramen} 
Dependence of the sign exponential scaling factor per particle $a/N$ of the warm dense paramagnetic UEG on the partition function parameter $\xi$ for $N=8$ at $r_s=3.23$ and $\Theta=1$ (blue), $\Theta=0.75$ (green), $\Theta=0.5$ (red). The diamonds and crosses have been obtained for the interacting UEG and ideal Fermi gas, respectively, while the squares have been obtained by adding the ideal correction $\Delta a_0(\xi)$ to the UEG data, see Eq.~(\ref{eq:correction}). The curves and dotted horizontal lines correspond to different fits that are discussed in the main text. The right panel corresponds to a magnified segment of the left panel.
}
\end{figure*}

\subsection{Uniform electron gas}\label{sec:UEG}

In Fig.~\ref{fig:I_want_ramen}, we investigate the sign exponential scaling factor $a(\xi)$ of the warm dense UEG with $N=8$ at $r_s=3.23$. Such a moderate coupling can be realized in experiments with hydrogen jets~\cite{Zastrau,Fletcher_Frontiers_2022,Hamann_PRR_2023}. The blue diamonds and crosses show results for the UEG and the ideal Fermi gas, respectively, at $\Theta=1$, i.e., for a moderate degree of quantum degeneracy. We find a negligible dependence of $a(\xi)$ on $\xi$ at these conditions for both systems, consistent with the findings from Fig.~\ref{fig:ideal_a_theta0p5}; see also the right panel of Fig.~\ref{fig:I_want_ramen} for a magnified segment around the UEG results at $\Theta=1$ (bottom points). The higher $a(\xi)$ values for the ideal Fermi gas compared to the interacting UEG are a consequence of the increased degeneracy and, consequently larger free energy difference or equivalently the smaller average sign in the former case. The green data points show a similar investigation at $\Theta=0.75$. In this case, we find a very small dependence of $a(\xi)$ on $\xi$ (of the order of $\sim1\%$ for small $\xi$), which can be fitted well by the empirical function
\begin{eqnarray}\label{eq:fit1}
    a(\xi) = c_0 + c_1|\xi|^{3/2}\ ,
\end{eqnarray}
see the solid green curves in Fig.~\ref{fig:I_want_ramen}. In fact, this fit has been constructed on the basis of data only for $\xi\geq-0.4$, yet it nicely reproduces the full PIMC data set. In other words, data in this limited $\xi$-range are sufficient to accurately estimate the true fermionic limit. For completeness, we have also obtained a dataset where we add 
the corresponding ideal Fermi gas correction to the PIMC results for $a(\xi)$ of the UEG 
\begin{eqnarray}\label{eq:correction}
    \Delta a_0(\xi) = a_0(-1) - a_0(\xi)\quad.
\end{eqnarray}
The results are included as the green squares into Fig.~\ref{fig:I_want_ramen}, but, in this case, they do not constitute an improvement over the raw PIMC results for the UEG. The red data points show results at $\Theta=0.5$. In this more strongly degenerate quantum case, we find a more substantial dependence of $a(\xi)$ on $\xi$ for both the ideal Fermi gas and the UEG that appear to qualitatively resemble each other. The addition of the Eq.~(\ref{eq:correction}) correction does constitute an improvement for these parameters, but a residual error of $\sim1\%$ remains e.g.~for $\xi=-0.2$. On the other hand, the advantage of this approach is that it only requires PIMC results for a single $\xi$-value as the correct fermionic limit of the ideal Fermi gas is a-priori known; no extrapolation over $\xi$ is needed. The use of Eq.~(\ref{eq:fit1}) for $\xi\geq-0.4$ to extrapolate to the fermionic limit of $\xi=-1$ gives the dashed black curve, which closely, but not exactly matches the correct limit. The addition of an extra term to the fit,
\begin{eqnarray}\label{eq:fit2}
    a(\xi) = c_0 + c_1|\xi|^{3/2}+c_2\sqrt{|\xi|}\ ,
\end{eqnarray}
also for $\xi\geq-0.4$, gives the solid red curve, which constitutes an even better match. In practice, we advocate that both fits are performed and that the resulting difference is employed as an empirical measure of the extrapolation error.

\begin{figure*}[t!]
\center
\includegraphics[width=0.499\textwidth]{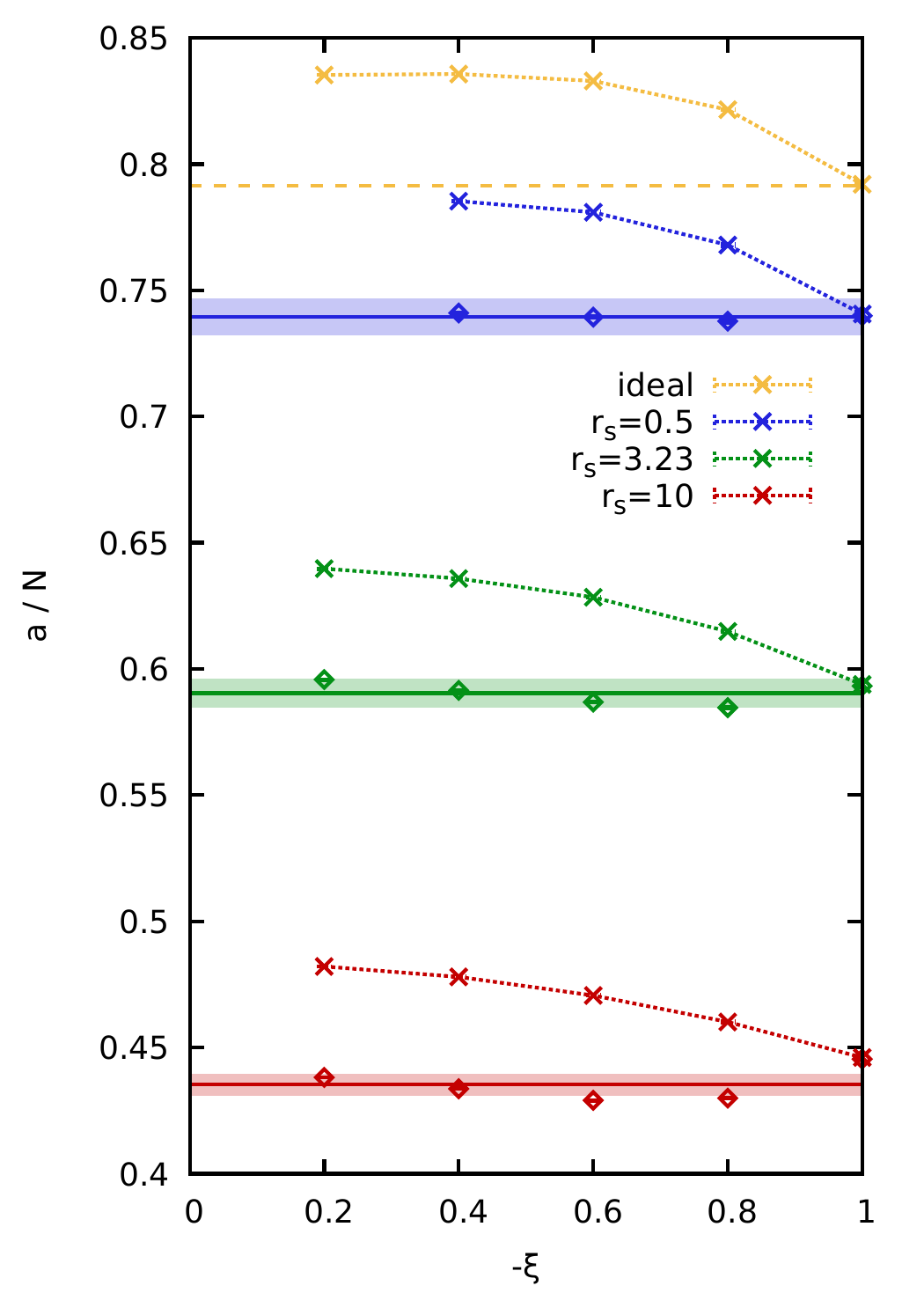}\includegraphics[width=0.499\textwidth]{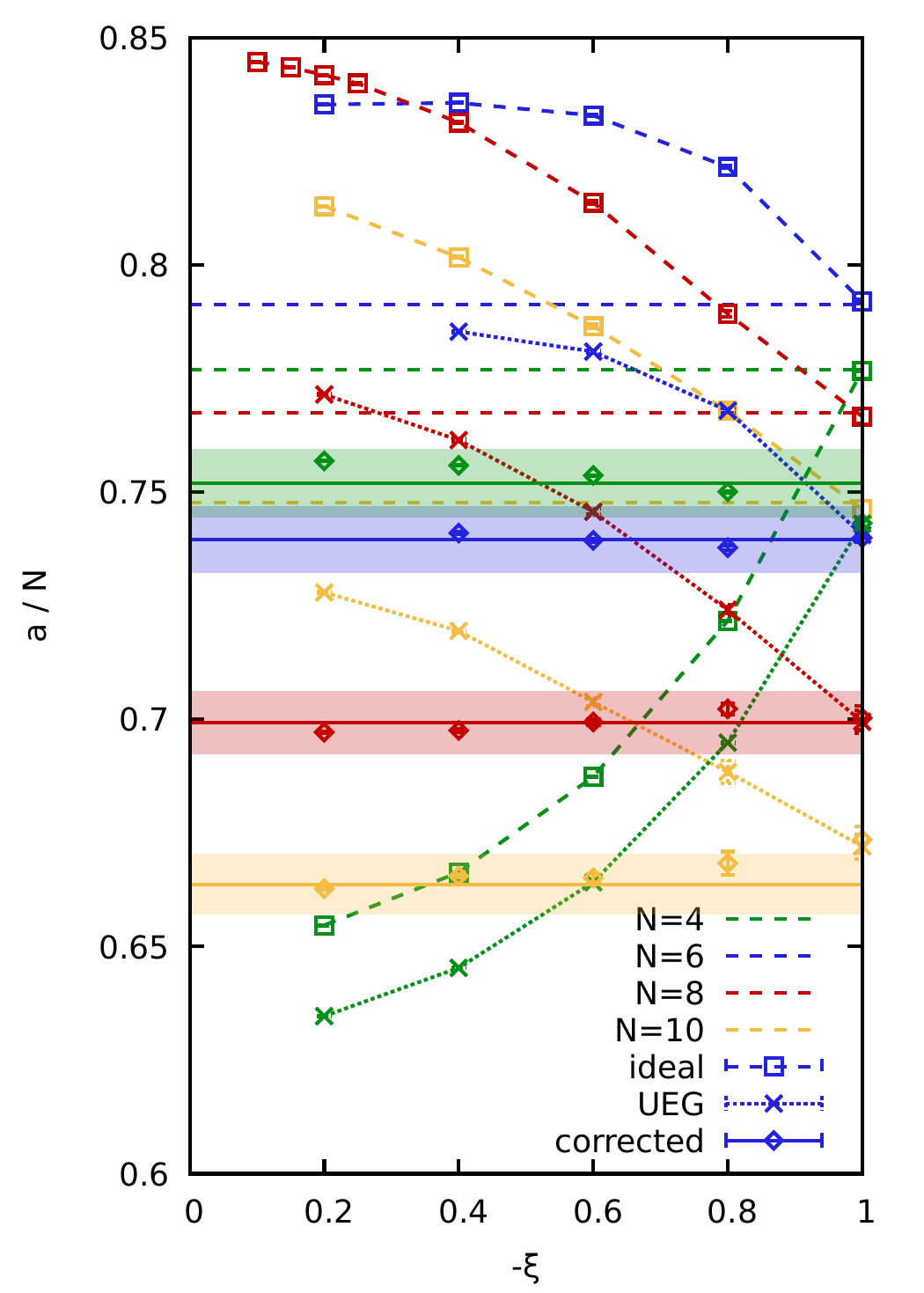}
\caption{ \label{fig:rs_and_N} 
Left panel: the sign exponential scaling factor per particle $a/N$ of the UEG for $N=6$ at $\Theta=0.5$ and
$r_s\to0$ (yellow), $r_s=0.5$ (blue), $r_s=3.23$ (green), $r_s=10$ (red). The crosses depict raw PIMC results, while the diamonds depict PIMC results obtained by adding the ideal correction $\Delta a_0(\xi)$ to the UEG data, see Eq.~(\ref{eq:correction}). Right panel: the sign exponential scaling factor per particle $a/N$ of the raw UEG (crosses), the ideal Fermi gas (squares) and the corrected UEG (diamonds) at $r_s=0.5$, $\Theta=0.5$ for different numbers of electrons ($N=4,6,8,10)$. In both panels, the shaded colored areas indicate intervals of $\pm1\%$ around the exact fermionic limit of $a(\xi=-1)$ that serve as a guide-to-the-eye.
}
\end{figure*}

Next we consider the effect of the density parameter $r_s$ on the behaviour of $a(\xi)$, which we investigate in the left panel of Fig.~\ref{fig:rs_and_N} for the most challenging case of $\Theta=0.5$ for $N=6$. The yellow crosses show results for the ideal Fermi gas which correspond to the $r_s\to0$ limit of the UEG; the dashed horizontal yellow line is the exact semi-analytical result for $a(-1)$, which matches the respective PIMC data point, as expected. The blue, green, and red data sets have been obtained from PIMC simulations of the UEG at $r_s=0.5$, $r_s=3.23$, and $r_s=10$. These conditions are respectively realized in laser compression experiments at the National Ignition Facility (NIF)~\cite{Moses_NIF,Tilo_Nature_2023}, the aforementioned hydrogen jet experiments and the margin of the strongly coupled electron liquid regime~\cite{dornheim_dynamic,dornheim_electron_liquid}. First, we find that $a(\xi)$ monotonically decreases with increasing $r_s$ as the formation of permutation cycles becomes increasingly suppressed by Coulomb repulsion~\cite{Dornheim_permutation_cycles}. Second, we find that $a(\xi)$ exhibits an increasingly sharp drop for large $|\xi|$ with a decrease in the degree of non-ideality. This indicates an increasing breakdown of the linear expression for the free energy difference between bosonic sector and fermionic sector images [cf.~Eq.~(\ref{eq:F_from_a_new})] with increasing $|\xi|$, as the system becomes shaped more profoundly by quantum statistics. Third, we consider the diamonds, which have been obtained by adding the ideal correction, Eq.~(\ref{eq:correction}), onto the raw PIMC results for $a(\xi)$. Overall, the correction works reasonably well and removes the bulk of the original dependence on $\xi$; yet, small residual errors of the order of $\sim1\%$ remain, see also the shaded areas indicating an interval of $\pm1\%$ around the true $a(-1)$ fermionic limit. Finally, the right panel of Fig.~\ref{fig:rs_and_N} compares PIMC simulation results for different particle numbers $N$ for $r_s=0.5$ and $\Theta=0.5$. Here, the squares, crosses and diamonds correspond to the ideal Fermi gas, the raw UEG, and the corrected [cf.~Eq.~(\ref{eq:correction})] UEG. As a general trend, we find that the $\xi$-dependence exhibits a less steep drop around large $|\xi|$ with increasing number of particles, which fits to the trends observed in the right panel of Fig.~\ref{fig:ideal_a_theta0p5}. In general, the correction works well and we obtain results within an interval of $\pm1\%$ of the true fermionic limit (shaded colored areas).

\subsection{Exchange--correlation free energy}\label{sec:Fxc}

Let us next apply our method to the estimation of the exchange--correlation free energy of the warm dense UEG, $F_\textnormal{xc}$. First and foremost, the advent of accurate, PIMC-based parametrizations of $F_\textnormal{xc}$~\cite{ksdt,groth_prl,review} has been of paramount importance for thermal density functional theory (DFT)~\cite{Mermin_DFT_1965} simulations of warm dense matter and directly allows for calculations on the level of the local density approximation (LDA)~\cite{karasiev_importance,kushal,new_POP,Sjostrom_PRB_2014,Moldabekov_JCTC_2024}. Moreover, they constitute the basis for more sophisticated functionals on higher rungs of Jacob's ladder of functional approximations~\cite{kozlowski2023generalized,Karasiev_PRL_2018,Karasiev_PRB_2022,Bonitz_POP_2024}. In addition, they are important input for a host of other applications~\cite{STARRETT201950,refId0,PhysRevE.62.8554} and determine both the short- and long-wavelength limit of the exchange--correlation kernel~\cite{quantum_theory,Hou_PRB_2022,dornheim2024shortwavelengthlimitdynamic,Dornheim_PRB_2024}.

\begin{figure*}[t!]
\center
\includegraphics[width=0.439\textwidth]{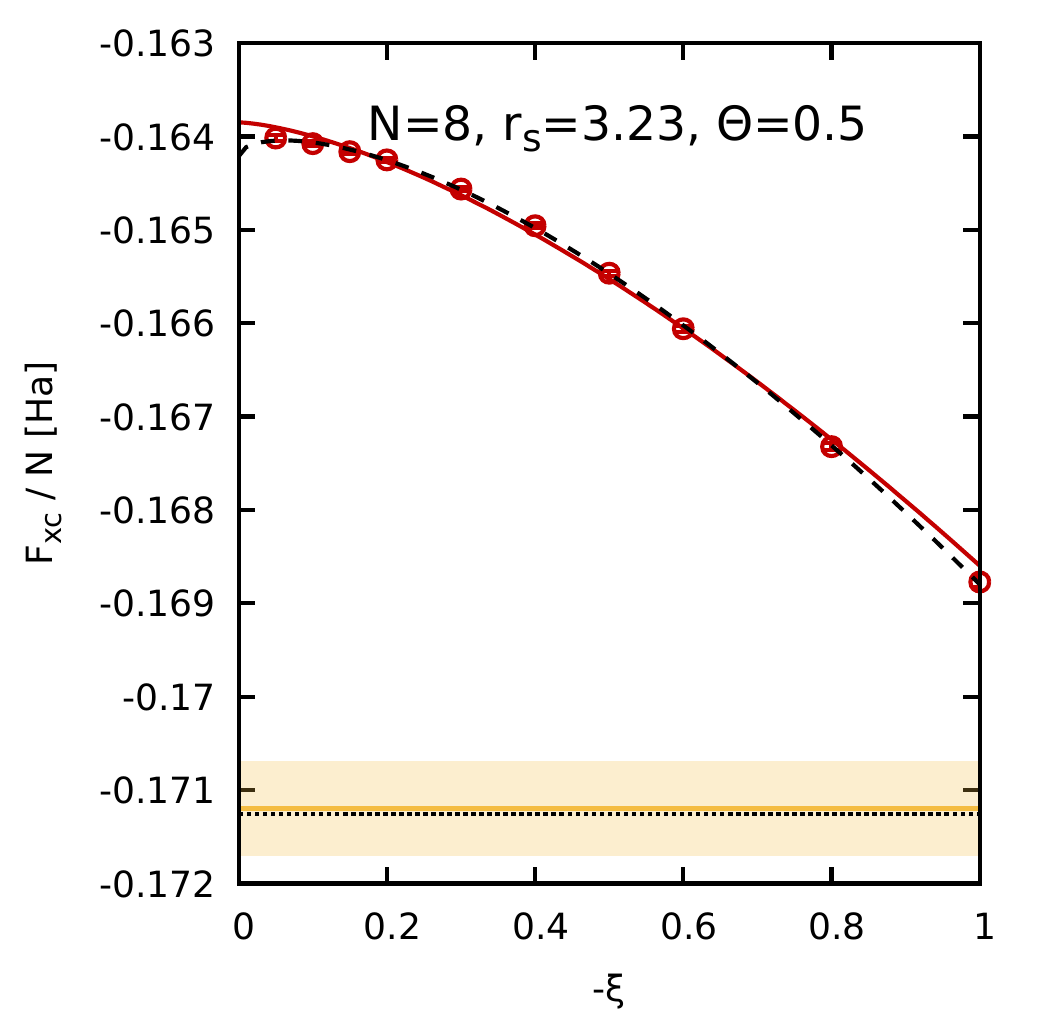}\includegraphics[width=0.439\textwidth]{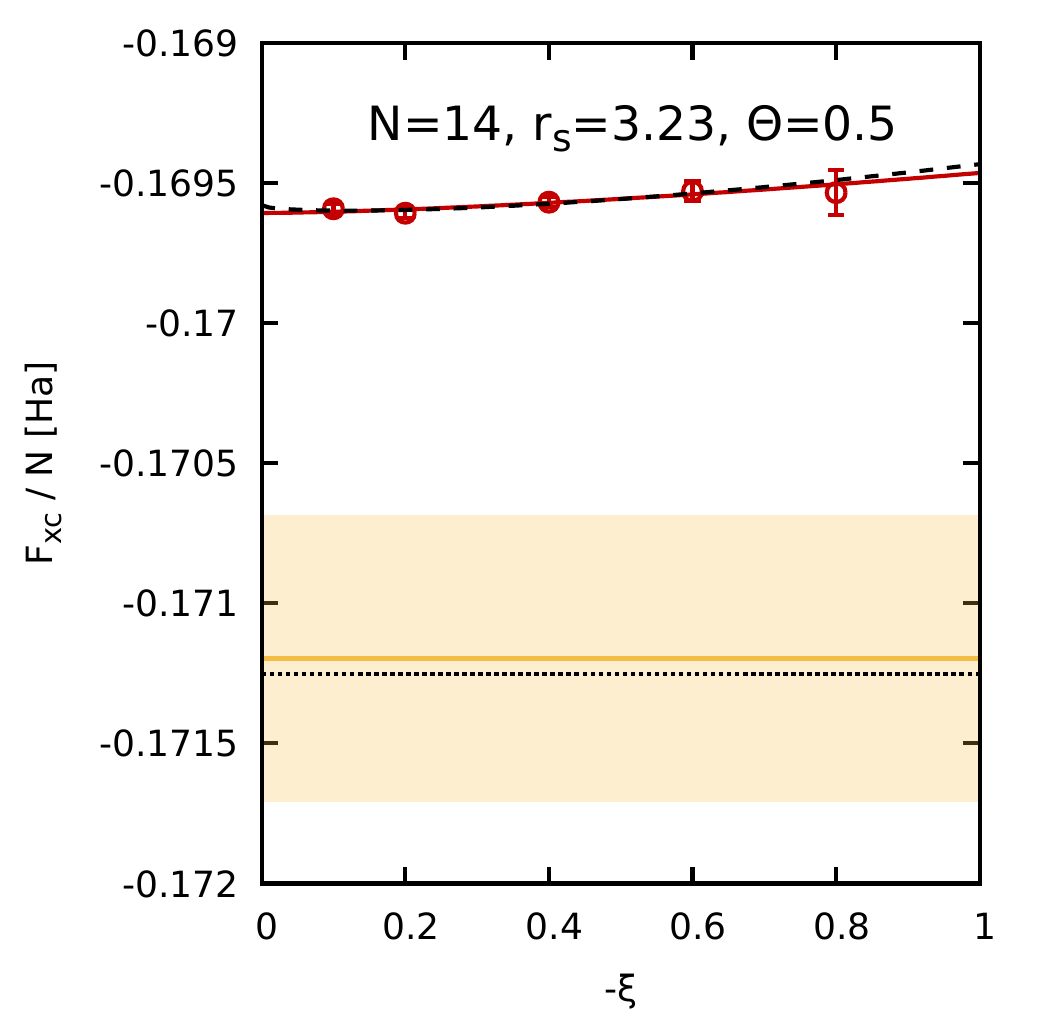}\\\includegraphics[width=0.439\textwidth]{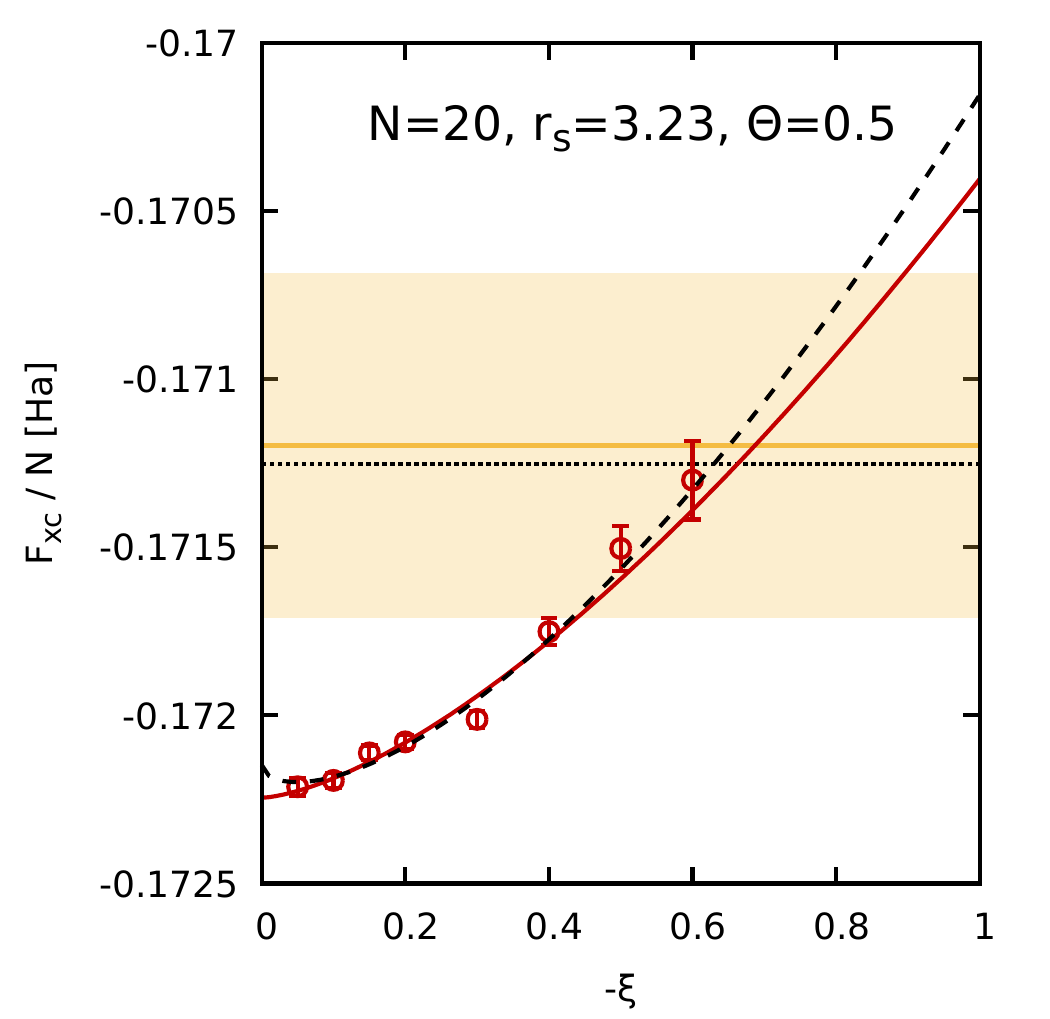}
\includegraphics[width=0.439\textwidth]{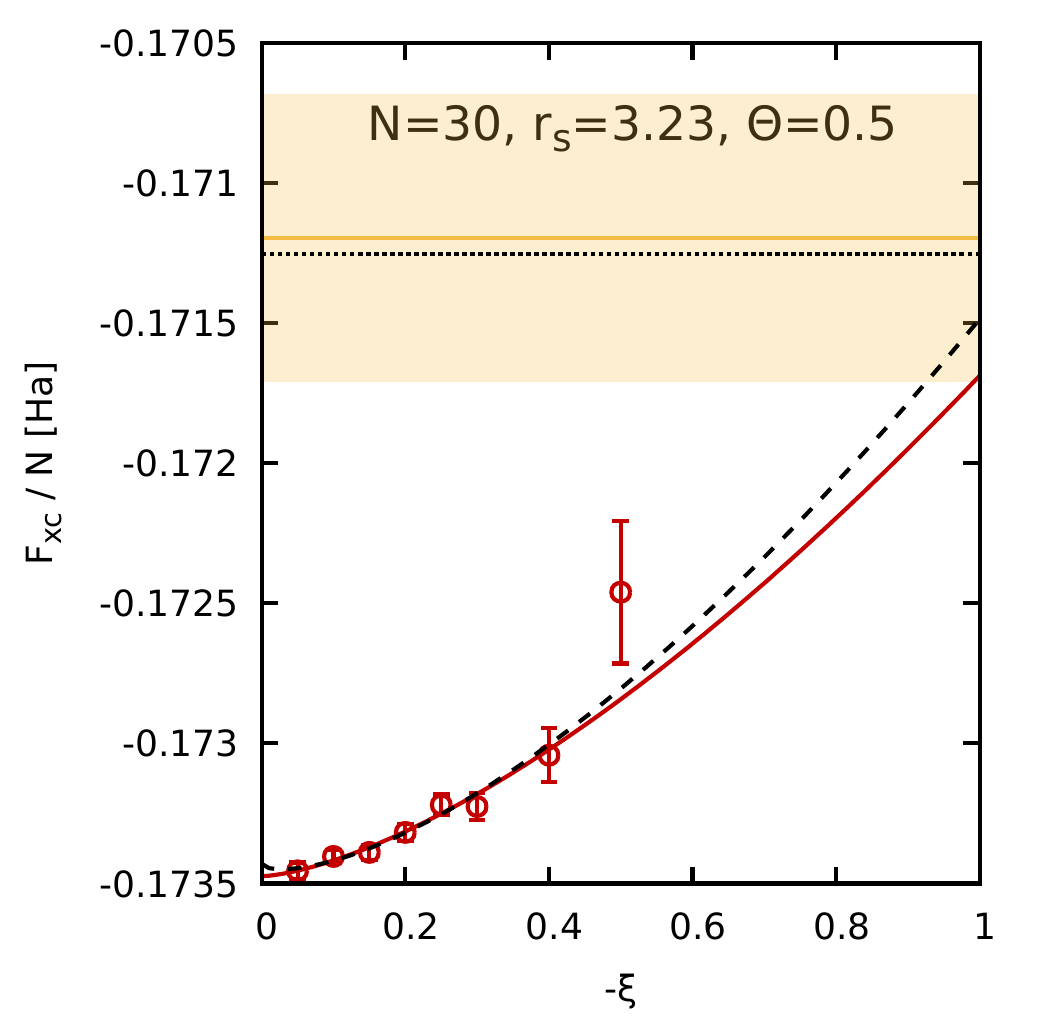}
\caption{ \label{fig:Fermionic_limit_rs3p23_theta0p5} 
Extrapolation of the XC-free energy per particle $F_\textnormal{XC}/N$ to the fermionic limit of $\xi=-1$ for different numbers of particles ($N=8,14,20,30$) at $r_s=3.23$ and $\Theta=0.5$. The red circles depict finite-size corrected PIMC results, while the solid red and dashed black curves correspond to the two-parameter, see Eq.~(\ref{eq:fit1}), and three-parameter, see Eq.~(\ref{eq:fit2}), empirical fits. The yellow vertical line and shaded yellow areas correspond to the GDSMFB parametrization and its respective nominal uncertainty interval of $\pm0.3\%$~\cite{groth_prl}.
}
\end{figure*}

A first important question is given by so-called finite-size effects, i.e., the difference between PIMC simulations for a finite number of particles and the thermodynamic limit that is defined by simultaneously taking the limits of $N\to\infty$ and $\Omega\to\infty$ with $n=N/\Omega$ being kept constant. Indeed, a vast literature has been dedicated to the study of finite-size effects in the UEG~\cite{Fraser_PRB_1996,Chiesa_PRL_2006,Drummond_PRB_2008,dornheim_prl,review,Dornheim_JCP_2021}. An accessible introduction to the finite-size corrections that are being applied in the presented work has been given in Ref.~\cite{review}, and we use the efficient and easy-to-use python implementation \texttt{UEGPY}~\cite{uegpy} by F.D.~Malone. In the following, all PIMC results have been obtained by subtracting from Eq.~(\ref{eq:F_total}) the ideal fermionic free energy of the $N$-body system $F_0(N) = -\textnormal{log}[S_0(N)]/\beta$ and subsequently adding the finite-size correction $\Delta F_\textnormal{xc}(N)$.

\begin{figure*}[t!]
\center
\includegraphics[width=0.439\textwidth]{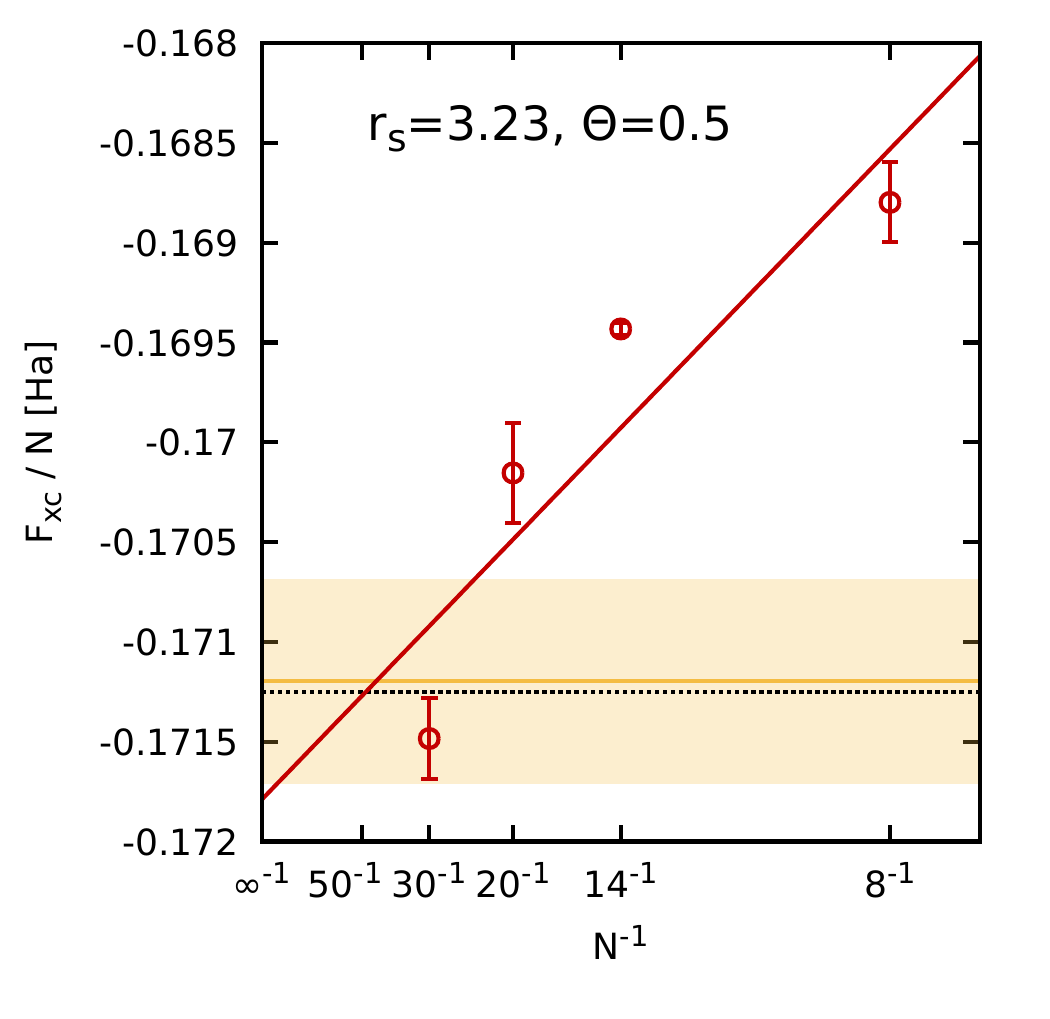}\includegraphics[width=0.439\textwidth]{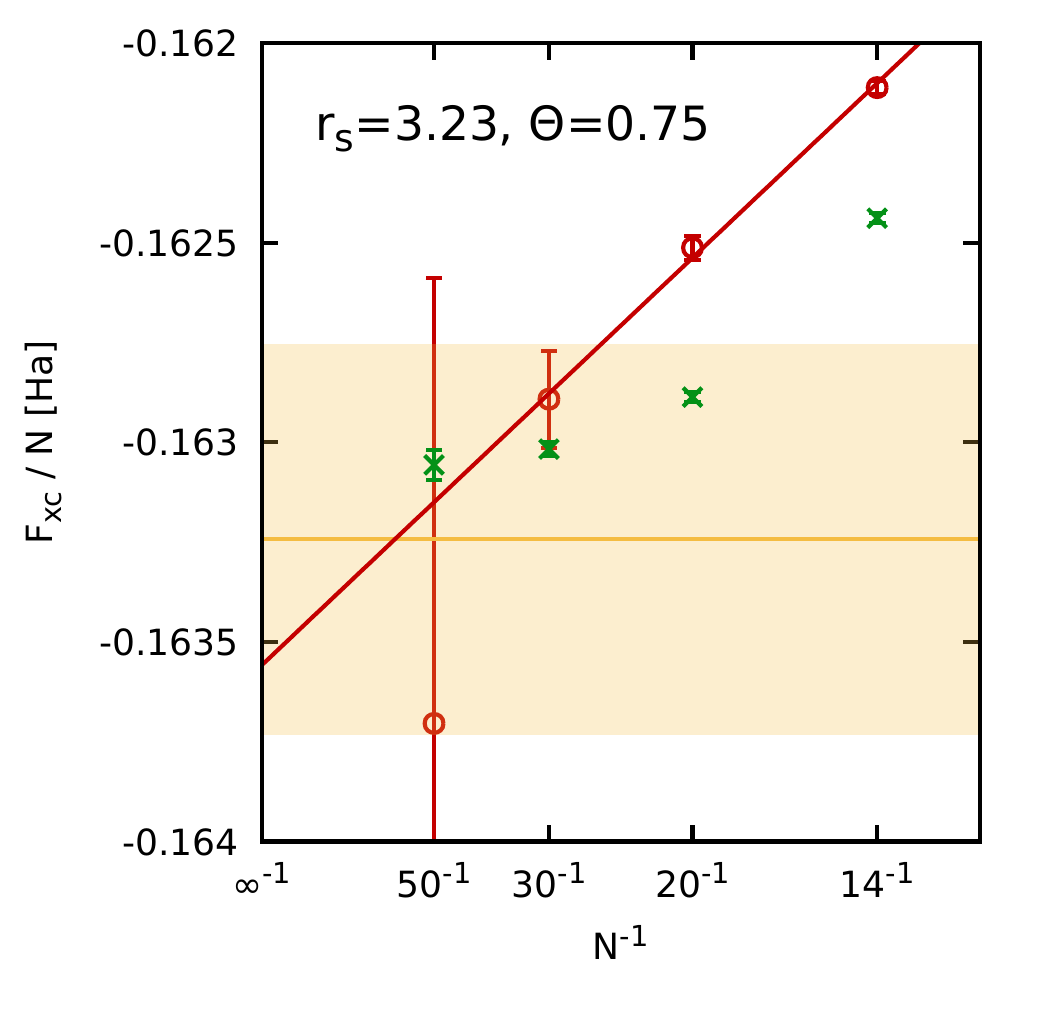}\\\includegraphics[width=0.439\textwidth]{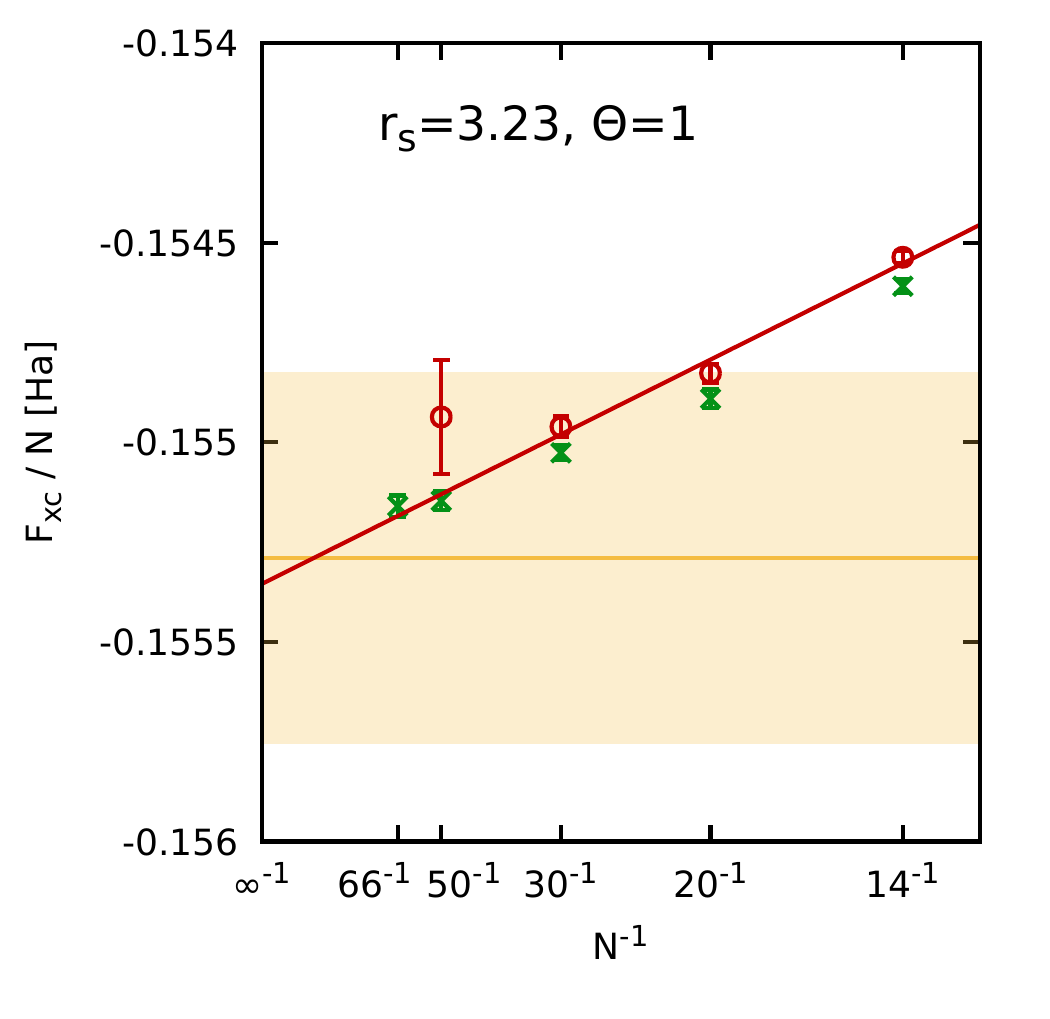}
\caption{ \label{fig:TDL_rs3p23} 
System-size dependence of the PIMC results for $F_\textnormal{xc}/N$ at $r_s=3.23$ and different temperatures ($\Theta=0.5,\,0.75,\,1$). The red symbols in the top left panel ($\Theta=0.5$) are obtained from an empirical extrapolation to the fermionic limit (see Fig.~\ref{fig:Fermionic_limit_rs3p23_theta0p5}), while the red symbols in the top right ($\Theta=0.75$) and bottom panels ($\Theta=1$) correspond to direct PIMC results at the fermionic limit $\xi=-1$. In addition, the green crosses correspond to PIMC results using $a(\xi)$ with $\xi=-0.2$. The yellow vertical line and shaded yellow areas correspond to the GDSMFB parametrization and its respective nominal uncertainty interval of $\pm0.3\%$~\cite{groth_prl}.
}
\end{figure*}

In Fig.~\ref{fig:Fermionic_limit_rs3p23_theta0p5}, we show our new PIMC results for $F_\textnormal{xc}/N$ for $r_s=3.23$ and $\Theta=0.5$ as a function of the partition function parameter $\xi$ for a number of different $N$. The vertical yellow lines and the surrounding shaded yellow area show the state-of-the-art parametrization by Groth, Dornheim, Sjostrom, Malone, Foulkes, Bonitz (GDSMFB)~\cite{groth_prl} and its nominal uncertainty interval of $\pm0.3\%$ that have been included as a reference. We find the familiar pattern of a significant dependence of $F_\textnormal{xc}$ on $\xi$, which manifests very differently depending on $N$. The solid red and dashed black curves show the two-parameter [cf.~Eq.~(\ref{eq:fit1})] and three-parameter [cf.~Eq.~(\ref{eq:fit2})] empirical fit functions. We find that a reliable extrapolation to the fermionic limit of $\xi=-1$ is possible in all cases. In the top left panel of Fig.~\ref{fig:TDL_rs3p23}, we show the thus extrapolated results as a function of the inverse system size. The residual finite-size errors are of the order of $\lesssim1\%$, which is consistent with previous investigations of similar properties~\cite{review,dornheim_prl,Dornheim_JCP_2021}. Overall, the PIMC results for different $N$ are well reproduced by a simple linear fit, which agrees with the given uncertainty range of the GDSMFB parametrization~\cite{groth_prl} (yellow area). In the top right and bottom panels, we show finite-size corrected PIMC results for $\Theta=0.75$ and $\Theta=1$ at the same density. For these parameters, we find that the full extrapolation procedure shown in Fig.~\ref{fig:Fermionic_limit_rs3p23_theta0p5} is not required, as direct fermionic PIMC simulations are feasible for many $N$, see the red circles. In addition, the green crosses have been obtained by using $a(\xi)$ for $\xi=-0.2$. The corresponding data have substantially lower error bars due to the less severe FSP. At the same time, we find significant differences for $\Theta=0.75$, whereas all data are in good agreement for $\Theta=1$. Generally, we find that the computationally much less demanding calculations with $\xi=-0.2$ allow for an accuracy of $\sim0.1\%$ in both cases, which is sufficient for many applications. Finally, we remark that we find excellent agreement with the GDSMFB parametrization in all cases~\cite{groth_prl} .

\begin{figure*}[t!]
\center
\includegraphics[width=0.439\textwidth]{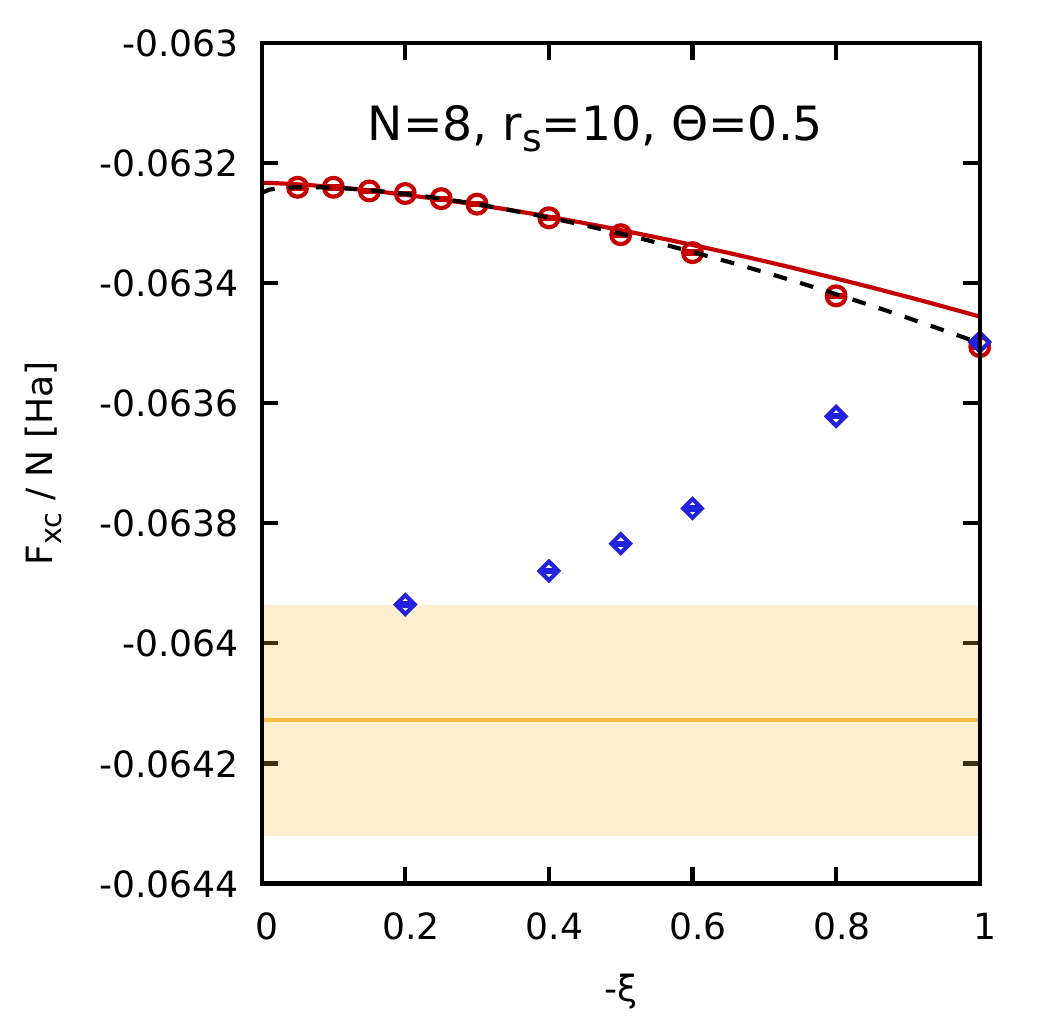}\includegraphics[width=0.439\textwidth]{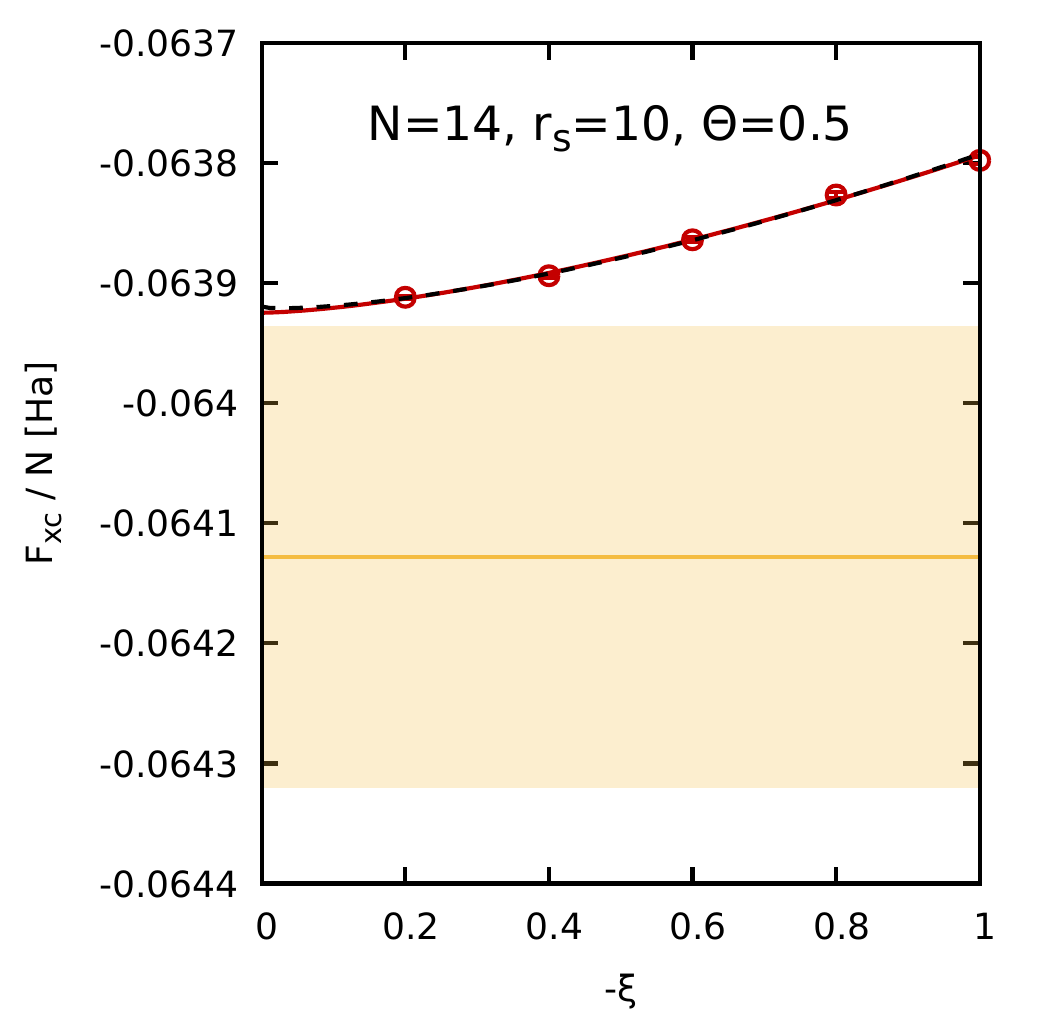}\\\includegraphics[width=0.439\textwidth]{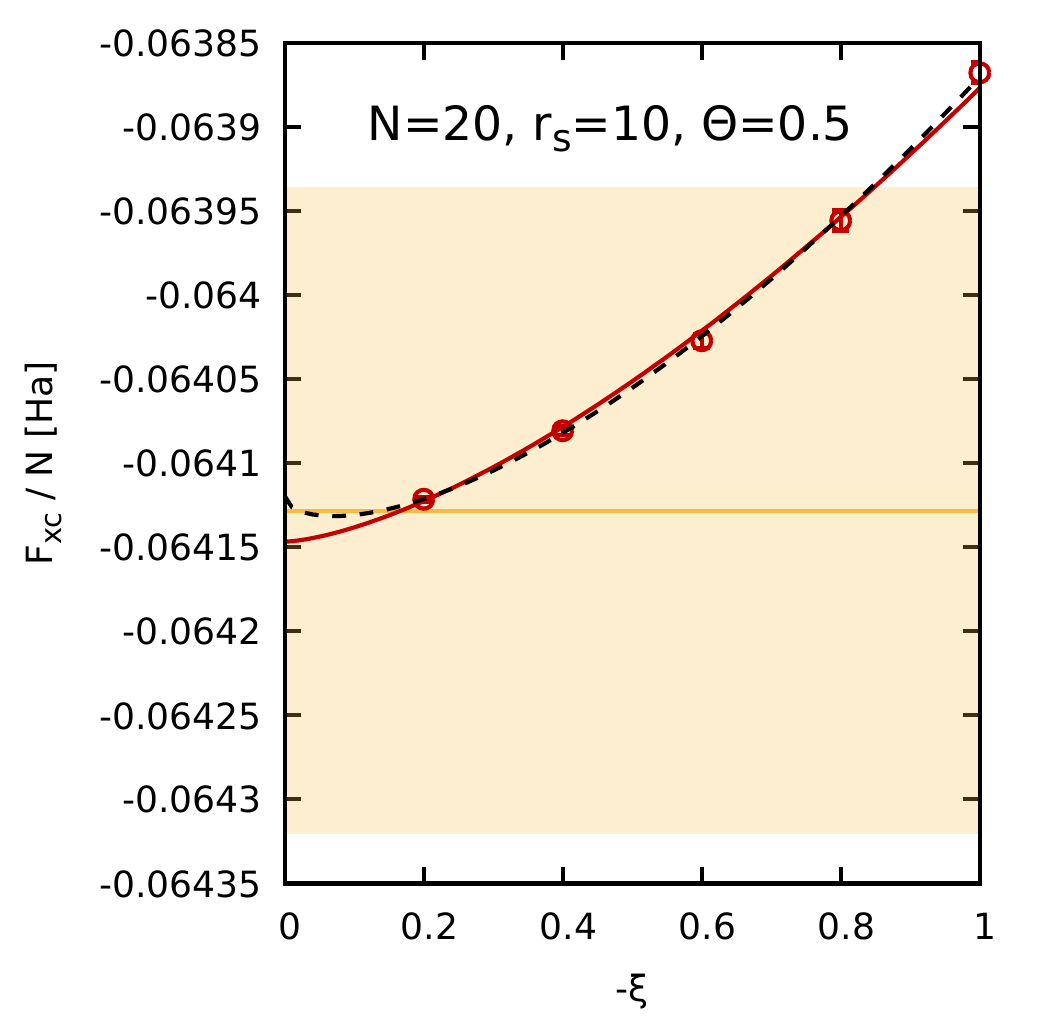}
\includegraphics[width=0.439\textwidth]{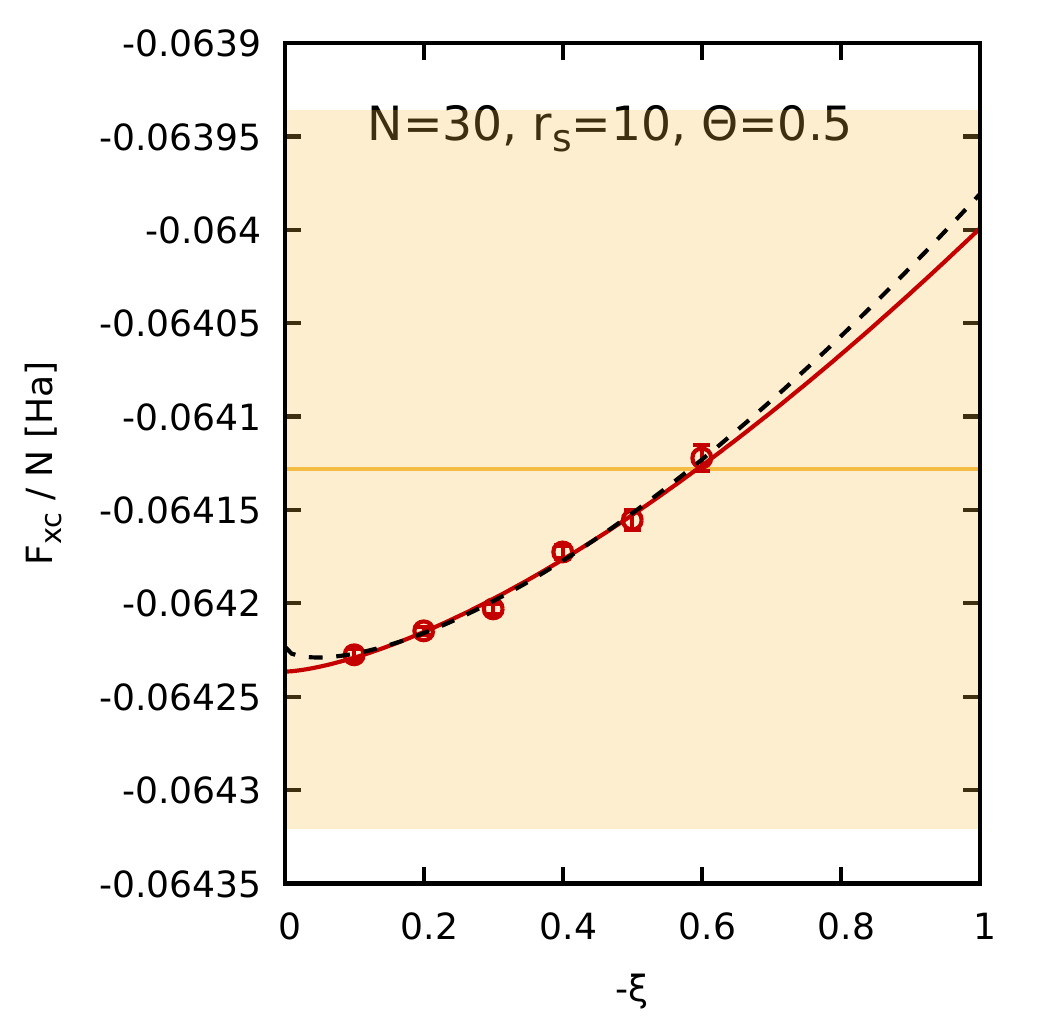}
\caption{ \label{fig:Fermionic_limit_rs10_theta0p5} 
Extrapolation of the XC-free energy per particle $F_\textnormal{XC}/N$ to the fermionic limit of $\xi=-1$ for different numbers of particles ($N=8,14,20,30$) at $r_s=10$ and $\Theta=0.5$. The red circles depict finite-size corrected PIMC results and the blue diamonds depict PIMC results obtained by adding the ideal correction $\Delta a_0(\xi)$ to the UEG data, see Eq.~(\ref{eq:correction}). The solid red and dashed black curves correspond to the two-parameter, see Eq.~(\ref{eq:fit1}), and three-parameter, see Eq.~(\ref{eq:fit2}), empirical fits. The yellow vertical line and shaded yellow areas correspond to the GDSMFB parametrization and its respective nominal uncertainty interval of $\pm0.3\%$~\cite{groth_prl}.
}
\end{figure*}

\begin{figure*}[t!]
\center
\includegraphics[width=0.439\textwidth]{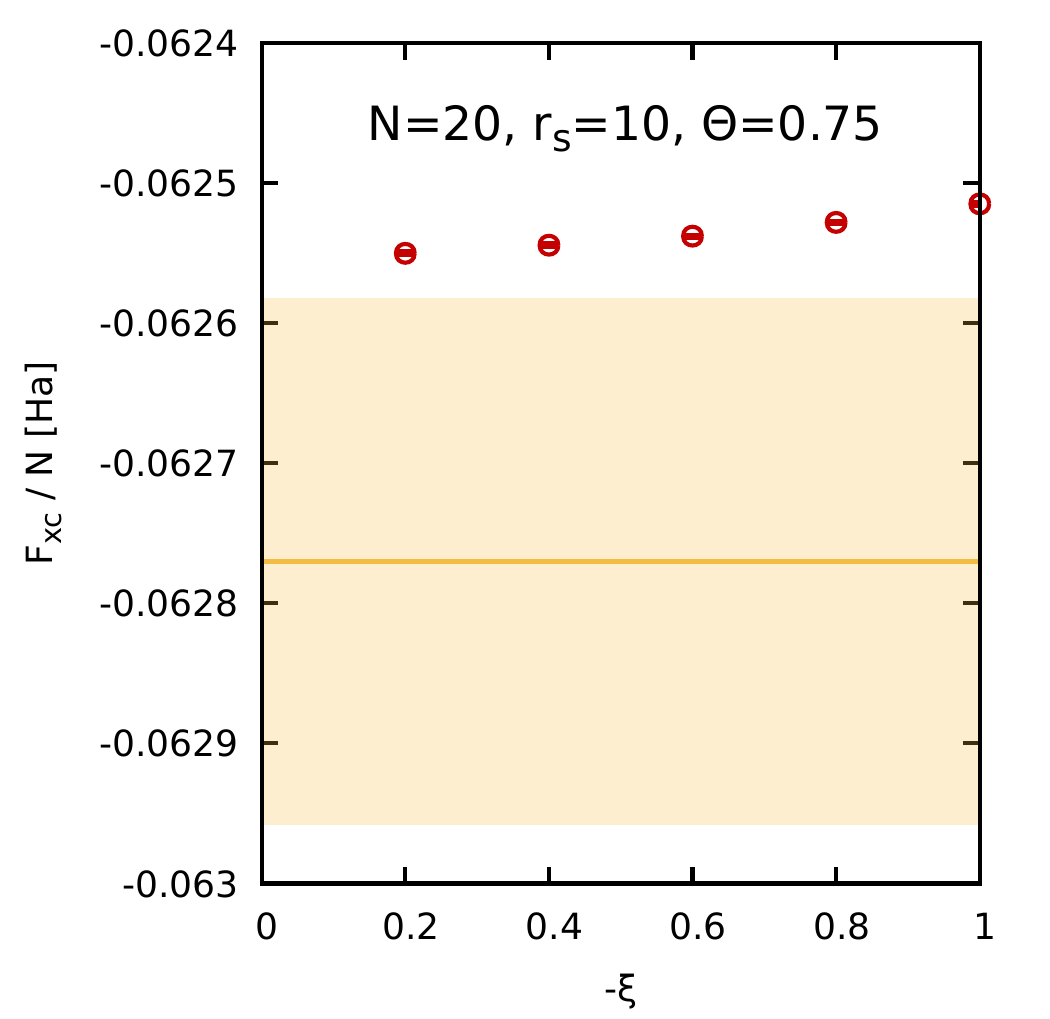}
\caption{ \label{fig:Fermionic_limit_rs10_theta0p75} 
Dependence of the XC-free energy per particle $F_\textnormal{XC}/N$ on the partition function parameter $\xi$ at $r_s=10$, $\Theta=0.75$, for different particle numbers $N$. The red circles depict finite-size corrected PIMC results. The yellow vertical line and shaded yellow areas correspond to the GDSMFB parametrization and its respective nominal uncertainty interval of $\pm0.3\%$~\cite{groth_prl}. 
}
\end{figure*}

\begin{figure*}[t!]
\center
\includegraphics[width=0.439\textwidth]{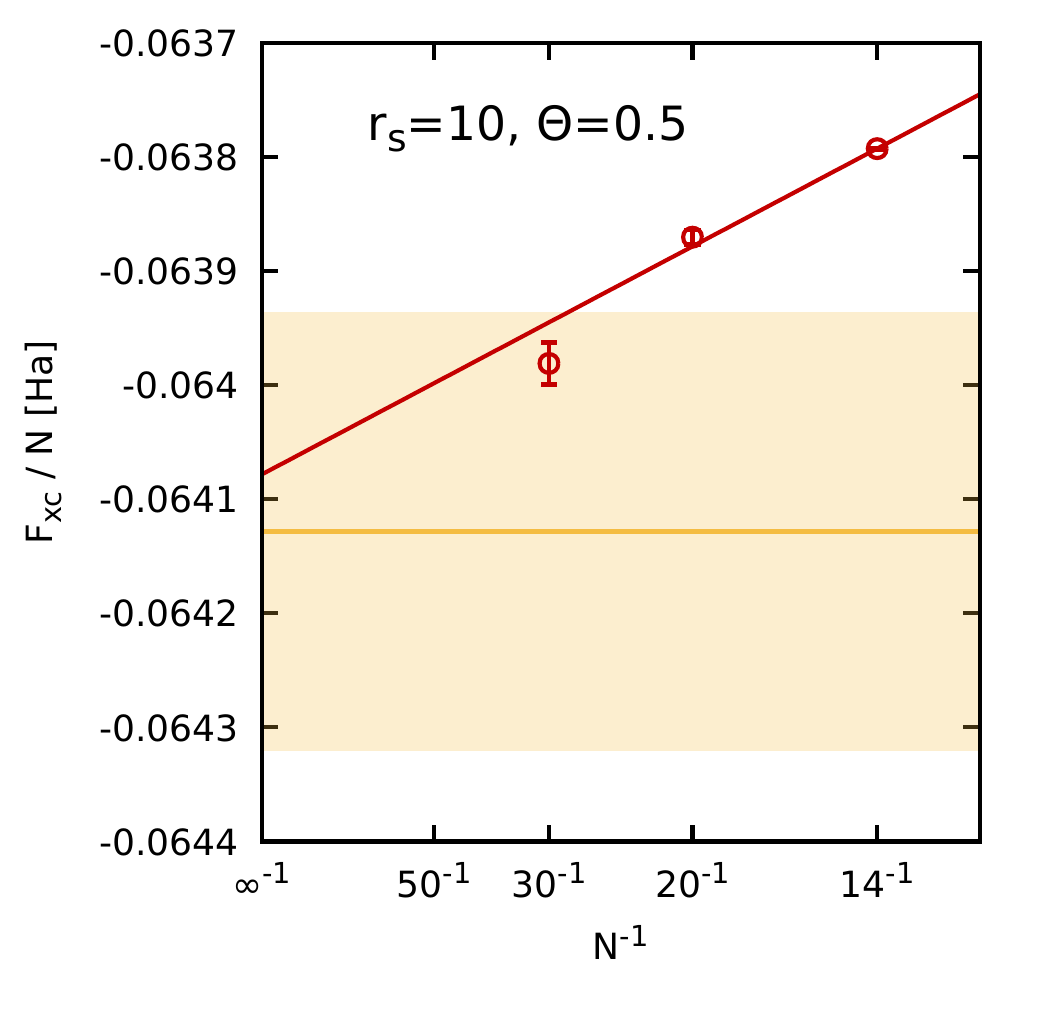}\includegraphics[width=0.439\textwidth]{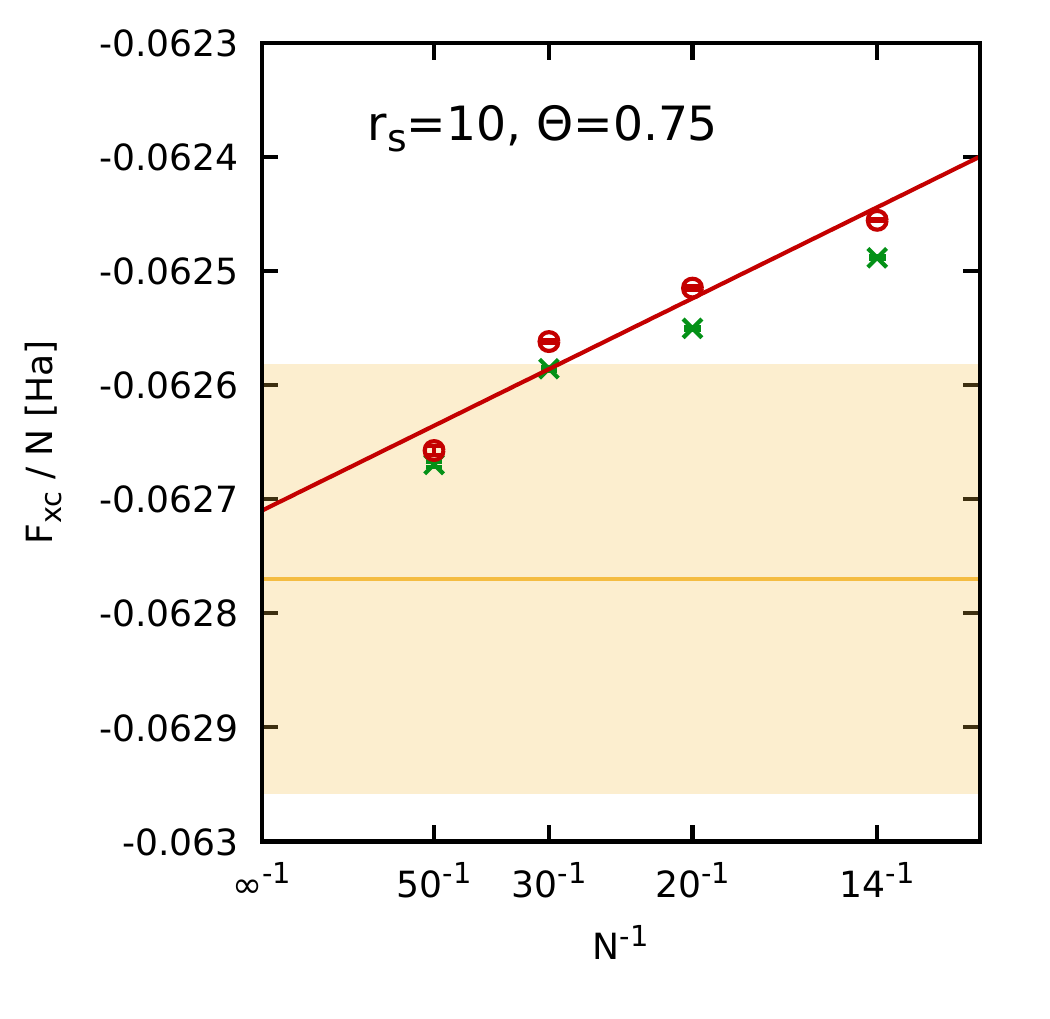}\\\includegraphics[width=0.439\textwidth]{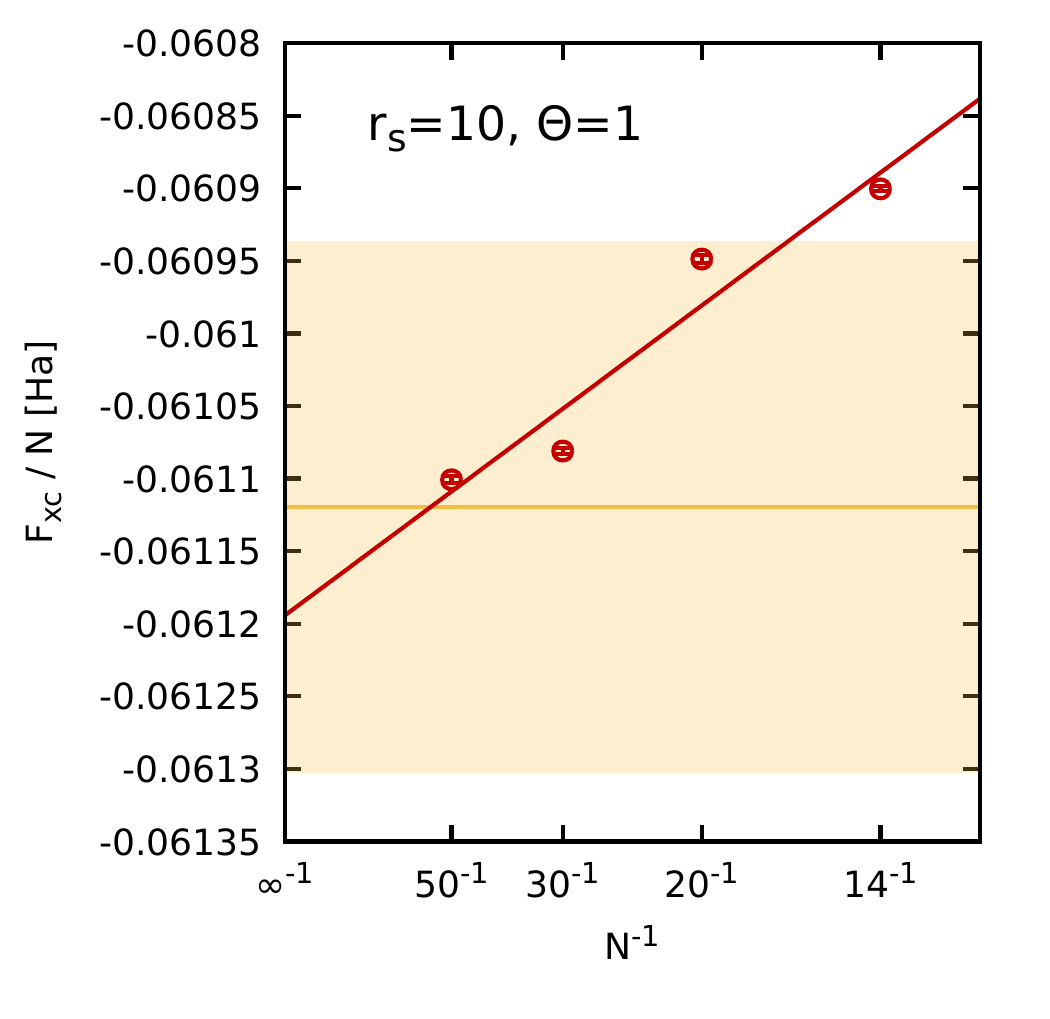}
\caption{ \label{fig:TDL_rs10} 
System-size dependence of the PIMC results for $F_\textnormal{xc}/N$ at $r_s=10$ and different temperatures ($\Theta=0.5,\,0.75,\,1$). The red symbols in the top left panel ($\Theta=0.5$) are obtained from an empirical extrapolation to the fermionic limit (see Fig.~\ref{fig:Fermionic_limit_rs10_theta0p5}), while the red symbols in the top right ($\Theta=0.75$) and bottom panels ($\Theta=1$) correspond to direct PIMC results at the fermionic limit $\xi=-1$. In addition, the green crosses correspond to PIMC results using $a(\xi)$ with $\xi=-0.2$. The yellow vertical line and shaded yellow areas correspond to the GDSMFB parametrization and its respective nominal uncertainty interval of $\pm0.3\%$~\cite{groth_prl}.
}
\end{figure*}

Let us next consider the case of $r_s=10$, i.e., the margin of the strongly coupled electron liquid regime. In Fig.~\ref{fig:Fermionic_limit_rs10_theta0p5}, we show the extrapolation of the finite-size corrected PIMC results for $F_\textnormal{xc} / N$ to the fermionic limit of $\xi=-1$ for different simulated numbers of electrons $N$ at the lowest considered temperature of $\Theta=0.5$. In addition, the blue diamonds have been obtained by applying the ideal correction [Eq.~(\ref{eq:correction})], which, however, does not constitute an improvement in the strongly coupled regime. Overall, we find that a reliable extrapolation to the limit of $\xi=-1$ is possible in all depicted cases. The corresponding extrapolation to the thermodynamic limit is shown in the top left panel of Fig.~\ref{fig:TDL_rs10} and is in excellent agreement with the GDSMFB parametrization~\cite{groth_prl}, as expected. Moreover, in Fig.~\ref{fig:Fermionic_limit_rs10_theta0p75}, we investigate the $\xi$-dependence of our PIMC results for $\Theta=0.75$, which is negligible. This observation is further substantiated by the top right panel of Fig.~\ref{fig:TDL_rs10} where the red circles and green crosses show PIMC results obtained using $a(-1)$ and $a(-0.2)$, respectively. Residual differences between the two datasets are much smaller than residual finite-size effects and can thus be neglected in practice. As before, we find excellent agreement between our new PIMC results for the thermodynamic limit and the GDSMFB parametrization~\cite{groth_prl} at both $\Theta=0.75$ and $\Theta=1$, as expected.

\begin{figure*}[t!]
\center
\includegraphics[width=0.439\textwidth]{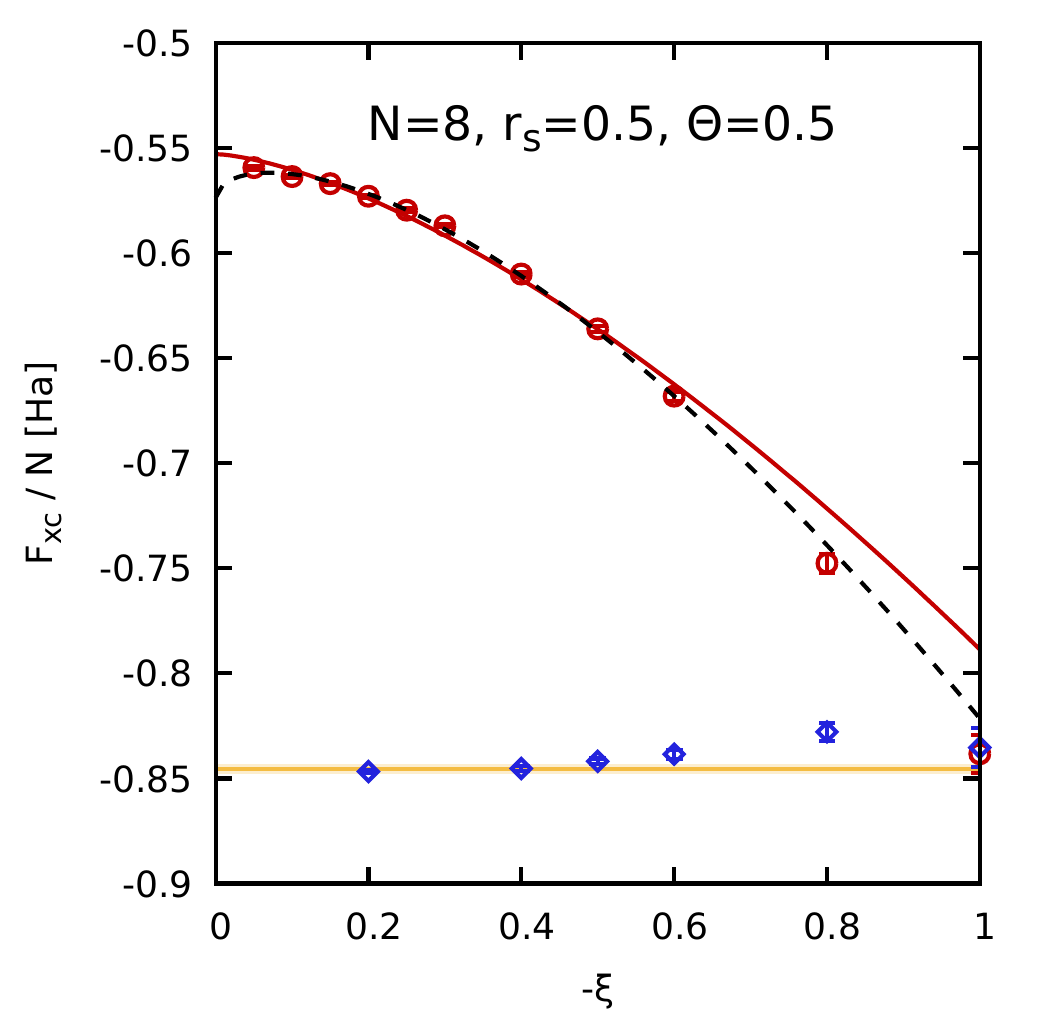}\includegraphics[width=0.439\textwidth]{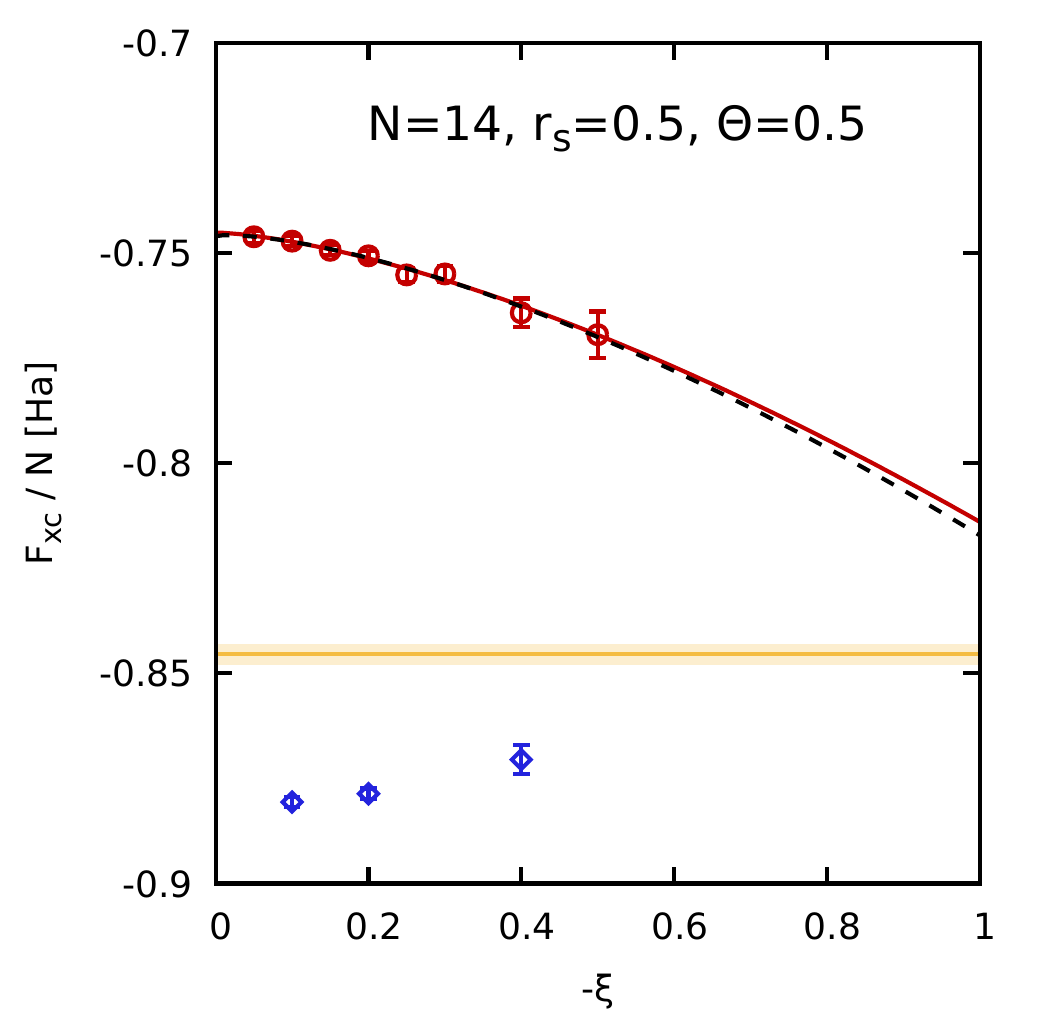}
\caption{ \label{fig:Fermionic_limit_rs0p5_theta0p5} 
Extrapolation of the XC-free energy per particle $F_\textnormal{XC}/N$ to the fermionic limit of $\xi=-1$ for different numbers of particles ($N=8,14$) at $r_s=0.5$ and $\Theta=0.5$. The red circles depict raw PIMC results and the blue diamonds depict PIMC results obtained by adding the ideal correction $\Delta a_0(\xi)$ to the UEG data, see Eq.~(\ref{eq:correction}). The solid red and dashed black curves correspond to the two-parameter, see Eq.~(\ref{eq:fit1}), and three-parameter, see Eq.~(\ref{eq:fit2}), empirical fits. The yellow vertical line and shaded yellow areas correspond to the GDSMFB parametrization and its respective nominal uncertainty interval of $\pm0.3\%$~\cite{groth_prl}.
}
\end{figure*}

\begin{figure*}[t!]
\center
\includegraphics[width=0.439\textwidth]{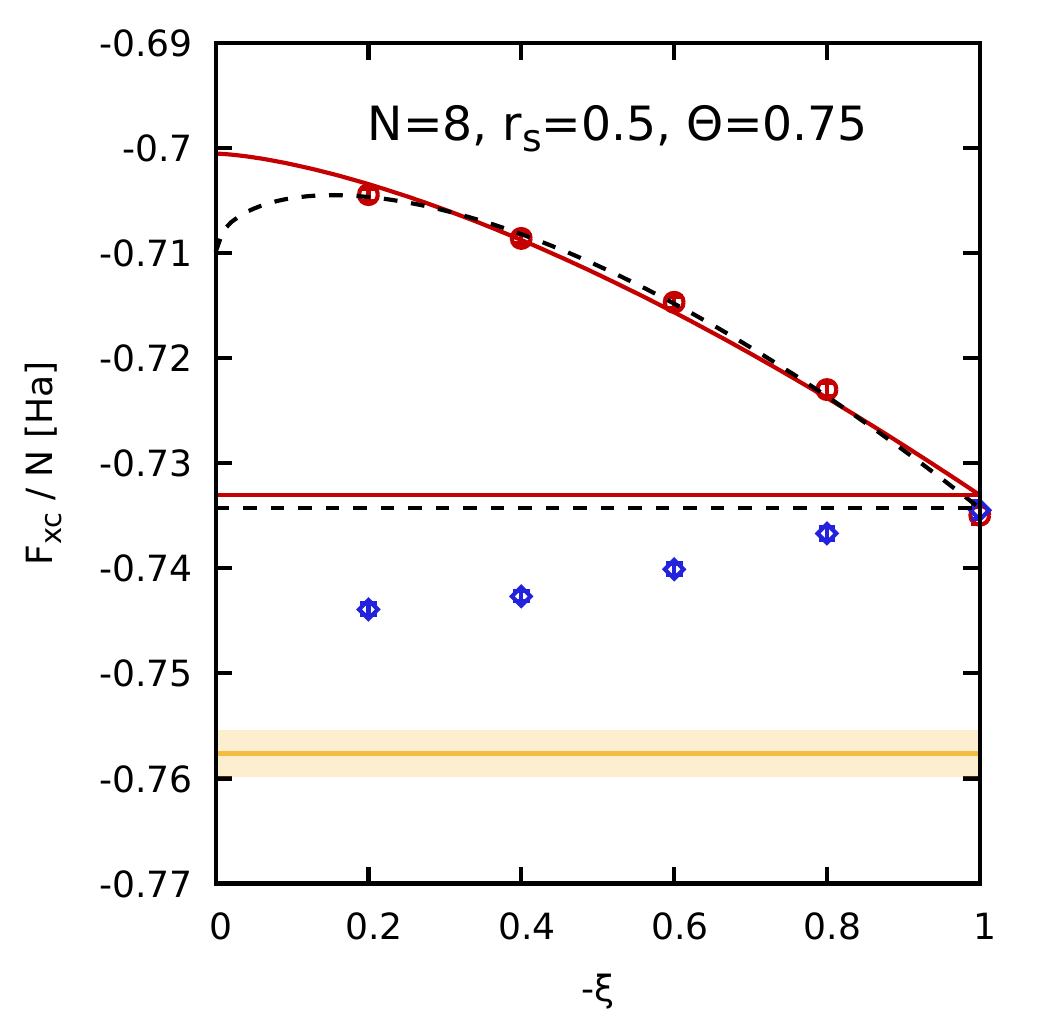}\includegraphics[width=0.439\textwidth]{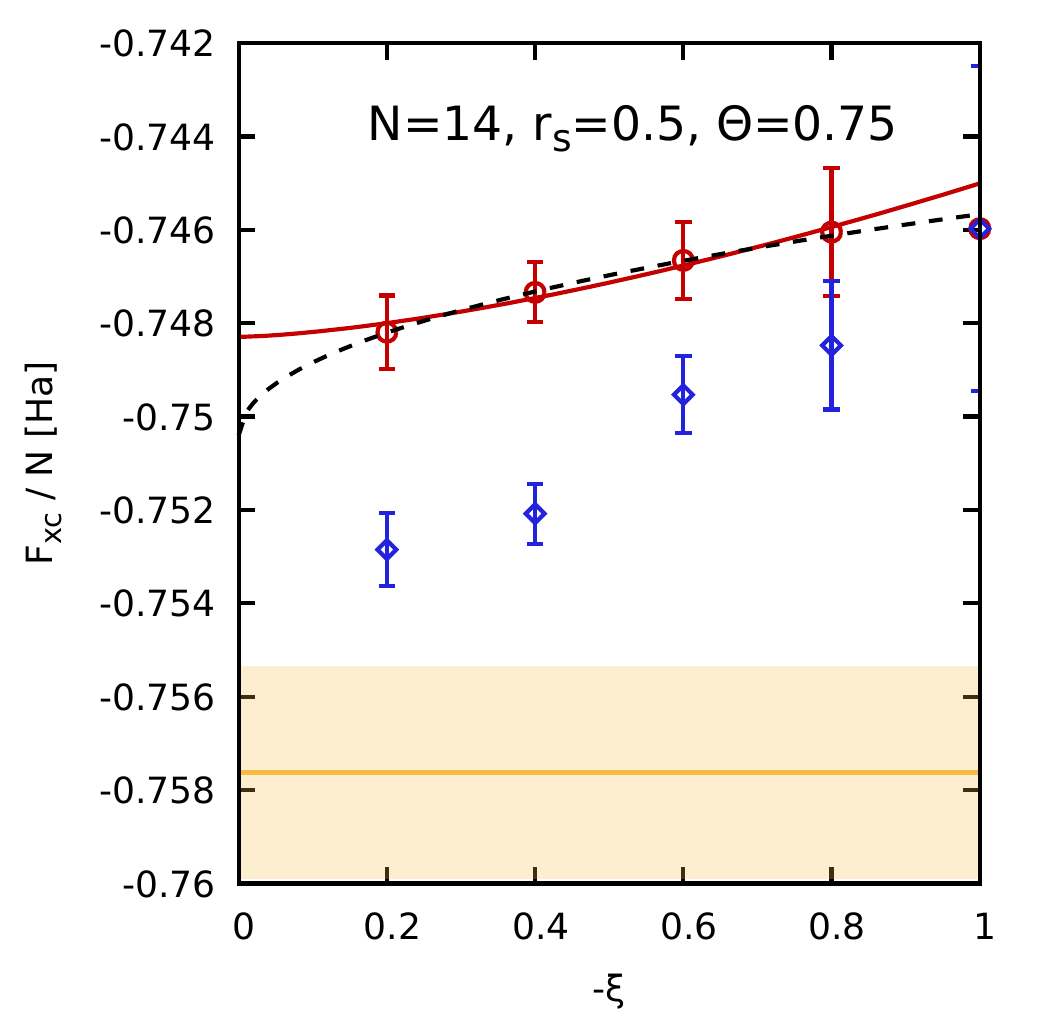}\\\includegraphics[width=0.439\textwidth]{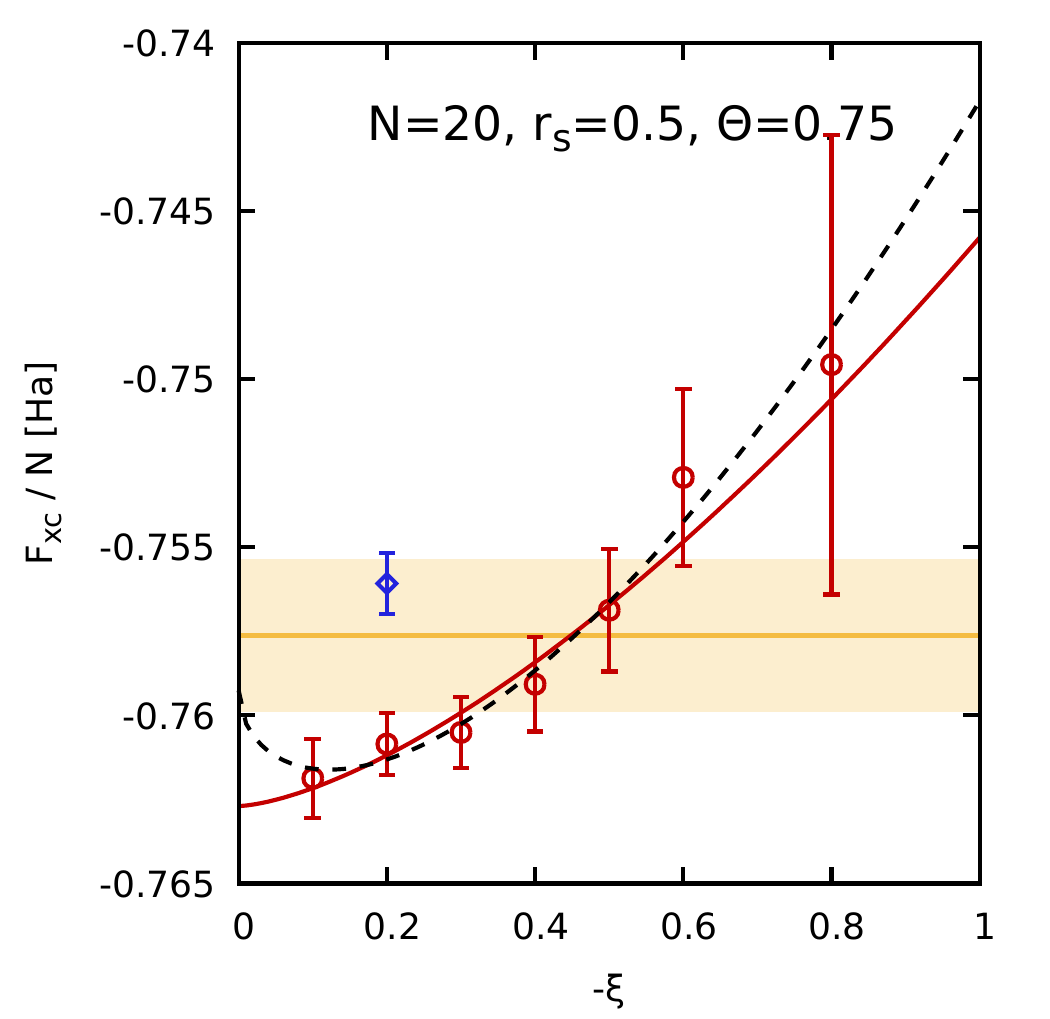}
\includegraphics[width=0.439\textwidth]{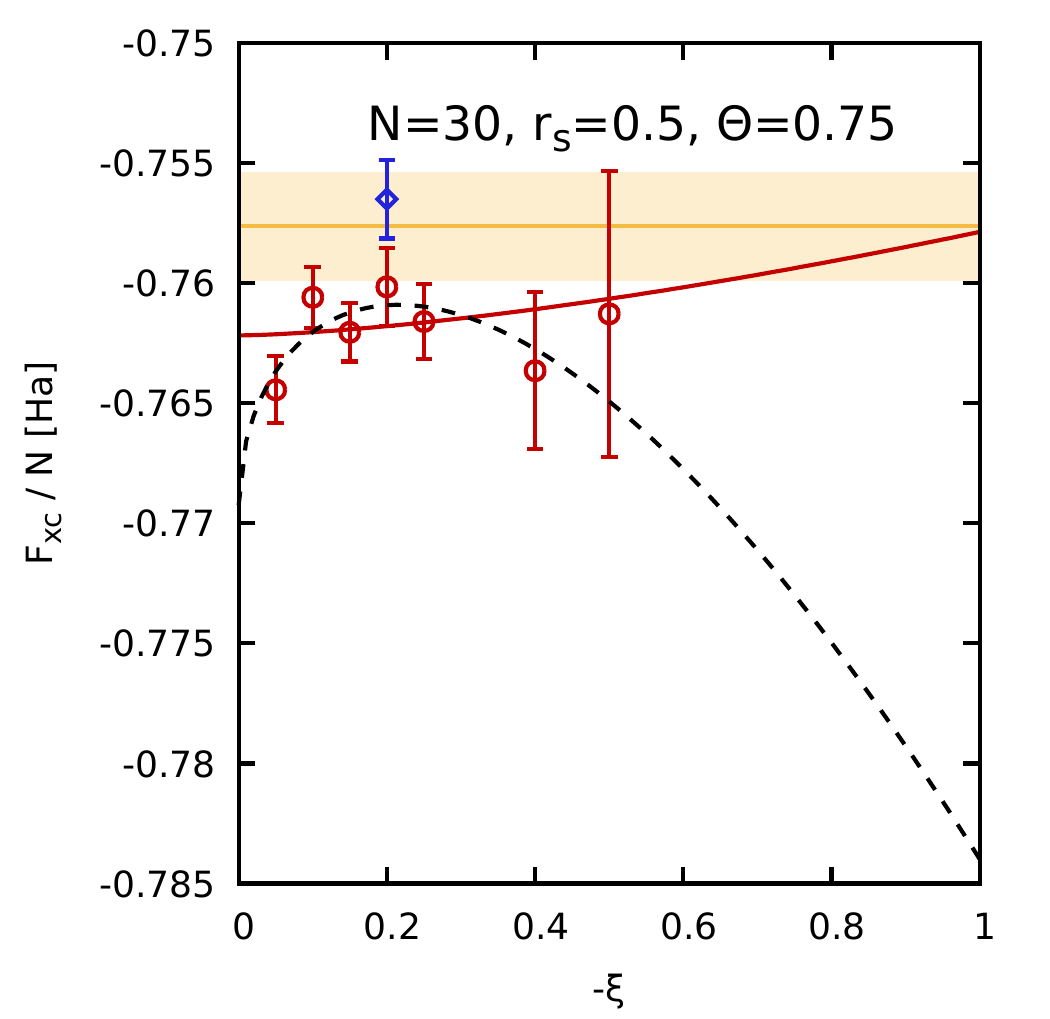}
\caption{ \label{fig:Fermionic_limit_rs0p5_theta0p75} 
Extrapolation of the XC-free energy per particle $F_\textnormal{XC}/N$ to the fermionic limit of $\xi=-1$ for different numbers of particles ($N=8,14,20,30$) at $r_s=0.5$ and $\Theta=0.75$. The red circles depict raw PIMC results and the blue diamonds depict PIMC results obtained by adding the ideal correction $\Delta a_0(\xi)$ to the UEG data, see Eq.~(\ref{eq:correction}). The solid red and dashed black curves correspond to the two-parameter, see Eq.~(\ref{eq:fit1}), and three-parameter, see Eq.~(\ref{eq:fit2}), empirical fits. The yellow vertical line and shaded yellow areas correspond to the GDSMFB parametrization and its respective nominal uncertainty interval of $\pm0.3\%$~\cite{groth_prl}.
}
\end{figure*}

\begin{figure*}[t!]
\center
\includegraphics[width=0.439\textwidth]{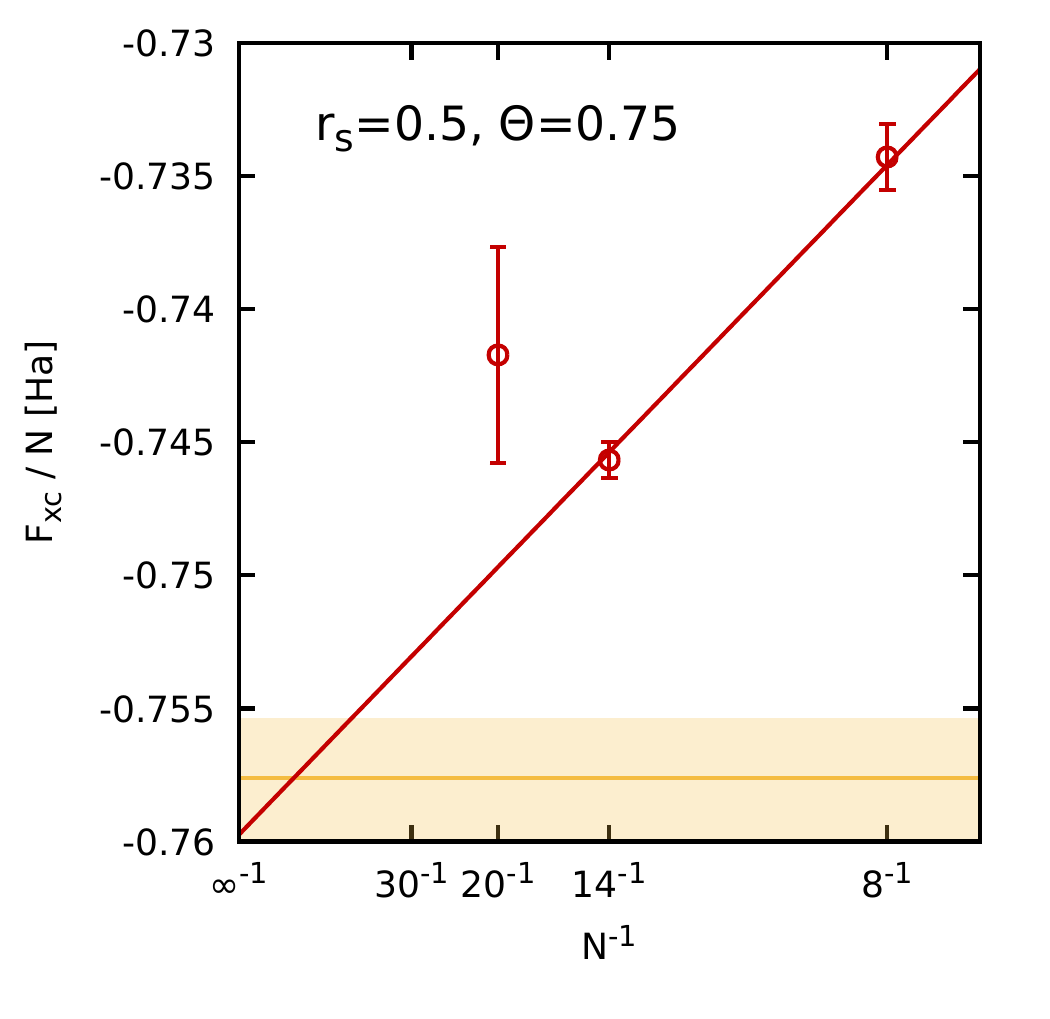}\includegraphics[width=0.439\textwidth]{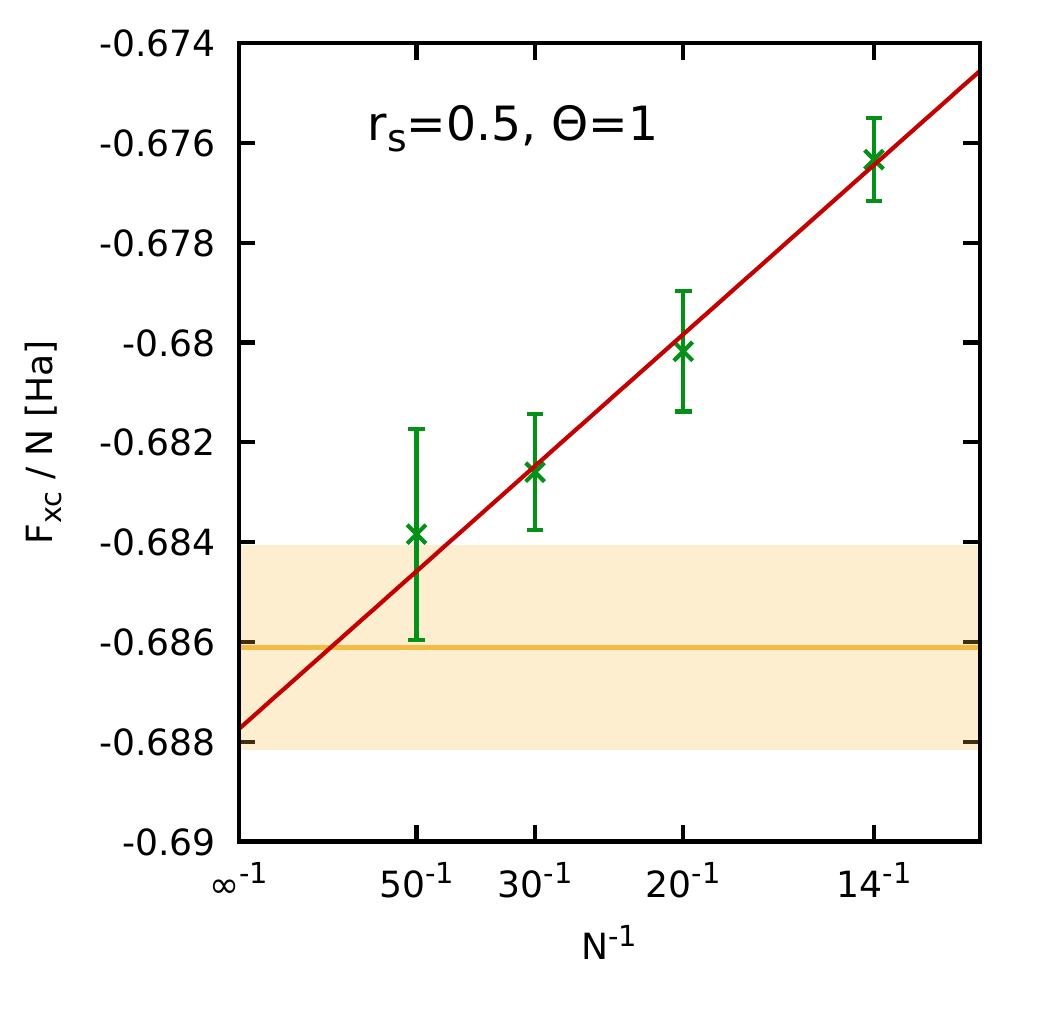}
\caption{ \label{fig:TDL_rs0p5} 
System-size dependence of the PIMC results for $F_\textnormal{xc}/N$ at $r_s=0.5$ and different temperatures ($\Theta=0.75,\,1$). The red symbols in the left panel ($\Theta=0.5$) are obtained from an empirical extrapolation to the fermionic limit (see Fig.~\ref{fig:Fermionic_limit_rs0p5_theta0p75}), while the green crosses in the right panel correspond to PIMC results using $a(\xi)$ with $\xi=-0.2$. The yellow vertical line and shaded yellow areas correspond to the GDSMFB parametrization and its respective nominal uncertainty interval of $\pm0.3\%$~\cite{groth_prl}.
}
\end{figure*}

\begin{figure*}[t!]
\center
\includegraphics[width=0.439\textwidth]{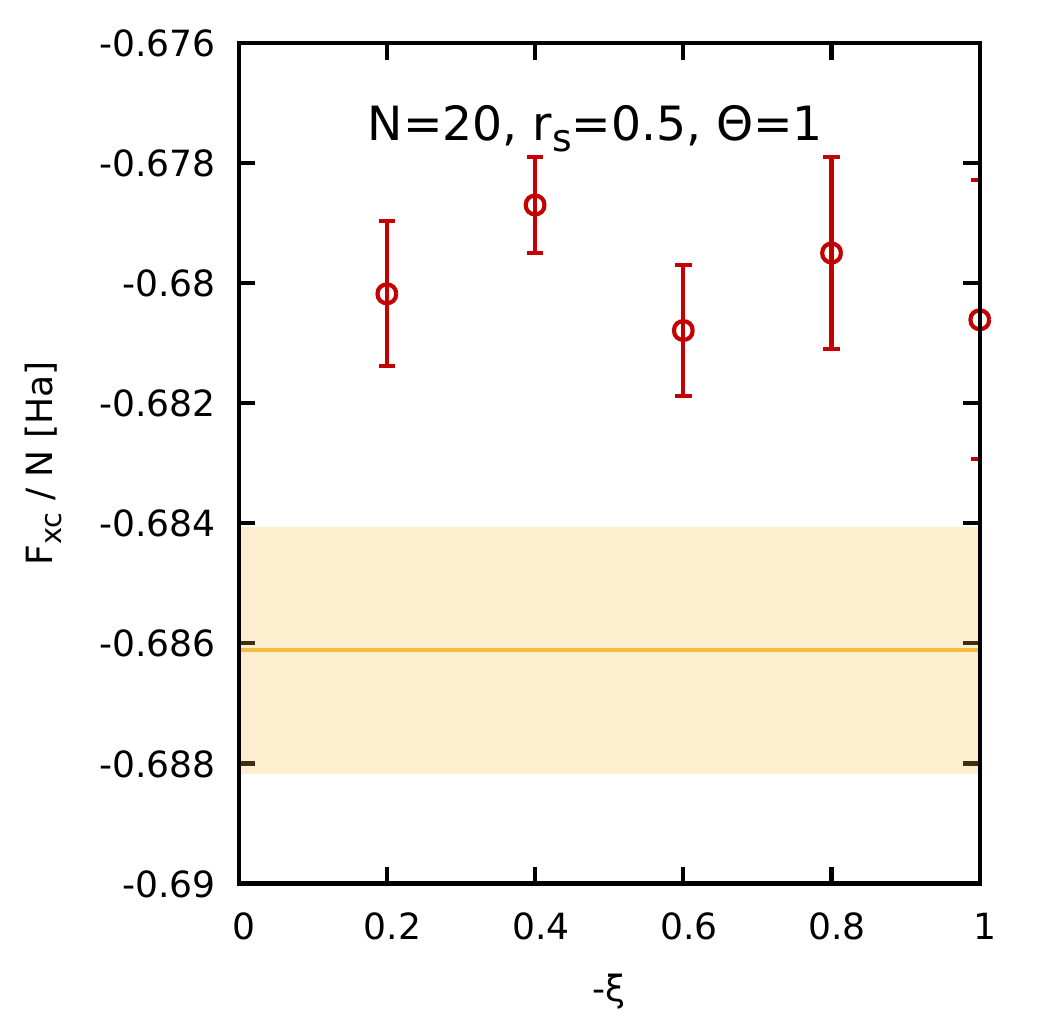}
\caption{ \label{fig:Fermionic_limit_rs0p5_theta1} 
Dependence of the XC-free energy per particle $F_\textnormal{XC}/N$ on the partition function parameter $\xi$ at $r_s=0.5$, $\Theta=1$, for $N=20$. The red circles depict finite-size corrected PIMC results. The yellow vertical line and shaded yellow areas correspond to the GDSMFB parametrization and its respective nominal uncertainty interval of $\pm0.3\%$~\cite{groth_prl}.
}
\end{figure*}

Let us continue our investigation by considering the most difficult high-density regime. In Fig.~\ref{fig:Fermionic_limit_rs0p5_theta0p5}, we investigate the $\xi$-dependence of our finite-size corrected PIMC results at $r_s=0.5$ and $\Theta=0.5$ for $N=8$ (left) and $N=14$ (right). Specifically, the red circles show raw PIMC results for different $\xi$ and the blue diamonds have been obtained by adding the ideal correction [Eq.~(\ref{eq:correction})]. Evidently, the correction removes some of the $\xi$-dependence, but the attained accuracy is not sufficient to meaningfully constrain existing models for $F_\textnormal{xc} / N$, cf.~the vertical yellow line corresponding to the GDSMFB parametrization~\cite{groth_prl}. Furthermore, we have performed two-parameter [cf.~Eq.~(\ref{eq:fit1})] and three-parameter [cf.~Eq.~(\ref{eq:fit2})] empirical fits within the range of $-0.4\leq\xi<0$ and the results are included as the solid red and dashed black curves, respectively. The latter fit performs particularly well and nicely reproduces our larger $\xi$ PIMC results that were not included into the fit input for $N=8$. In summary, our new approach is thus indeed capable of giving reasonable results for the free energy in the high-density low-temperature regime, but the accuracy is not high enough for the quantification of the comparably small XC-contribution to the full free energy. In Fig.~\ref{fig:Fermionic_limit_rs0p5_theta0p75}, we perform a similar analysis at $r_s=0.5$ and $\Theta=0.75$. Overall, we find that i) the ideal correction [Eq.~(\ref{eq:correction})] reduces the dependence on $\xi$ but is not decisive, ii) the extrapolation to the fermionic limit works well. The sole exception to the latter observation is given by the dashed black curve that corresponds to the three-parameter empirical fit [cf.~Eq.~(\ref{eq:fit2})] for $N=30$, which is evidently spurious.  The corresponding extrapolation of the residual finite-size errors to the thermodynamic limit is shown in the left panel of Fig.~\ref{fig:TDL_rs0p5}, where we again find good agreement with the result of the GDSMFB parametrization~\cite{groth_prl}. Finally, in Fig.~\ref{fig:Fermionic_limit_rs0p5_theta1}, we consider the dependence of our PIMC results for $F_\textnormal{xc} / N$ on $\xi$ for $\Theta=1$. At these conditions, the proposed linear expression for the free energy difference between boson and fermions [Eq.~(\ref{eq:F_from_a_new})] is very accurate and our results for $a(\xi)$ are independent of $\xi$. Corresponding results utilizing $a(-0.2)$ for different $N$ are shown in the right panel of Fig.~\ref{fig:TDL_rs0p5} and lead to the expected thermodynamic limit.

\section{Discussion}\label{sec:outlook}

We have combined the recent $\eta-$ensemble PIMC approach to the free energy~\cite{Dornheim_PRBL_2025,dornheim2024etaensemblepathintegralmonte} with the $\xi-$extrapolation technique~\cite{Xiong_JCP_2022,Dornheim_JCP_xi_2023} to deal with the fermion sign problem. This has allowed us to directly compute the XC-free energy of the UEG over a broad range of densities and temperatures, leading to the following conclusions. For moderate levels of quantum degeneracy ($\Theta\gtrsim1$), the average sign exhibits a simple exponential decay with $\xi$, leading to a straightforward linear expression for the corresponding free-energy contribution. As a consequence, it is possible to obtain highly accurate results for the free energy in the fermionic limit with substantially reduced computational cost. For higher levels of quantum degeneracy, i.e., $\Theta\lesssim0.75$, the linearity breaks down. Nevertheless, we find that an empirical extrapolation over $\xi$ works well in most cases. As a consequence, we have been able to obtain accurate results for $F_\textnormal{xc} / N$ down to $\Theta=0.5$, which was not possible with previous coordinate space PIMC methods without using approximate nodal restrictions.

In addition to being interesting in their own right, our results constitute an independent cross check of available parametrizations of the warm dense UEG in a regime where input data had been particularly sparse~\cite{review,status}. The excellent agreement with the GDSMFB parametrization~\cite{groth_prl} thus further substantiates the high quality of our state-of-the-art description of this archetypal model system. Future efforts might include the estimation of the free energy of real WDM systems starting with hydrogen~\cite{Bonitz_POP_2024,Dornheim_JCP_2024,Dornheim_MRE_2024} and other light elements potentially up to beryllium~\cite{Dornheim_JCP_2024,Dornheim_Science_2024}. We note that both ingredients to the present method---the $\eta-$ensemble approach and the $\xi$-extrapolation technique---can be applied to any Fermi system, including ultracold $^3$He~\cite{morresi2025normalliquid3hestudied,Dornheim_SciRep_2022,Ceperley_PRL_1992} and electrons in quantum dots~\cite{Egger_PRL_1999,Dornheim_NJP_2015,Xiong_JCP_2022,Dornheim_JCP_xi_2023}. From a methodological perspective, we mention possible future efficiency gains by utilizing clever treatments of the long-range Coulomb interaction~\cite{PhysRevX.13.031006} or implemented path contraction schemes~\cite{PhysRevE.93.043305} that are common practice in path-integral molecular dynamics simulations.

\section*{Data Availability}

The data supporting the findings of this study will be maide available on the Rossendorf Data Repository (RODARE) upon publication.

\section*{Acknowledgments}

This work was partially supported by the Center for Advanced Systems Understanding (CASUS) which is financed by Germany’s Federal Ministry of Education and Research (BMBF) and by the Saxon state government out of the State budget approved by the Saxon State Parliament.
This work has received funding from the European Research Council (ERC) under the European Union’s Horizon 2022 research and innovation programme
(Grant agreement No. 101076233, "PREXTREME").
The PIMC calculations were carried out at the Norddeutscher Verbund f\"ur Hoch- und H\"ochstleistungsrechnen (HLRN) under grant mvp00024 and on a Bull Cluster at the Center for Information Services and High Performance Computing (ZIH) at Technische Universit\"at Dresden.


\bibliography{acs-achemso}

\providecommand{\latin}[1]{#1}
\makeatletter
\providecommand{\doi}
  {\begingroup\let\do\@makeother\dospecials
  \catcode`\{=1 \catcode`\}=2 \doi@aux}
\providecommand{\doi@aux}[1]{\endgroup\texttt{#1}}
\makeatother
\providecommand*\mcitethebibliography{\thebibliography}
\csname @ifundefined\endcsname{endmcitethebibliography}  {\let\endmcitethebibliography\endthebibliography}{}
\begin{mcitethebibliography}{96}
\providecommand*\natexlab[1]{#1}
\providecommand*\mciteSetBstSublistMode[1]{}
\providecommand*\mciteSetBstMaxWidthForm[2]{}
\providecommand*\mciteBstWouldAddEndPuncttrue
  {\def\EndOfBibitem{\unskip.}}
\providecommand*\mciteBstWouldAddEndPunctfalse
  {\let\EndOfBibitem\relax}
\providecommand*\mciteSetBstMidEndSepPunct[3]{}
\providecommand*\mciteSetBstSublistLabelBeginEnd[3]{}
\providecommand*\EndOfBibitem{}
\mciteSetBstSublistMode{f}
\mciteSetBstMaxWidthForm{subitem}{(\alph{mcitesubitemcount})}
\mciteSetBstSublistLabelBeginEnd
  {\mcitemaxwidthsubitemform\space}
  {\relax}
  {\relax}

\bibitem[Ceperley(1995)]{cep}
Ceperley,~D.~M. Path integrals in the theory of condensed helium. \emph{Rev. Mod. Phys} \textbf{1995}, \emph{67}, 279\relax
\mciteBstWouldAddEndPuncttrue
\mciteSetBstMidEndSepPunct{\mcitedefaultmidpunct}
{\mcitedefaultendpunct}{\mcitedefaultseppunct}\relax
\EndOfBibitem
\bibitem[Kleinert(2009)]{kleinert2009path}
Kleinert,~H. \emph{Path Integrals in Quantum Mechanics, Statistics, Polymer Physics, and Financial Markets}; EBL-Schweitzer; World Scientific, 2009\relax
\mciteBstWouldAddEndPuncttrue
\mciteSetBstMidEndSepPunct{\mcitedefaultmidpunct}
{\mcitedefaultendpunct}{\mcitedefaultseppunct}\relax
\EndOfBibitem
\bibitem[Boninsegni \latin{et~al.}(2006)Boninsegni, Prokofev, and Svistunov]{boninsegni1}
Boninsegni,~M.; Prokofev,~N.~V.; Svistunov,~B.~V. Worm algorithm and diagrammatic {M}onte {C}arlo: A new approach to continuous-space path integral {M}onte {C}arlo simulations. \emph{Phys. Rev. E} \textbf{2006}, \emph{74}, 036701\relax
\mciteBstWouldAddEndPuncttrue
\mciteSetBstMidEndSepPunct{\mcitedefaultmidpunct}
{\mcitedefaultendpunct}{\mcitedefaultseppunct}\relax
\EndOfBibitem
\bibitem[Boninsegni \latin{et~al.}(2006)Boninsegni, Prokofev, and Svistunov]{boninsegni2}
Boninsegni,~M.; Prokofev,~N.~V.; Svistunov,~B.~V. Worm Algorithm for Continuous-Space Path Integral {M}onte {C}arlo Simulations. \emph{Phys. Rev. Lett} \textbf{2006}, \emph{96}, 070601\relax
\mciteBstWouldAddEndPuncttrue
\mciteSetBstMidEndSepPunct{\mcitedefaultmidpunct}
{\mcitedefaultendpunct}{\mcitedefaultseppunct}\relax
\EndOfBibitem
\bibitem[Filinov \latin{et~al.}(2010)Filinov, Prokof'ev, and Bonitz]{Filinov_PRL_2010}
Filinov,~A.; Prokof'ev,~N.~V.; Bonitz,~M. Berezinskii-Kosterlitz-Thouless Transition in Two-Dimensional Dipole Systems. \emph{Phys. Rev. Lett.} \textbf{2010}, \emph{105}, 070401\relax
\mciteBstWouldAddEndPuncttrue
\mciteSetBstMidEndSepPunct{\mcitedefaultmidpunct}
{\mcitedefaultendpunct}{\mcitedefaultseppunct}\relax
\EndOfBibitem
\bibitem[Morresi and Garberoglio(2025)Morresi, and Garberoglio]{morresi2025revisitingpropertiessuperfluidnormal}
Morresi,~T.; Garberoglio,~G. Revisiting the properties of superfluid and normal liquid ${}^4$He using ab initio potentials. 2025; \url{https://arxiv.org/abs/2501.08730}\relax
\mciteBstWouldAddEndPuncttrue
\mciteSetBstMidEndSepPunct{\mcitedefaultmidpunct}
{\mcitedefaultendpunct}{\mcitedefaultseppunct}\relax
\EndOfBibitem
\bibitem[Brown \latin{et~al.}(2013)Brown, Clark, DuBois, and Ceperley]{Brown_PRL_2013}
Brown,~E.~W.; Clark,~B.~K.; DuBois,~J.~L.; Ceperley,~D.~M. Path-Integral Monte Carlo Simulation of the Warm Dense Homogeneous Electron Gas. \emph{Phys. Rev. Lett.} \textbf{2013}, \emph{110}, 146405\relax
\mciteBstWouldAddEndPuncttrue
\mciteSetBstMidEndSepPunct{\mcitedefaultmidpunct}
{\mcitedefaultendpunct}{\mcitedefaultseppunct}\relax
\EndOfBibitem
\bibitem[Dornheim \latin{et~al.}(2016)Dornheim, Groth, Sjostrom, Malone, Foulkes, and Bonitz]{dornheim_prl}
Dornheim,~T.; Groth,~S.; Sjostrom,~T.; Malone,~F.~D.; Foulkes,~W. M.~C.; Bonitz,~M. Ab Initio Quantum {M}onte {C}arlo Simulation of the Warm Dense Electron Gas in the Thermodynamic Limit. \emph{Phys. Rev. Lett.} \textbf{2016}, \emph{117}, 156403\relax
\mciteBstWouldAddEndPuncttrue
\mciteSetBstMidEndSepPunct{\mcitedefaultmidpunct}
{\mcitedefaultendpunct}{\mcitedefaultseppunct}\relax
\EndOfBibitem
\bibitem[Militzer \latin{et~al.}(2021)Militzer, Gonz\'alez-Cataldo, Zhang, Driver, and Soubiran]{Militzer2021EOS}
Militzer,~B.; Gonz\'alez-Cataldo,~F.; Zhang,~S.; Driver,~K.~P.; Soubiran,~F. m.~c. First-principles equation of state database for warm dense matter computation. \emph{Phys. Rev. E} \textbf{2021}, \emph{103}, 013203\relax
\mciteBstWouldAddEndPuncttrue
\mciteSetBstMidEndSepPunct{\mcitedefaultmidpunct}
{\mcitedefaultendpunct}{\mcitedefaultseppunct}\relax
\EndOfBibitem
\bibitem[Bonitz \latin{et~al.}(2024)Bonitz, Vorberger, Bethkenhagen, Böhme, Ceperley, Filinov, Gawne, Graziani, Gregori, Hamann, Hansen, Holzmann, Hu, Kählert, Karasiev, Kleinschmidt, Kordts, Makait, Militzer, Moldabekov, Pierleoni, Preising, Ramakrishna, Redmer, Schwalbe, Svensson, and Dornheim]{Bonitz_POP_2024}
Bonitz,~M.; Vorberger,~J.; Bethkenhagen,~M.; Böhme,~M.~P.; Ceperley,~D.~M.; Filinov,~A.; Gawne,~T.; Graziani,~F.; Gregori,~G.; Hamann,~P.; Hansen,~S.~B.; Holzmann,~M.; Hu,~S.~X.; Kählert,~H.; Karasiev,~V.~V.; Kleinschmidt,~U.; Kordts,~L.; Makait,~C.; Militzer,~B.; Moldabekov,~Z.~A.; Pierleoni,~C.; Preising,~M.; Ramakrishna,~K.; Redmer,~R.; Schwalbe,~S.; Svensson,~P.; Dornheim,~T. Toward first principles-based simulations of dense hydrogen. \emph{Phys. Plasmas} \textbf{2024}, \emph{31}, 110501\relax
\mciteBstWouldAddEndPuncttrue
\mciteSetBstMidEndSepPunct{\mcitedefaultmidpunct}
{\mcitedefaultendpunct}{\mcitedefaultseppunct}\relax
\EndOfBibitem
\bibitem[Dornheim \latin{et~al.}(2019)Dornheim, Vorberger, Groth, Hoffmann, Moldabekov, and Bonitz]{dornheim_ML}
Dornheim,~T.; Vorberger,~J.; Groth,~S.; Hoffmann,~N.; Moldabekov,~Z.; Bonitz,~M. The Static Local Field Correction of the Warm Dense Electron Gas: An ab Initio Path Integral {M}onte {C}arlo Study and Machine Learning Representation. \emph{J. Chem. Phys} \textbf{2019}, \emph{151}, 194104\relax
\mciteBstWouldAddEndPuncttrue
\mciteSetBstMidEndSepPunct{\mcitedefaultmidpunct}
{\mcitedefaultendpunct}{\mcitedefaultseppunct}\relax
\EndOfBibitem
\bibitem[Dornheim \latin{et~al.}(2021)Dornheim, Moldabekov, and Vorberger]{Dornheim_JCP_ITCF_2021}
Dornheim,~T.; Moldabekov,~Z.~A.; Vorberger,~J. Nonlinear density response from imaginary-time correlation functions: Ab initio path integral Monte Carlo simulations of the warm dense electron gas. \emph{J. Chem. Phys.} \textbf{2021}, \emph{155}, 054110\relax
\mciteBstWouldAddEndPuncttrue
\mciteSetBstMidEndSepPunct{\mcitedefaultmidpunct}
{\mcitedefaultendpunct}{\mcitedefaultseppunct}\relax
\EndOfBibitem
\bibitem[Dornheim \latin{et~al.}(2020)Dornheim, Vorberger, and Bonitz]{Dornheim_PRL_2020}
Dornheim,~T.; Vorberger,~J.; Bonitz,~M. Nonlinear Electronic Density Response in Warm Dense Matter. \emph{Phys. Rev. Lett.} \textbf{2020}, \emph{125}, 085001\relax
\mciteBstWouldAddEndPuncttrue
\mciteSetBstMidEndSepPunct{\mcitedefaultmidpunct}
{\mcitedefaultendpunct}{\mcitedefaultseppunct}\relax
\EndOfBibitem
\bibitem[Dornheim \latin{et~al.}(2023)Dornheim, Moldabekov, Ramakrishna, Tolias, Baczewski, Kraus, Preston, Chapman, Böhme, Döppner, Graziani, Bonitz, Cangi, and Vorberger]{Dornheim_review}
Dornheim,~T.; Moldabekov,~Z.~A.; Ramakrishna,~K.; Tolias,~P.; Baczewski,~A.~D.; Kraus,~D.; Preston,~T.~R.; Chapman,~D.~A.; Böhme,~M.~P.; Döppner,~T.; Graziani,~F.; Bonitz,~M.; Cangi,~A.; Vorberger,~J. {Electronic density response of warm dense matter}. \emph{Phys. Plasmas} \textbf{2023}, \emph{30}, 032705\relax
\mciteBstWouldAddEndPuncttrue
\mciteSetBstMidEndSepPunct{\mcitedefaultmidpunct}
{\mcitedefaultendpunct}{\mcitedefaultseppunct}\relax
\EndOfBibitem
\bibitem[Kora and Boninsegni(2018)Kora, and Boninsegni]{Boninsegni_maximum_entropy}
Kora,~Y.; Boninsegni,~M. Dynamic structure factor of superfluid $^{4}\mathrm{He}$ from quantum Monte Carlo: Maximum entropy revisited. \emph{Phys. Rev. B} \textbf{2018}, \emph{98}, 134509\relax
\mciteBstWouldAddEndPuncttrue
\mciteSetBstMidEndSepPunct{\mcitedefaultmidpunct}
{\mcitedefaultendpunct}{\mcitedefaultseppunct}\relax
\EndOfBibitem
\bibitem[Filinov and Bonitz(2012)Filinov, and Bonitz]{Filinov_PRA_2012}
Filinov,~A.; Bonitz,~M. Collective and single-particle excitations in two-dimensional dipolar Bose gases. \emph{Phys. Rev. A} \textbf{2012}, \emph{86}, 043628\relax
\mciteBstWouldAddEndPuncttrue
\mciteSetBstMidEndSepPunct{\mcitedefaultmidpunct}
{\mcitedefaultendpunct}{\mcitedefaultseppunct}\relax
\EndOfBibitem
\bibitem[Ferr\'e and Boronat(2016)Ferr\'e, and Boronat]{Ferre_PRB_2016}
Ferr\'e,~G.; Boronat,~J. Dynamic structure factor of liquid $^{4}\mathrm{He}$ across the normal-superfluid transition. \emph{Phys. Rev. B} \textbf{2016}, \emph{93}, 104510\relax
\mciteBstWouldAddEndPuncttrue
\mciteSetBstMidEndSepPunct{\mcitedefaultmidpunct}
{\mcitedefaultendpunct}{\mcitedefaultseppunct}\relax
\EndOfBibitem
\bibitem[Dornheim \latin{et~al.}(2018)Dornheim, Groth, Vorberger, and Bonitz]{dornheim_dynamic}
Dornheim,~T.; Groth,~S.; Vorberger,~J.; Bonitz,~M. Ab initio Path Integral {M}onte {C}arlo Results for the Dynamic Structure Factor of Correlated Electrons: From the Electron Liquid to Warm Dense Matter. \emph{Phys. Rev. Lett.} \textbf{2018}, \emph{121}, 255001\relax
\mciteBstWouldAddEndPuncttrue
\mciteSetBstMidEndSepPunct{\mcitedefaultmidpunct}
{\mcitedefaultendpunct}{\mcitedefaultseppunct}\relax
\EndOfBibitem
\bibitem[Chuna \latin{et~al.}(2025)Chuna, Barnfield, Dornheim, Friedlander, and Hoheisel]{chuna2025dualformulationmaximumentropy}
Chuna,~T.; Barnfield,~N.; Dornheim,~T.; Friedlander,~M.~P.; Hoheisel,~T. Dual formulation of the maximum entropy method applied to analytic continuation of quantum Monte Carlo data. 2025; \url{https://arxiv.org/abs/2501.01869}\relax
\mciteBstWouldAddEndPuncttrue
\mciteSetBstMidEndSepPunct{\mcitedefaultmidpunct}
{\mcitedefaultendpunct}{\mcitedefaultseppunct}\relax
\EndOfBibitem
\bibitem[Vitali \latin{et~al.}(2010)Vitali, Rossi, Reatto, and Galli]{Vitali_PRB_2010}
Vitali,~E.; Rossi,~M.; Reatto,~L.; Galli,~D.~E. Ab initio low-energy dynamics of superfluid and solid $^{4}\textnormal{H}\textnormal{e}$. \emph{Phys. Rev. B} \textbf{2010}, \emph{82}, 174510\relax
\mciteBstWouldAddEndPuncttrue
\mciteSetBstMidEndSepPunct{\mcitedefaultmidpunct}
{\mcitedefaultendpunct}{\mcitedefaultseppunct}\relax
\EndOfBibitem
\bibitem[Jarrell and Gubernatis(1996)Jarrell, and Gubernatis]{Jarrel_PhysReports_1996}
Jarrell,~M.; Gubernatis,~J. Bayesian inference and the analytic continuation of imaginary-time quantum Monte Carlo data. \emph{Phys. Rep.} \textbf{1996}, \emph{269}, 133--195\relax
\mciteBstWouldAddEndPuncttrue
\mciteSetBstMidEndSepPunct{\mcitedefaultmidpunct}
{\mcitedefaultendpunct}{\mcitedefaultseppunct}\relax
\EndOfBibitem
\bibitem[Lyubartsev \latin{et~al.}(1992)Lyubartsev, Martsinovski, Shevkunov, and Vorontsov‐Velyaminov]{Lyubartsev_JCP_1992}
Lyubartsev,~A.~P.; Martsinovski,~A.~A.; Shevkunov,~S.~V.; Vorontsov‐Velyaminov,~P.~N. New approach to Monte Carlo calculation of the free energy: Method of expanded ensembles. \emph{J. Chem. Phys.} \textbf{1992}, \emph{96}, 1776--1783\relax
\mciteBstWouldAddEndPuncttrue
\mciteSetBstMidEndSepPunct{\mcitedefaultmidpunct}
{\mcitedefaultendpunct}{\mcitedefaultseppunct}\relax
\EndOfBibitem
\bibitem[Dornheim \latin{et~al.}(2025)Dornheim, Moldabekov, Schwalbe, and Vorberger]{Dornheim_PRBL_2025}
Dornheim,~T.; Moldabekov,~Z.; Schwalbe,~S.; Vorberger,~J. Direct free energy calculation from ab initio path integral Monte Carlo simulations of warm dense matter. \emph{Phys. Rev. B} \textbf{2025}, \emph{111}, L041114\relax
\mciteBstWouldAddEndPuncttrue
\mciteSetBstMidEndSepPunct{\mcitedefaultmidpunct}
{\mcitedefaultendpunct}{\mcitedefaultseppunct}\relax
\EndOfBibitem
\bibitem[Dornheim \latin{et~al.}(2024)Dornheim, Tolias, Moldabekov, and Vorberger]{dornheim2024etaensemblepathintegralmonte}
Dornheim,~T.; Tolias,~P.; Moldabekov,~Z.; Vorberger,~J. eta-ensemble path integral Monte Carlo approach to the free energy of the warm dense electron gas and the uniform electron liquid. 2024; \url{https://arxiv.org/abs/2412.13596}\relax
\mciteBstWouldAddEndPuncttrue
\mciteSetBstMidEndSepPunct{\mcitedefaultmidpunct}
{\mcitedefaultendpunct}{\mcitedefaultseppunct}\relax
\EndOfBibitem
\bibitem[Zhou and Dai(2018)Zhou, and Dai]{Zhou_2018}
Zhou,~C.-C.; Dai,~W.-S. Canonical partition functions: ideal quantum gases, interacting classical gases, and interacting quantum gases. \emph{J. Stat. Mech.: Theory Exp.} \textbf{2018}, \emph{2018}, 023105\relax
\mciteBstWouldAddEndPuncttrue
\mciteSetBstMidEndSepPunct{\mcitedefaultmidpunct}
{\mcitedefaultendpunct}{\mcitedefaultseppunct}\relax
\EndOfBibitem
\bibitem[Troyer and Wiese(2005)Troyer, and Wiese]{troyer}
Troyer,~M.; Wiese,~U.~J. Computational Complexity and Fundamental Limitations to Fermionic Quantum {M}onte {C}arlo Simulations. \emph{Phys. Rev. Lett} \textbf{2005}, \emph{94}, 170201\relax
\mciteBstWouldAddEndPuncttrue
\mciteSetBstMidEndSepPunct{\mcitedefaultmidpunct}
{\mcitedefaultendpunct}{\mcitedefaultseppunct}\relax
\EndOfBibitem
\bibitem[Dornheim(2019)]{dornheim_sign_problem}
Dornheim,~T. Fermion sign problem in path integral {M}onte {C}arlo simulations: Quantum dots, ultracold atoms, and warm dense matter. \emph{Phys. Rev. E} \textbf{2019}, \emph{100}, 023307\relax
\mciteBstWouldAddEndPuncttrue
\mciteSetBstMidEndSepPunct{\mcitedefaultmidpunct}
{\mcitedefaultendpunct}{\mcitedefaultseppunct}\relax
\EndOfBibitem
\bibitem[Dornheim(2021)]{Dornheim_JPA_2021}
Dornheim,~T. Fermion sign problem in path integral Monte Carlo simulations: grand-canonical ensemble. \emph{J. Phys. A: Math. Theor.} \textbf{2021}, \emph{54}, 335001\relax
\mciteBstWouldAddEndPuncttrue
\mciteSetBstMidEndSepPunct{\mcitedefaultmidpunct}
{\mcitedefaultendpunct}{\mcitedefaultseppunct}\relax
\EndOfBibitem
\bibitem[Egger \latin{et~al.}(1999)Egger, H\"ausler, Mak, and Grabert]{Egger_PRL_1999}
Egger,~R.; H\"ausler,~W.; Mak,~C.~H.; Grabert,~H. Crossover from Fermi Liquid to Wigner Molecule Behavior in Quantum Dots. \emph{Phys. Rev. Lett.} \textbf{1999}, \emph{82}, 3320--3323\relax
\mciteBstWouldAddEndPuncttrue
\mciteSetBstMidEndSepPunct{\mcitedefaultmidpunct}
{\mcitedefaultendpunct}{\mcitedefaultseppunct}\relax
\EndOfBibitem
\bibitem[Reusch \latin{et~al.}(2001)Reusch, H\"ausler, and Grabert]{Reimann_PRB_2000}
Reusch,~B.; H\"ausler,~W.; Grabert,~H. Wigner molecules in quantum dots. \emph{Phys. Rev. B} \textbf{2001}, \emph{63}, 113313\relax
\mciteBstWouldAddEndPuncttrue
\mciteSetBstMidEndSepPunct{\mcitedefaultmidpunct}
{\mcitedefaultendpunct}{\mcitedefaultseppunct}\relax
\EndOfBibitem
\bibitem[Clark \latin{et~al.}(2009)Clark, Casula, and Ceperley]{Clark_PRL_2009}
Clark,~B.~K.; Casula,~M.; Ceperley,~D.~M. Hexatic and Mesoscopic Phases in a 2D Quantum Coulomb System. \emph{Phys. Rev. Lett.} \textbf{2009}, \emph{103}, 055701\relax
\mciteBstWouldAddEndPuncttrue
\mciteSetBstMidEndSepPunct{\mcitedefaultmidpunct}
{\mcitedefaultendpunct}{\mcitedefaultseppunct}\relax
\EndOfBibitem
\bibitem[Azadi and Drummond(2022)Azadi, and Drummond]{Azadi_PRB_2022}
Azadi,~S.; Drummond,~N.~D. Low-density phase diagram of the three-dimensional electron gas. \emph{Phys. Rev. B} \textbf{2022}, \emph{105}, 245135\relax
\mciteBstWouldAddEndPuncttrue
\mciteSetBstMidEndSepPunct{\mcitedefaultmidpunct}
{\mcitedefaultendpunct}{\mcitedefaultseppunct}\relax
\EndOfBibitem
\bibitem[Takada(2016)]{Takada_PRB_2016}
Takada,~Y. Emergence of an excitonic collective mode in the dilute electron gas. \emph{Phys. Rev. B} \textbf{2016}, \emph{94}, 245106\relax
\mciteBstWouldAddEndPuncttrue
\mciteSetBstMidEndSepPunct{\mcitedefaultmidpunct}
{\mcitedefaultendpunct}{\mcitedefaultseppunct}\relax
\EndOfBibitem
\bibitem[Koskelo \latin{et~al.}(2025)Koskelo, Reining, and Gatti]{koskelo2023shortrange}
Koskelo,~J.; Reining,~L.; Gatti,~M. Short-range excitonic phenomena in low-density metals. \emph{Phys. Rev. Lett.} \textbf{2025}, \emph{134}, 046402\relax
\mciteBstWouldAddEndPuncttrue
\mciteSetBstMidEndSepPunct{\mcitedefaultmidpunct}
{\mcitedefaultendpunct}{\mcitedefaultseppunct}\relax
\EndOfBibitem
\bibitem[Dornheim \latin{et~al.}(2022)Dornheim, Moldabekov, Vorberger, K{\"a}hlert, and Bonitz]{Dornheim_Nature_2022}
Dornheim,~T.; Moldabekov,~Z.; Vorberger,~J.; K{\"a}hlert,~H.; Bonitz,~M. Electronic pair alignment and roton feature in the warm dense electron gas. \emph{Communications Physics} \textbf{2022}, \emph{5}, 304\relax
\mciteBstWouldAddEndPuncttrue
\mciteSetBstMidEndSepPunct{\mcitedefaultmidpunct}
{\mcitedefaultendpunct}{\mcitedefaultseppunct}\relax
\EndOfBibitem
\bibitem[Godfrin \latin{et~al.}(2012)Godfrin, Meschke, Lauter, Sultan, B{\"o}hm, Krotscheck, and Panholzer]{Godfrin2012}
Godfrin,~H.; Meschke,~M.; Lauter,~H.-J.; Sultan,~A.; B{\"o}hm,~H.~M.; Krotscheck,~E.; Panholzer,~M. Observation of a roton collective mode in a two-dimensional Fermi liquid. \emph{Nature} \textbf{2012}, \emph{483}, 576--579\relax
\mciteBstWouldAddEndPuncttrue
\mciteSetBstMidEndSepPunct{\mcitedefaultmidpunct}
{\mcitedefaultendpunct}{\mcitedefaultseppunct}\relax
\EndOfBibitem
\bibitem[Militzer and Driver(2015)Militzer, and Driver]{Militzer_PRL_2015}
Militzer,~B.; Driver,~K.~P. Development of Path Integral Monte Carlo Simulations with Localized Nodal Surfaces for Second-Row Elements. \emph{Phys. Rev. Lett.} \textbf{2015}, \emph{115}, 176403\relax
\mciteBstWouldAddEndPuncttrue
\mciteSetBstMidEndSepPunct{\mcitedefaultmidpunct}
{\mcitedefaultendpunct}{\mcitedefaultseppunct}\relax
\EndOfBibitem
\bibitem[Schoof \latin{et~al.}(2015)Schoof, Groth, Vorberger, and Bonitz]{Schoof_PRL_2015}
Schoof,~T.; Groth,~S.; Vorberger,~J.; Bonitz,~M. Ab Initio Thermodynamic Results for the Degenerate Electron Gas at Finite Temperature. \emph{Phys. Rev. Lett.} \textbf{2015}, \emph{115}, 130402\relax
\mciteBstWouldAddEndPuncttrue
\mciteSetBstMidEndSepPunct{\mcitedefaultmidpunct}
{\mcitedefaultendpunct}{\mcitedefaultseppunct}\relax
\EndOfBibitem
\bibitem[Dornheim \latin{et~al.}(2015)Dornheim, Groth, Filinov, and Bonitz]{Dornheim_NJP_2015}
Dornheim,~T.; Groth,~S.; Filinov,~A.; Bonitz,~M. Permutation blocking path integral Monte Carlo: a highly efficient approach to the simulation of strongly degenerate non-ideal fermions. \emph{New J. Phys.} \textbf{2015}, \emph{17}, 073017\relax
\mciteBstWouldAddEndPuncttrue
\mciteSetBstMidEndSepPunct{\mcitedefaultmidpunct}
{\mcitedefaultendpunct}{\mcitedefaultseppunct}\relax
\EndOfBibitem
\bibitem[Malone \latin{et~al.}(2016)Malone, Blunt, Brown, Lee, Spencer, Foulkes, and Shepherd]{Malone_PRL_2016}
Malone,~F.~D.; Blunt,~N.~S.; Brown,~E.~W.; Lee,~D. K.~K.; Spencer,~J.~S.; Foulkes,~W. M.~C.; Shepherd,~J.~J. Accurate Exchange-Correlation Energies for the Warm Dense Electron Gas. \emph{Phys. Rev. Lett.} \textbf{2016}, \emph{117}, 115701\relax
\mciteBstWouldAddEndPuncttrue
\mciteSetBstMidEndSepPunct{\mcitedefaultmidpunct}
{\mcitedefaultendpunct}{\mcitedefaultseppunct}\relax
\EndOfBibitem
\bibitem[Lee \latin{et~al.}(2021)Lee, Morales, and Malone]{Joonho_JCP_2021}
Lee,~J.; Morales,~M.~A.; Malone,~F.~D. A phaseless auxiliary-field quantum Monte Carlo perspective on the uniform electron gas at finite temperatures: Issues, observations, and benchmark study. \emph{J. Chem. Phys.} \textbf{2021}, \emph{154}, 064109\relax
\mciteBstWouldAddEndPuncttrue
\mciteSetBstMidEndSepPunct{\mcitedefaultmidpunct}
{\mcitedefaultendpunct}{\mcitedefaultseppunct}\relax
\EndOfBibitem
\bibitem[Hirshberg \latin{et~al.}(2020)Hirshberg, Invernizzi, and Parrinello]{Hirshberg_JCP_2020}
Hirshberg,~B.; Invernizzi,~M.; Parrinello,~M. Path integral molecular dynamics for fermions: Alleviating the sign problem with the Bogoliubov inequality. \emph{J. Chem. Phys.} \textbf{2020}, \emph{152}, 171102\relax
\mciteBstWouldAddEndPuncttrue
\mciteSetBstMidEndSepPunct{\mcitedefaultmidpunct}
{\mcitedefaultendpunct}{\mcitedefaultseppunct}\relax
\EndOfBibitem
\bibitem[Dornheim \latin{et~al.}(2020)Dornheim, Invernizzi, Vorberger, and Hirshberg]{Dornheim_Bogoliubov_2020}
Dornheim,~T.; Invernizzi,~M.; Vorberger,~J.; Hirshberg,~B. Attenuating the fermion sign problem in path integral Monte Carlo simulations using the Bogoliubov inequality and thermodynamic integration. \emph{J. Chem. Phys.} \textbf{2020}, \emph{153}, 234104\relax
\mciteBstWouldAddEndPuncttrue
\mciteSetBstMidEndSepPunct{\mcitedefaultmidpunct}
{\mcitedefaultendpunct}{\mcitedefaultseppunct}\relax
\EndOfBibitem
\bibitem[Yilmaz \latin{et~al.}(2020)Yilmaz, Hunger, Dornheim, Groth, and Bonitz]{Yilmaz_JCP_2020}
Yilmaz,~A.; Hunger,~K.; Dornheim,~T.; Groth,~S.; Bonitz,~M. Restricted configuration path integral Monte Carlo. \emph{J. Chem. Phys.} \textbf{2020}, \emph{153}, 124114\relax
\mciteBstWouldAddEndPuncttrue
\mciteSetBstMidEndSepPunct{\mcitedefaultmidpunct}
{\mcitedefaultendpunct}{\mcitedefaultseppunct}\relax
\EndOfBibitem
\bibitem[Filinov and Bonitz(2023)Filinov, and Bonitz]{Filinov_PRE_2023}
Filinov,~A.~V.; Bonitz,~M. Equation of state of partially ionized hydrogen and deuterium plasma revisited. \emph{Phys. Rev. E} \textbf{2023}, \emph{108}, 055212\relax
\mciteBstWouldAddEndPuncttrue
\mciteSetBstMidEndSepPunct{\mcitedefaultmidpunct}
{\mcitedefaultendpunct}{\mcitedefaultseppunct}\relax
\EndOfBibitem
\bibitem[Xiong and Xiong(2022)Xiong, and Xiong]{Xiong_JCP_2022}
Xiong,~Y.; Xiong,~H. {On the thermodynamic properties of fictitious identical particles and the application to fermion sign problem}. \emph{J. Chem. Phys.} \textbf{2022}, \emph{157}, 094112\relax
\mciteBstWouldAddEndPuncttrue
\mciteSetBstMidEndSepPunct{\mcitedefaultmidpunct}
{\mcitedefaultendpunct}{\mcitedefaultseppunct}\relax
\EndOfBibitem
\bibitem[Xiong and Xiong(2023)Xiong, and Xiong]{Xiong_PRE_2023}
Xiong,~Y.; Xiong,~H. Thermodynamics of fermions at any temperature based on parametrized partition function. \emph{Phys. Rev. E} \textbf{2023}, \emph{107}, 055308\relax
\mciteBstWouldAddEndPuncttrue
\mciteSetBstMidEndSepPunct{\mcitedefaultmidpunct}
{\mcitedefaultendpunct}{\mcitedefaultseppunct}\relax
\EndOfBibitem
\bibitem[Xiong \latin{et~al.}(2024)Xiong, Liu, and Xiong]{Xiong_PRE_2024}
Xiong,~Y.; Liu,~S.; Xiong,~H. Quadratic scaling path integral molecular dynamics for fictitious identical particles and its application to fermion systems. \emph{Phys. Rev. E} \textbf{2024}, \emph{110}, 065303\relax
\mciteBstWouldAddEndPuncttrue
\mciteSetBstMidEndSepPunct{\mcitedefaultmidpunct}
{\mcitedefaultendpunct}{\mcitedefaultseppunct}\relax
\EndOfBibitem
\bibitem[Dornheim \latin{et~al.}(2023)Dornheim, Tolias, Groth, Moldabekov, Vorberger, and Hirshberg]{Dornheim_JCP_xi_2023}
Dornheim,~T.; Tolias,~P.; Groth,~S.; Moldabekov,~Z.~A.; Vorberger,~J.; Hirshberg,~B. {Fermionic physics from ab initio path integral Monte Carlo simulations of fictitious identical particles}. \emph{J.~Chem.~Phys.} \textbf{2023}, \emph{159}, 164113\relax
\mciteBstWouldAddEndPuncttrue
\mciteSetBstMidEndSepPunct{\mcitedefaultmidpunct}
{\mcitedefaultendpunct}{\mcitedefaultseppunct}\relax
\EndOfBibitem
\bibitem[Dornheim \latin{et~al.}(2024)Dornheim, Schwalbe, Moldabekov, Vorberger, and Tolias]{Dornheim_JPCL_2024}
Dornheim,~T.; Schwalbe,~S.; Moldabekov,~Z.~A.; Vorberger,~J.; Tolias,~P. Ab Initio Path Integral {Monte Carlo} Simulations of the Uniform Electron Gas on Large Length Scales. \emph{J. Phys. Chem. Lett.} \textbf{2024}, \emph{15}, 1305--1313\relax
\mciteBstWouldAddEndPuncttrue
\mciteSetBstMidEndSepPunct{\mcitedefaultmidpunct}
{\mcitedefaultendpunct}{\mcitedefaultseppunct}\relax
\EndOfBibitem
\bibitem[Dornheim \latin{et~al.}(2024)Dornheim, Schwalbe, Böhme, Moldabekov, Vorberger, and Tolias]{Dornheim_JCP_2024}
Dornheim,~T.; Schwalbe,~S.; Böhme,~M.~P.; Moldabekov,~Z.~A.; Vorberger,~J.; Tolias,~P. {Ab initio path integral Monte Carlo simulations of warm dense two-component systems without fixed nodes: Structural properties}. \emph{J. Chem. Phys.} \textbf{2024}, \emph{160}, 164111\relax
\mciteBstWouldAddEndPuncttrue
\mciteSetBstMidEndSepPunct{\mcitedefaultmidpunct}
{\mcitedefaultendpunct}{\mcitedefaultseppunct}\relax
\EndOfBibitem
\bibitem[Dornheim \latin{et~al.}(2024)Dornheim, Döppner, Tolias, Böhme, Fletcher, Gawne, Graziani, Kraus, MacDonald, Moldabekov, Schwalbe, Gericke, and Vorberger]{Dornheim_Science_2024}
Dornheim,~T.; Döppner,~T.; Tolias,~P.; Böhme,~M.; Fletcher,~L.; Gawne,~T.; Graziani,~F.; Kraus,~D.; MacDonald,~M.; Moldabekov,~Z.; Schwalbe,~S.; Gericke,~D.; Vorberger,~J. Unraveling electronic correlations in warm dense quantum plasmas. 2024\relax
\mciteBstWouldAddEndPuncttrue
\mciteSetBstMidEndSepPunct{\mcitedefaultmidpunct}
{\mcitedefaultendpunct}{\mcitedefaultseppunct}\relax
\EndOfBibitem
\bibitem[Dornheim \latin{et~al.}(2024)Dornheim, Bellenbaum, Bethkenhagen, Hansen, Böhme, Döppner, Fletcher, Gawne, Gericke, Hamel, Kraus, MacDonald, Moldabekov, Preston, Redmer, Schörner, Schwalbe, Tolias, and Vorberger]{dornheim2024modelfreerayleighweightxray}
Dornheim,~T.; Bellenbaum,~H.~M.; Bethkenhagen,~M.; Hansen,~S.~B.; Böhme,~M.~P.; Döppner,~T.; Fletcher,~L.~B.; Gawne,~T.; Gericke,~D.~O.; Hamel,~S.; Kraus,~D.; MacDonald,~M.~J.; Moldabekov,~Z.~A.; Preston,~T.~R.; Redmer,~R.; Schörner,~M.; Schwalbe,~S.; Tolias,~P.; Vorberger,~J. Model-free Rayleigh weight from x-ray Thomson scattering measurements. 2024; \url{https://arxiv.org/abs/2409.08591}\relax
\mciteBstWouldAddEndPuncttrue
\mciteSetBstMidEndSepPunct{\mcitedefaultmidpunct}
{\mcitedefaultendpunct}{\mcitedefaultseppunct}\relax
\EndOfBibitem
\bibitem[Morresi and Garberoglio(2025)Morresi, and Garberoglio]{morresi2025normalliquid3hestudied}
Morresi,~T.; Garberoglio,~G. Normal liquid $^3$He studied by Path Integral Monte Carlo with a parametrized partition function. \emph{Phys. Rev. B} \textbf{2025}, \emph{111}, 014521\relax
\mciteBstWouldAddEndPuncttrue
\mciteSetBstMidEndSepPunct{\mcitedefaultmidpunct}
{\mcitedefaultendpunct}{\mcitedefaultseppunct}\relax
\EndOfBibitem
\bibitem[Dornheim \latin{et~al.}(2018)Dornheim, Groth, and Bonitz]{review}
Dornheim,~T.; Groth,~S.; Bonitz,~M. The uniform electron gas at warm dense matter conditions. \emph{Phys. Rep.} \textbf{2018}, \emph{744}, 1--86\relax
\mciteBstWouldAddEndPuncttrue
\mciteSetBstMidEndSepPunct{\mcitedefaultmidpunct}
{\mcitedefaultendpunct}{\mcitedefaultseppunct}\relax
\EndOfBibitem
\bibitem[Groth \latin{et~al.}(2017)Groth, Dornheim, Sjostrom, Malone, Foulkes, and Bonitz]{groth_prl}
Groth,~S.; Dornheim,~T.; Sjostrom,~T.; Malone,~F.~D.; Foulkes,~W. M.~C.; Bonitz,~M. Ab initio Exchange--Correlation Free Energy of the Uniform Electron Gas at Warm Dense Matter Conditions. \emph{Phys. Rev. Lett.} \textbf{2017}, \emph{119}, 135001\relax
\mciteBstWouldAddEndPuncttrue
\mciteSetBstMidEndSepPunct{\mcitedefaultmidpunct}
{\mcitedefaultendpunct}{\mcitedefaultseppunct}\relax
\EndOfBibitem
\bibitem[Karasiev \latin{et~al.}(2014)Karasiev, Sjostrom, Dufty, and Trickey]{ksdt}
Karasiev,~V.~V.; Sjostrom,~T.; Dufty,~J.; Trickey,~S.~B. Accurate Homogeneous Electron Gas Exchange-Correlation Free Energy for Local Spin-Density Calculations. \emph{Phys. Rev. Lett.} \textbf{2014}, \emph{112}, 076403\relax
\mciteBstWouldAddEndPuncttrue
\mciteSetBstMidEndSepPunct{\mcitedefaultmidpunct}
{\mcitedefaultendpunct}{\mcitedefaultseppunct}\relax
\EndOfBibitem
\bibitem[Karasiev \latin{et~al.}(2019)Karasiev, Trickey, and Dufty]{status}
Karasiev,~V.~V.; Trickey,~S.~B.; Dufty,~J.~W. Status of free-energy representations for the homogeneous electron gas. \emph{Phys. Rev. B} \textbf{2019}, \emph{99}, 195134\relax
\mciteBstWouldAddEndPuncttrue
\mciteSetBstMidEndSepPunct{\mcitedefaultmidpunct}
{\mcitedefaultendpunct}{\mcitedefaultseppunct}\relax
\EndOfBibitem
\bibitem[Graziani \latin{et~al.}(2014)Graziani, Desjarlais, Redmer, and Trickey]{wdm_book}
Graziani,~F., Desjarlais,~M.~P., Redmer,~R., Trickey,~S.~B., Eds. \emph{Frontiers and Challenges in Warm Dense Matter}; Springer: International Publishing, 2014\relax
\mciteBstWouldAddEndPuncttrue
\mciteSetBstMidEndSepPunct{\mcitedefaultmidpunct}
{\mcitedefaultendpunct}{\mcitedefaultseppunct}\relax
\EndOfBibitem
\bibitem[Dornheim \latin{et~al.}(2022)Dornheim, Moldabekov, Vorberger, and Militzer]{Dornheim_SciRep_2022}
Dornheim,~T.; Moldabekov,~Z.~A.; Vorberger,~J.; Militzer,~B. Path integral Monte Carlo approach to the structural properties and collective excitations of liquid $^3$He without fixed nodes. \emph{Scientific Reports} \textbf{2022}, \emph{12}, 708\relax
\mciteBstWouldAddEndPuncttrue
\mciteSetBstMidEndSepPunct{\mcitedefaultmidpunct}
{\mcitedefaultendpunct}{\mcitedefaultseppunct}\relax
\EndOfBibitem
\bibitem[Ceperley(1992)]{Ceperley_PRL_1992}
Ceperley,~D.~M. Path-integral calculations of normal liquid $^{3}\mathrm{He}$. \emph{Phys. Rev. Lett.} \textbf{1992}, \emph{69}, 331--334\relax
\mciteBstWouldAddEndPuncttrue
\mciteSetBstMidEndSepPunct{\mcitedefaultmidpunct}
{\mcitedefaultendpunct}{\mcitedefaultseppunct}\relax
\EndOfBibitem
\bibitem[Chandler and Wolynes(1981)Chandler, and Wolynes]{Chandler_JCP_1981}
Chandler,~D.; Wolynes,~P.~G. Exploiting the isomorphism between quantum theory and classical statistical mechanics of polyatomic fluids. \emph{J. Chem. Phys.} \textbf{1981}, \emph{74}, 4078--4095\relax
\mciteBstWouldAddEndPuncttrue
\mciteSetBstMidEndSepPunct{\mcitedefaultmidpunct}
{\mcitedefaultendpunct}{\mcitedefaultseppunct}\relax
\EndOfBibitem
\bibitem[Dornheim \latin{et~al.}(2019)Dornheim, Groth, Filinov, and Bonitz]{Dornheim_permutation_cycles}
Dornheim,~T.; Groth,~S.; Filinov,~A.~V.; Bonitz,~M. Path integral Monte Carlo simulation of degenerate electrons: Permutation-cycle properties. \emph{J. Chem. Phys.} \textbf{2019}, \emph{151}, 014108\relax
\mciteBstWouldAddEndPuncttrue
\mciteSetBstMidEndSepPunct{\mcitedefaultmidpunct}
{\mcitedefaultendpunct}{\mcitedefaultseppunct}\relax
\EndOfBibitem
\bibitem[Dornheim \latin{et~al.}(2024)Dornheim, Böhme, and Schwalbe]{ISHTAR}
Dornheim,~T.; Böhme,~M.; Schwalbe,~S. {ISHTAR - Imaginary-time Stochastic High- performance Tool for Ab initio Research}. 2024; \url{https://doi.org/10.5281/zenodo.10497098}\relax
\mciteBstWouldAddEndPuncttrue
\mciteSetBstMidEndSepPunct{\mcitedefaultmidpunct}
{\mcitedefaultendpunct}{\mcitedefaultseppunct}\relax
\EndOfBibitem
\bibitem[rep()]{repo}
A link to a repository containing all PIMC raw data will be made available upon publication.\relax
\mciteBstWouldAddEndPunctfalse
\mciteSetBstMidEndSepPunct{\mcitedefaultmidpunct}
{}{\mcitedefaultseppunct}\relax
\EndOfBibitem
\bibitem[Zastrau \latin{et~al.}(2014)Zastrau, Sperling, Harmand, Becker, Bornath, Bredow, Dziarzhytski, Fennel, Fletcher, F{"o}rster, G{"o}de, Gregori, Hilbert, Hochhaus, Holst, Laarmann, Lee, Ma, Mithen, Mitzner, Murphy, Nakatsutsumi, Neumayer, Przystawik, Roling, Schulz, Siemer, Skruszewicz, Tiggesb{"a}umker, Toleikis, Tschentscher, White, W{"o}stmann, Zacharias, D{"o}ppner, Glenzer, and Redmer]{Zastrau}
Zastrau,~U.; Sperling,~P.; Harmand,~M.; Becker,~A.; Bornath,~T.; Bredow,~R.; Dziarzhytski,~S.; Fennel,~T.; Fletcher,~L.~B.; F{"o}rster,~E.; G{"o}de,~S.; Gregori,~G.; Hilbert,~V.; Hochhaus,~D.; Holst,~B.; Laarmann,~T.; Lee,~H.~J.; Ma,~T.; Mithen,~J.~P.; Mitzner,~R.; Murphy,~C.~D.; Nakatsutsumi,~M.; Neumayer,~P.; Przystawik,~A.; Roling,~S.; Schulz,~M.; Siemer,~B.; Skruszewicz,~S.; Tiggesb{"a}umker,~J.; Toleikis,~S.; Tschentscher,~T.; White,~T.; W{"o}stmann,~M.; Zacharias,~H.; D{"o}ppner,~T.; Glenzer,~S.~H.; Redmer,~R. Resolving ultrafast heating of dense cryogenic hydrogen. \emph{Phys. Rev. Lett} \textbf{2014}, \emph{112}, 105002\relax
\mciteBstWouldAddEndPuncttrue
\mciteSetBstMidEndSepPunct{\mcitedefaultmidpunct}
{\mcitedefaultendpunct}{\mcitedefaultseppunct}\relax
\EndOfBibitem
\bibitem[Fletcher \latin{et~al.}(2022)Fletcher, Vorberger, Schumaker, Ruyer, Goede, Galtier, Zastrau, Alves, Baalrud, Baggott, Barbrel, Chen, Döppner, Gauthier, Granados, Kim, Kraus, Lee, MacDonald, Mishra, Pelka, Ravasio, Roedel, Fry, Redmer, Fiuza, Gericke, and Glenzer]{Fletcher_Frontiers_2022}
Fletcher,~L.~B.; Vorberger,~J.; Schumaker,~W.; Ruyer,~C.; Goede,~S.; Galtier,~E.; Zastrau,~U.; Alves,~E.~P.; Baalrud,~S.~D.; Baggott,~R.~A.; Barbrel,~B.; Chen,~Z.; Döppner,~T.; Gauthier,~M.; Granados,~E.; Kim,~J.~B.; Kraus,~D.; Lee,~H.~J.; MacDonald,~M.~J.; Mishra,~R.; Pelka,~A.; Ravasio,~A.; Roedel,~C.; Fry,~A.~R.; Redmer,~R.; Fiuza,~F.; Gericke,~D.~O.; Glenzer,~S.~H. Electron-Ion Temperature Relaxation in Warm Dense Hydrogen Observed With Picosecond Resolved X-Ray Scattering. \emph{Frontiers in Physics} \textbf{2022}, \emph{10}\relax
\mciteBstWouldAddEndPuncttrue
\mciteSetBstMidEndSepPunct{\mcitedefaultmidpunct}
{\mcitedefaultendpunct}{\mcitedefaultseppunct}\relax
\EndOfBibitem
\bibitem[Hamann \latin{et~al.}(2023)Hamann, Kordts, Filinov, Bonitz, Dornheim, and Vorberger]{Hamann_PRR_2023}
Hamann,~P.; Kordts,~L.; Filinov,~A.; Bonitz,~M.; Dornheim,~T.; Vorberger,~J. Prediction of a roton-type feature in warm dense hydrogen. \emph{Phys. Rev. Res.} \textbf{2023}, \emph{5}, 033039\relax
\mciteBstWouldAddEndPuncttrue
\mciteSetBstMidEndSepPunct{\mcitedefaultmidpunct}
{\mcitedefaultendpunct}{\mcitedefaultseppunct}\relax
\EndOfBibitem
\bibitem[Moses \latin{et~al.}(2009)Moses, Boyd, Remington, Keane, and Al-Ayat]{Moses_NIF}
Moses,~E.~I.; Boyd,~R.~N.; Remington,~B.~A.; Keane,~C.~J.; Al-Ayat,~R. The National Ignition Facility: Ushering in a new age for high energy density science. \emph{Phys. Plasmas} \textbf{2009}, \emph{16}, 041006\relax
\mciteBstWouldAddEndPuncttrue
\mciteSetBstMidEndSepPunct{\mcitedefaultmidpunct}
{\mcitedefaultendpunct}{\mcitedefaultseppunct}\relax
\EndOfBibitem
\bibitem[D{\"o}ppner \latin{et~al.}(2023)D{\"o}ppner, Bethkenhagen, Kraus, Neumayer, Chapman, Bachmann, Baggott, B{\"o}hme, Divol, Falcone, Fletcher, Landen, MacDonald, Saunders, Sch{\"o}rner, Sterne, Vorberger, Witte, Yi, Redmer, Glenzer, and Gericke]{Tilo_Nature_2023}
D{\"o}ppner,~T.; Bethkenhagen,~M.; Kraus,~D.; Neumayer,~P.; Chapman,~D.~A.; Bachmann,~B.; Baggott,~R.~A.; B{\"o}hme,~M.~P.; Divol,~L.; Falcone,~R.~W.; Fletcher,~L.~B.; Landen,~O.~L.; MacDonald,~M.~J.; Saunders,~A.~M.; Sch{\"o}rner,~M.; Sterne,~P.~A.; Vorberger,~J.; Witte,~B. B.~L.; Yi,~A.; Redmer,~R.; Glenzer,~S.~H.; Gericke,~D.~O. Observing the onset of pressure-driven K-shell delocalization. \emph{Nature} \textbf{2023}, \emph{618}, 270–275\relax
\mciteBstWouldAddEndPuncttrue
\mciteSetBstMidEndSepPunct{\mcitedefaultmidpunct}
{\mcitedefaultendpunct}{\mcitedefaultseppunct}\relax
\EndOfBibitem
\bibitem[Dornheim \latin{et~al.}(2020)Dornheim, Sjostrom, Tanaka, and Vorberger]{dornheim_electron_liquid}
Dornheim,~T.; Sjostrom,~T.; Tanaka,~S.; Vorberger,~J. Strongly coupled electron liquid: Ab initio path integral Monte Carlo simulations and dielectric theories. \emph{Phys. Rev. B} \textbf{2020}, \emph{101}, 045129\relax
\mciteBstWouldAddEndPuncttrue
\mciteSetBstMidEndSepPunct{\mcitedefaultmidpunct}
{\mcitedefaultendpunct}{\mcitedefaultseppunct}\relax
\EndOfBibitem
\bibitem[Mermin(1965)]{Mermin_DFT_1965}
Mermin,~N.~D. Thermal Properties of the Inhomogeneous Electron Gas. \emph{Phys. Rev.} \textbf{1965}, \emph{137}, A1441--A1443\relax
\mciteBstWouldAddEndPuncttrue
\mciteSetBstMidEndSepPunct{\mcitedefaultmidpunct}
{\mcitedefaultendpunct}{\mcitedefaultseppunct}\relax
\EndOfBibitem
\bibitem[Karasiev \latin{et~al.}(2016)Karasiev, Calderin, and Trickey]{karasiev_importance}
Karasiev,~V.~V.; Calderin,~L.; Trickey,~S.~B. Importance of finite-temperature exchange correlation for warm dense matter calculations. \emph{Phys. Rev. E} \textbf{2016}, \emph{93}, 063207\relax
\mciteBstWouldAddEndPuncttrue
\mciteSetBstMidEndSepPunct{\mcitedefaultmidpunct}
{\mcitedefaultendpunct}{\mcitedefaultseppunct}\relax
\EndOfBibitem
\bibitem[Ramakrishna \latin{et~al.}(2020)Ramakrishna, Dornheim, and Vorberger]{kushal}
Ramakrishna,~K.; Dornheim,~T.; Vorberger,~J. Influence of finite temperature exchange-correlation effects in hydrogen. \emph{Phys. Rev. B} \textbf{2020}, \emph{101}, 195129\relax
\mciteBstWouldAddEndPuncttrue
\mciteSetBstMidEndSepPunct{\mcitedefaultmidpunct}
{\mcitedefaultendpunct}{\mcitedefaultseppunct}\relax
\EndOfBibitem
\bibitem[Bonitz \latin{et~al.}(2020)Bonitz, Dornheim, Moldabekov, Zhang, Hamann, Kählert, Filinov, Ramakrishna, and Vorberger]{new_POP}
Bonitz,~M.; Dornheim,~T.; Moldabekov,~Z.~A.; Zhang,~S.; Hamann,~P.; Kählert,~H.; Filinov,~A.; Ramakrishna,~K.; Vorberger,~J. Ab initio simulation of warm dense matter. \emph{Phys. Plasmas} \textbf{2020}, \emph{27}, 042710\relax
\mciteBstWouldAddEndPuncttrue
\mciteSetBstMidEndSepPunct{\mcitedefaultmidpunct}
{\mcitedefaultendpunct}{\mcitedefaultseppunct}\relax
\EndOfBibitem
\bibitem[Sjostrom and Daligault(2014)Sjostrom, and Daligault]{Sjostrom_PRB_2014}
Sjostrom,~T.; Daligault,~J. Gradient corrections to the exchange-correlation free energy. \emph{Phys. Rev. B} \textbf{2014}, \emph{90}, 155109\relax
\mciteBstWouldAddEndPuncttrue
\mciteSetBstMidEndSepPunct{\mcitedefaultmidpunct}
{\mcitedefaultendpunct}{\mcitedefaultseppunct}\relax
\EndOfBibitem
\bibitem[Moldabekov \latin{et~al.}(2024)Moldabekov, Schwalbe, B{\"o}hme, Vorberger, Shao, Pavanello, Graziani, and Dornheim]{Moldabekov_JCTC_2024}
Moldabekov,~Z.; Schwalbe,~S.; B{\"o}hme,~M.~P.; Vorberger,~J.; Shao,~X.; Pavanello,~M.; Graziani,~F.~R.; Dornheim,~T. Bound-State Breaking and the Importance of Thermal Exchange--Correlation Effects in Warm Dense Hydrogen. \emph{J. Chem. Theory Comput.} \textbf{2024}, \emph{20}, 68--78\relax
\mciteBstWouldAddEndPuncttrue
\mciteSetBstMidEndSepPunct{\mcitedefaultmidpunct}
{\mcitedefaultendpunct}{\mcitedefaultseppunct}\relax
\EndOfBibitem
\bibitem[Kozlowski \latin{et~al.}(2023)Kozlowski, Perchak, and Burke]{kozlowski2023generalized}
Kozlowski,~J.; Perchak,~D.; Burke,~K. Generalized Gradient Approximation Made Thermal. 2023\relax
\mciteBstWouldAddEndPuncttrue
\mciteSetBstMidEndSepPunct{\mcitedefaultmidpunct}
{\mcitedefaultendpunct}{\mcitedefaultseppunct}\relax
\EndOfBibitem
\bibitem[Karasiev \latin{et~al.}(2018)Karasiev, Dufty, and Trickey]{Karasiev_PRL_2018}
Karasiev,~V.~V.; Dufty,~J.~W.; Trickey,~S.~B. Nonempirical Semilocal Free-Energy Density Functional for Matter under Extreme Conditions. \emph{Phys. Rev. Lett.} \textbf{2018}, \emph{120}, 076401\relax
\mciteBstWouldAddEndPuncttrue
\mciteSetBstMidEndSepPunct{\mcitedefaultmidpunct}
{\mcitedefaultendpunct}{\mcitedefaultseppunct}\relax
\EndOfBibitem
\bibitem[Karasiev \latin{et~al.}(2022)Karasiev, Mihaylov, and Hu]{Karasiev_PRB_2022}
Karasiev,~V.~V.; Mihaylov,~D.~I.; Hu,~S.~X. Meta-GGA exchange-correlation free energy density functional to increase the accuracy of warm dense matter simulations. \emph{Phys. Rev. B} \textbf{2022}, \emph{105}, L081109\relax
\mciteBstWouldAddEndPuncttrue
\mciteSetBstMidEndSepPunct{\mcitedefaultmidpunct}
{\mcitedefaultendpunct}{\mcitedefaultseppunct}\relax
\EndOfBibitem
\bibitem[Starrett \latin{et~al.}(2019)Starrett, Gill, Sjostrom, and Greeff]{STARRETT201950}
Starrett,~C.; Gill,~N.; Sjostrom,~T.; Greeff,~C. Wide ranging equation of state with Tartarus: A hybrid Green’s function/orbital based average atom code. \emph{Comput. Phys. Commun.} \textbf{2019}, \emph{235}, 50--62\relax
\mciteBstWouldAddEndPuncttrue
\mciteSetBstMidEndSepPunct{\mcitedefaultmidpunct}
{\mcitedefaultendpunct}{\mcitedefaultseppunct}\relax
\EndOfBibitem
\bibitem[Potekhin and Chabrier(2013)Potekhin, and Chabrier]{refId0}
Potekhin,~A.~Y.; Chabrier,~G. Equation of state for magnetized Coulomb plasmas. \emph{A\&A} \textbf{2013}, \emph{550}, A43\relax
\mciteBstWouldAddEndPuncttrue
\mciteSetBstMidEndSepPunct{\mcitedefaultmidpunct}
{\mcitedefaultendpunct}{\mcitedefaultseppunct}\relax
\EndOfBibitem
\bibitem[Potekhin and Chabrier(2000)Potekhin, and Chabrier]{PhysRevE.62.8554}
Potekhin,~A.~Y.; Chabrier,~G. Equation of state of fully ionized electron-ion plasmas. II. Extension to relativistic densities and to the solid phase. \emph{Phys. Rev. E} \textbf{2000}, \emph{62}, 8554--8563\relax
\mciteBstWouldAddEndPuncttrue
\mciteSetBstMidEndSepPunct{\mcitedefaultmidpunct}
{\mcitedefaultendpunct}{\mcitedefaultseppunct}\relax
\EndOfBibitem
\bibitem[Giuliani and Vignale(2008)Giuliani, and Vignale]{quantum_theory}
Giuliani,~G.; Vignale,~G. \emph{Quantum Theory of the Electron Liquid}; Cambridge University Press: Cambridge, 2008\relax
\mciteBstWouldAddEndPuncttrue
\mciteSetBstMidEndSepPunct{\mcitedefaultmidpunct}
{\mcitedefaultendpunct}{\mcitedefaultseppunct}\relax
\EndOfBibitem
\bibitem[Hou \latin{et~al.}(2022)Hou, Wang, Haule, Deng, and Chen]{Hou_PRB_2022}
Hou,~P.-C.; Wang,~B.-Z.; Haule,~K.; Deng,~Y.; Chen,~K. Exchange-correlation effect in the charge response of a warm dense electron gas. \emph{Phys. Rev. B} \textbf{2022}, \emph{106}, L081126\relax
\mciteBstWouldAddEndPuncttrue
\mciteSetBstMidEndSepPunct{\mcitedefaultmidpunct}
{\mcitedefaultendpunct}{\mcitedefaultseppunct}\relax
\EndOfBibitem
\bibitem[Dornheim \latin{et~al.}(2024)Dornheim, Tolias, Moldabekov, and Vorberger]{dornheim2024shortwavelengthlimitdynamic}
Dornheim,~T.; Tolias,~P.; Moldabekov,~Z.; Vorberger,~J. Short wavelength limit of the dynamic {Matsubara} local field correction. 2024; \url{https://arxiv.org/abs/2408.04669}\relax
\mciteBstWouldAddEndPuncttrue
\mciteSetBstMidEndSepPunct{\mcitedefaultmidpunct}
{\mcitedefaultendpunct}{\mcitedefaultseppunct}\relax
\EndOfBibitem
\bibitem[Dornheim \latin{et~al.}(2024)Dornheim, Tolias, Kalkavouras, Moldabekov, and Vorberger]{Dornheim_PRB_2024}
Dornheim,~T.; Tolias,~P.; Kalkavouras,~F.; Moldabekov,~Z.~A.; Vorberger,~J. Dynamic exchange correlation effects in the strongly coupled electron liquid. \emph{Phys. Rev. B} \textbf{2024}, \emph{110}, 075137\relax
\mciteBstWouldAddEndPuncttrue
\mciteSetBstMidEndSepPunct{\mcitedefaultmidpunct}
{\mcitedefaultendpunct}{\mcitedefaultseppunct}\relax
\EndOfBibitem
\bibitem[Fraser \latin{et~al.}(1996)Fraser, Foulkes, Rajagopal, Needs, Kenny, and Williamson]{Fraser_PRB_1996}
Fraser,~L.~M.; Foulkes,~W. M.~C.; Rajagopal,~G.; Needs,~R.~J.; Kenny,~S.~D.; Williamson,~A.~J. Finite-size effects and Coulomb interactions in quantum Monte Carlo calculations for homogeneous systems with periodic boundary conditions. \emph{Phys. Rev. B} \textbf{1996}, \emph{53}, 1814--1832\relax
\mciteBstWouldAddEndPuncttrue
\mciteSetBstMidEndSepPunct{\mcitedefaultmidpunct}
{\mcitedefaultendpunct}{\mcitedefaultseppunct}\relax
\EndOfBibitem
\bibitem[Chiesa \latin{et~al.}(2006)Chiesa, Ceperley, Martin, and Holzmann]{Chiesa_PRL_2006}
Chiesa,~S.; Ceperley,~D.~M.; Martin,~R.~M.; Holzmann,~M. Finite-Size Error in Many-Body Simulations with Long-Range Interactions. \emph{Phys. Rev. Lett.} \textbf{2006}, \emph{97}, 076404\relax
\mciteBstWouldAddEndPuncttrue
\mciteSetBstMidEndSepPunct{\mcitedefaultmidpunct}
{\mcitedefaultendpunct}{\mcitedefaultseppunct}\relax
\EndOfBibitem
\bibitem[Drummond \latin{et~al.}(2008)Drummond, Needs, Sorouri, and Foulkes]{Drummond_PRB_2008}
Drummond,~N.~D.; Needs,~R.~J.; Sorouri,~A.; Foulkes,~W. M.~C. Finite-size errors in continuum quantum Monte Carlo calculations. \emph{Phys. Rev. B} \textbf{2008}, \emph{78}, 125106\relax
\mciteBstWouldAddEndPuncttrue
\mciteSetBstMidEndSepPunct{\mcitedefaultmidpunct}
{\mcitedefaultendpunct}{\mcitedefaultseppunct}\relax
\EndOfBibitem
\bibitem[Dornheim and Vorberger(2021)Dornheim, and Vorberger]{Dornheim_JCP_2021}
Dornheim,~T.; Vorberger,~J. Overcoming finite-size effects in electronic structure simulations at extreme conditions. \emph{J. Chem. Phys.} \textbf{2021}, \emph{154}, 144103\relax
\mciteBstWouldAddEndPuncttrue
\mciteSetBstMidEndSepPunct{\mcitedefaultmidpunct}
{\mcitedefaultendpunct}{\mcitedefaultseppunct}\relax
\EndOfBibitem
\bibitem[ueg()]{uegpy}
See https://github.com/fdmalone/uegpy\relax
\mciteBstWouldAddEndPuncttrue
\mciteSetBstMidEndSepPunct{\mcitedefaultmidpunct}
{\mcitedefaultendpunct}{\mcitedefaultseppunct}\relax
\EndOfBibitem
\bibitem[Dornheim \latin{et~al.}(2024)Dornheim, Schwalbe, Tolias, Böhme, Moldabekov, and Vorberger]{Dornheim_MRE_2024}
Dornheim,~T.; Schwalbe,~S.; Tolias,~P.; Böhme,~M.~P.; Moldabekov,~Z.~A.; Vorberger,~J. Ab initio density response and local field factor of warm dense hydrogen. \emph{Matter Radiat. Extrem.} \textbf{2024}, \emph{9}, 057401\relax
\mciteBstWouldAddEndPuncttrue
\mciteSetBstMidEndSepPunct{\mcitedefaultmidpunct}
{\mcitedefaultendpunct}{\mcitedefaultseppunct}\relax
\EndOfBibitem
\bibitem[M\"uller \latin{et~al.}(2023)M\"uller, Christiansen, Schnabel, and Janke]{PhysRevX.13.031006}
M\"uller,~F.; Christiansen,~H.; Schnabel,~S.; Janke,~W. Fast, Hierarchical, and Adaptive Algorithm for Metropolis Monte Carlo Simulations of Long-Range Interacting Systems. \emph{Phys. Rev. X} \textbf{2023}, \emph{13}, 031006\relax
\mciteBstWouldAddEndPuncttrue
\mciteSetBstMidEndSepPunct{\mcitedefaultmidpunct}
{\mcitedefaultendpunct}{\mcitedefaultseppunct}\relax
\EndOfBibitem
\bibitem[John \latin{et~al.}(2016)John, Spura, Habershon, and K\"uhne]{PhysRevE.93.043305}
John,~C.; Spura,~T.; Habershon,~S.; K\"uhne,~T.~D. Quantum ring-polymer contraction method: Including nuclear quantum effects at no additional computational cost in comparison to ab initio molecular dynamics. \emph{Phys. Rev. E} \textbf{2016}, \emph{93}, 043305\relax
\mciteBstWouldAddEndPuncttrue
\mciteSetBstMidEndSepPunct{\mcitedefaultmidpunct}
{\mcitedefaultendpunct}{\mcitedefaultseppunct}\relax
\EndOfBibitem
\end{mcitethebibliography}


\end{document}